\documentclass[11pt,a4paper]{article}
\pdfoutput=1
 
\usepackage[nosort]{cite}

\usepackage[british]{babel} %dates in british format

\usepackage{epsfig,rotating}
\usepackage{amsmath,amssymb,amsthm}
\usepackage{amsfonts}
\usepackage{mathrsfs}
\usepackage{bbm}
\usepackage{bm}
\usepackage[normalem]{ulem}
\usepackage[all,knot]{xy}
\xyoption{arc}

\usepackage{tikz}
\usetikzlibrary{arrows}
\usetikzlibrary{patterns}
\usetikzlibrary{decorations.markings}

\usepackage{mathabx}

\usepackage{xcolor}
\usepackage[textwidth = 430 pt, textheight = 630 pt]{geometry}
\definecolor{MyDarkBlue}{rgb}{0.15,0.25,0.45}
 
\usepackage{enumerate}
 
\usepackage[linktocpage=true,hypertexnames=false]{hyperref}
\hypersetup{
colorlinks=true,
citecolor=MyDarkBlue,
linkcolor=MyDarkBlue,
urlcolor=MyDarkBlue,
pdfauthor={Branislav Jurco, Christian S\"amann, Martin Wolf},
pdftitle={Higher Groupoid Bundles, Higher Spaces, and Self-Dual Tensor Field Equations},
pdfsubject={hep-th}
breaklinks=true
}

\let\fn\footnote
\renewcommand{\footnote}[1]{\linespread{1.1}\fn{#1}\linespread{1.29}}

\makeatletter
\newcommand{\xRightarrow}[2][]{\ext@arrow 0359\Rightarrowfill@{#1}{#2}}
\renewcommand{\section}{\@startsection
{section}{1}{\z@}{-3.5ex plus -1ex minus
    -.2ex}{2.3ex plus .2ex}{\bf\mathversion{bold} }}
\renewcommand{\subsection}{\@startsection{subsection}{2}{\z@}{-3.25ex
plus -1ex minus
   -.2ex}{1.5ex plus .2ex}{\bf\mathversion{bold} }}
\renewcommand{\subsubsection}{\@startsection{subsubsection}{3}{-2.45ex}{-3.25ex
\makeatother
plus -1ex minus -.2ex}{1.5ex plus .2ex}{\it }}
\renewcommand{\thesection}{\arabic{section}}
\renewcommand{\thesubsection}{\arabic{section}.\arabic{subsection}}
\renewcommand{\@seccntformat}[1]{\@nameuse{the#1}.~~}

\renewcommand{\theequation}{\thesection.\arabic{equation}}
\makeatletter \@addtoreset{equation}{section}

\renewcommand*\l@section{\@dottedtocline{1}{0em}{2em}}
\renewcommand*\l@subsection{\@dottedtocline{2}{2em}{2.4em}}
\renewcommand*\l@subsubsection{\@dottedtocline{4}{3.8em}{3.7em}}

\renewcommand\tableofcontents{%
    \section*{\large\contentsname
        \@mkboth{%
          \MakeUppercase\contentsname}{\MakeUppercase\contentsname}}%
       {\baselineskip=15pt plus 2pt minus 1pt
    \@starttoc{toc}}%
}

\renewenvironment{thebibliography}[1]
     {\baselineskip=16pt plus 2pt minus 1pt
      \section*{\large\refname
        \@mkboth{\MakeUppercase\refname}{\MakeUppercase\refname}}%
     \list{\@biblabel{\@arabic\c@enumiv}}%
           {\settowidth\labelwidth{\@biblabel{#1}}%
            \leftmargin\labelwidth
            \advance\leftmargin\labelsep
            \@openbib@code
            \usecounter{enumiv}%
            \let\p@enumiv\@empty
            \renewcommand\theenumiv{\@arabic\c@enumiv}}%
      \sloppy
      \clubpenalty4000
      \@clubpenalty \clubpenalty
      \widowpenalty4000%
      \sfcode`\.\@m
 \catcode`\^^M=10%
}

\newcommand{\acknowledgements}{\section*{Acknowledgements}
\addcontentsline{toc}{section}{Acknowledgements}}

\newcommand{\datamanagement}{\section*{Data Management}
\addcontentsline{toc}{section}{Data Management}}

\newcommand{\appendices}{
\section*{Appendix}\label{appendices}\setcounter{subsection}{0}
\addcontentsline{toc}{section}{Appendix}
\setcounter{equation}{0}
\setcounter{thm}{0}
\makeatletter
\renewcommand{\theequation}{\Alph{subsection}.\arabic{equation}}
\renewcommand{\thesubsection}{\Alph{subsection}}
\renewcommand{\thethm}{\Alph{subsection}.\arabic{thm}}
\@addtoreset{equation}{subsection}
\@addtoreset{thm}{subsection}
\makeatother
}

\newtheorem{thm}{Theorem}[section]
\renewcommand{\thethm}{\thesection.\arabic{thm}}
\newtheorem{lemma}[thm]{Lemma}
\newtheorem{definition}[thm]{Definition}

\newtheorem{prop}[thm]{Proposition}
\newtheorem{cor}[thm]{Corollary}
\newtheorem{rem}[thm]{Remark}
\newtheorem{exam}[thm]{Example}

\def\periodb#1{\setbox0=\hbox{$#1$}#1\hskip-\wd0\hbox to\wd0{-}}

\newcommand{\unit}{\mathbbm{1}}   			% identity map/matrix
\newcommand{\id}{\mathrm{id}}   			% identity map/matrix
\newcommand{\CA}{\mathcal{A}}    			% cal-letters

\newcommand{\CB}{\mathcal{B}}
\newcommand{\CCB}{\mathscr{B}}
\newcommand{\CC}{\mathcal{C}}
\newcommand{\CCC}{\mathscr{C}}

\newcommand{\CD}{\mathcal{D}}
\newcommand{\CCD}{\mathscr{D}}

\newcommand{\CF}{\mathcal{F}}

\newcommand{\CG}{\mathcal{G}}
\newcommand{\CCG}{\mathscr{G}}

\newcommand{\CN}{\mathcal{N}}

\newcommand{\CO}{\mathcal{O}}

\newcommand{\CS}{\mathcal{S}}

\newcommand{\CCX}{\mathscr{X}}
\newcommand{\CCU}{\mathscr{U}}

\newcommand{\CCY}{\mathscr{Y}}

\newcommand{\CCZ}{\mathscr{Z}}
\newcommand{\CE}{\mathcal{E}}
\newcommand{\frg}{\mathfrak{g}}				% frak-letters
\newcommand{\frU}{\mathfrak{U}}
\newcommand{\FR}{\mathbbm{R}}     			% field of real numbers
\newcommand{\FC}{\mathbbm{C}}     			% field of complex numbers
\newcommand{\NN}{\mathbbm{N}}     			% set of natural numbers
\newcommand{\RZ}{\mathbbm{Z}}     			% ring of integers
\newcommand{\PP}{{\mathbbm{P}}}    			% complex projective plane
\newcommand{\dd}{\mathrm{d}}     			% total differential
\newcommand{\dpar}{\partial}     			% partial differential
\newcommand{\embd}{{\hookrightarrow}}     		% embedded
\newcommand{\eps}{{\varepsilon}}			% antisymmetric tensors
\newcommand{\eand}{{~~~\mbox{and}~~~}}     		% and etc. in equations
\newcommand{\ewith}{{~~~\mbox{with}~~~}}
\newcommand{\efor}{{~~~\mbox{for}~~~}}

\newcommand{\der}[1]{\frac{\dpar}{\dpar #1}}   		% partielle ableitung, 1 argument
\newcommand{\dder}[1]{\frac{\dd}{\dd #1}}   		% partielle ableitung, 1 argument
\newcommand{\sSO}{\mathsf{SO}}
\newcommand{\sLie}{\mathsf{Lie}}

\newcommand{\sG}{\mathsf{G}}

\newcommand{\sN}{\mathsf{N}}
\newcommand{\sM}{\mathsf{M}}

\newcommand{\Kan}{\mathrm{Kan}}     			% trace

\newcommand{\remark}[1]{}     				% remark
\newcommand{\myxymatrix}[1]{\vcenter{\vbox{\xymatrix{#1}}}}
\def\tyng(#1){\hbox{\tiny$\yng(#1)$}}			% small Young diagram
\def\tyoung(#1){\hbox{\tiny$\young(#1)$}}			% small Young diagram
\renewcommand{\uline}[1]{\underline{\it #1}}

\newcommand{\sft}{\mathsf{t}}
\newcommand{\sff}{\mathsf{f}}

\newcommand{\sfa}{\mathsf{a}}

\newcommand{\sfr}{\mathsf{r}}
\newcommand{\sfl}{\mathsf{l}}

\newcommand{\sfs}{\mathsf{s}}

\newcommand{\Dec}{\mathsf{Dec}_0}

\newcommand{\sfd}{\mathsf{d}}
\newcommand{\sB}{\mathsf{B}}

\newenvironment{pf}{\begin{proof}}{\end{proof}}

\newcommand{\CatCat}{\mathsf{Cat}}
\newcommand{\CatSet}{\mathsf{Set}}
\newcommand{\CatsSet}{\mathsf{sSet}}
\newcommand{\CatFun}{\mathsf{Fun}}
\newcommand{\CatTop}{\mathsf{Top}}
\newcommand{\CatDiff}{\mathsf{Mfd}}
\newcommand{\CatsDiff}{\mathsf{sMfd}}

\newcommand{\CatSurSub}{\mathsf{SurSub}}
\newcommand{\CatSMfd}{\mathsf{SMfd}}
\newcommand{\CatsSMfd}{\mathsf{sSMfd}}

\newcommand{\shom}{\mathsf{hom}}
\newcommand{\inthom}{\underline{\mathsf{hom}}}
\newcommand{\CatVect}{\mathsf{VectBun}}
\newcommand{\CatGrp}{\mathsf{Grp}}

\newcommand{\doublearrow}{\mathrel{\substack{\longrightarrow\\[-0.6ex]
                      \longrightarrow}}}
\newcommand{\triplearrow}{\mathrel{\substack{\longrightarrow\\[-0.6ex]
                      \longrightarrow \\[-0.6ex]
                      \longrightarrow}}}
\newcommand{\quadarrow}{\mathrel{\substack{\longrightarrow\\[-0.6ex]
                      \longrightarrow \\[-0.6ex]
                      \longrightarrow\\[-0.6ex]
                      \longrightarrow}}}

\begin{document}

\begin{titlepage}

\setcounter{page}{0}
\renewcommand{\thefootnote}{\fnsymbol{footnote}}

\begin{flushright}
 EMPG--16--09\\ DMUS--MP--16/06
\end{flushright}

\begin{center}

{\LARGE\textbf{\mathversion{bold}Higher Groupoid Bundles, Higher Spaces, \\[-6pt]
and Self-Dual Tensor Field Equations}\par}

\vspace{.5cm}

{\large
Branislav Jur\v co$^{a}$, Christian S\"amann$^{b}$, and Martin Wolf$^{\,c}$
\footnote{{\it E-mail addresses:\/}
\href{mailto:branislav.jurco@gmail.com}{\ttfamily branislav.jurco@gmail.com},
\href{mailto:c.saemann@hw.ac.uk}{\ttfamily c.saemann@hw.ac.uk}, 
\href{mailto:m.wolf@surrey.ac.uk}{\ttfamily m.wolf@surrey.ac.uk}
}}

\vspace{.3cm}

{\it
$^a$ 
Charles University in Prague\\
Faculty of Mathematics and Physics, Mathematical Institute\\
Prague 186 75, Czech Republic\\[.3cm]

$^b$ Maxwell Institute for Mathematical Sciences\\
Department of Mathematics,
Heriot--Watt University\\
Edinburgh EH14 4AS, United Kingdom\\[.3cm]

$^c$
Department of Mathematics,
University of Surrey\\
Guildford GU2 7XH, United Kingdom\\[.3cm]

}

\vspace{.1cm}

{\bf Abstract}
\end{center}
\vspace{-.5cm}
\begin{quote}
We develop a description of higher gauge theory with higher groupoids as gauge structure from first principles. This approach captures ordinary gauge theories and gauged sigma models as well as their categorifications on a very general class of (higher) spaces comprising presentable differentiable stacks, as e.g.\ orbifolds. We start off with a self-contained review on simplicial sets as models of $(\infty,1)$-categories. We then discuss principal bundles in terms of simplicial maps and their homotopies. We explain in detail a differentiation procedure, suggested by {\v S}evera, that maps higher groupoids to $L_\infty$-algebroids. Generalising this procedure, we define connections for higher groupoid bundles. As an application, we obtain six-dimensional superconformal field theories via a Penrose--Ward transform of higher groupoid bundles over a twistor space. This construction reduces the search for non-Abelian self-dual tensor field equations in six dimensions to a search for the appropriate (higher) gauge structure. The treatment aims to be accessible to theoretical physicists.

\vfill\noindent 24th August 2016

\end{quote}

\setcounter{footnote}{0}\renewcommand{\thefootnote}{\arabic{thefootnote}}

\end{titlepage}

\tableofcontents

\bigskip
\bigskip
\hrule
\bigskip
\bigskip

\section{Introduction and results}

In M-theory, interactions between parallel M5-branes are mediated by M2-branes ending on them. The boundaries of these M2-branes are tensionless and essentially massless strings. Therefore, supergravity decouples and it makes sense to study the isolated dynamics of these so-called self-dual strings. The physical theory describing this dynamics is a $\CN=(2,0)$ superconformal field theory in six-dimensions \cite{Witten:1995zh}, which is often simply referred to as the $(2,0)$-theory. Its field content comprises a self-dual 3-form curvature of a 2-form potential as well as five scalar fields, parametrising transverse fluctuations of the M5-brane, and four chiral fermions.

While the case of a single M5-brane is known and reasonably well understood, the case of multiple M5-branes is an open problem which has been attracting much attention over the last few years. From a mathematical perspective, the 2-form potential describing a single M5-brane is part of the connective structure of an Abelian gerbe. Following an analogous reasoning as for D-branes, it is natural to expect non-Abelian analogues of gerbes to underlie the $(2,0)$-theory for multiple M5-branes. Such gerbes are better known as principal 2-bundles with connections, and they provide a description of the kinematical data of categorified or higher gauge theory \cite{Breen:math0106083,Baez:2002jn,Aschieri:2003mw,Aschieri:2004yz,Bartels:2004aa,Baez:2004in,Jurco:2005qj,Schreiber:2008aa}, see also \cite{Baez:2010ya}.

By now, it has become clear that the language of higher gauge theory is indeed the appropriate one for describing M5-branes, see e.g.\ \cite{Fiorenza:2012tb,Fiorenza:2013nha,Fiorenza:2015gla}. A number of natural actions and equations of motion for higher gauge theory have been discussed \cite{Kotov:2010wr,Fiorenza:2011jr,Soncini:2014ara,Zucchini:2015ohw}, but these are higher analogues of Chern--Simons theory and therefore not suited for a classical description of the $(2,0)$-theory. There are reasons to believe that such a classical description does not exist, such as e.g.\ the necessary absence of continuous parameters in the $(2,0)$-theory. However, most of these reasons also apply to classical descriptions of multiple M2-branes, and such models have indeed been found \cite{Bagger:2007jr,Gustavsson:2007vu}. 

A rather natural way of constructing a classical candidate $(2,0)$-theory is to combine higher gauge theory with twistor geometry. Recall that there is a twistor space which carries a useful representation of solutions to the $\CN=3$ supersymmetric Yang--Mills equations in terms of holomorphic principal bundles \cite{Witten:1978xx,Isenberg:1978kk}. Analogously, there is a twistor space that carries a representation of Abelian self-dual 3-forms in six dimensions in terms of holomorphic gerbes, see \cite{springerlink:10.1007/BF00132253,Hughston:1986hb,Hughston:1979TN,Hughston:1987aa,Hughston:1988nz,0198535651} as well as \cite{Saemann:2011nb,Mason:2011nw,Mason:2012va}. It is now not difficult to supersymmetrically extend this twistor space and to examine an interpretation of the non-Abelian holomorphic principal 2-bundles it can support \cite{Saemann:2012uq}. The result is indeed a set of non-Abelian and superconformal field equations in six-dimensions with all the desired features such as field content, symmetries and field equations on the Abelian sector.

A drawback of these equations is that they appear to be somewhat restrictive. In particular, no solutions are known which are not given by an embedding of free Abelian configurations into the non-Abelian setting. For example, one would expect to see truly non-Abelian self-dual strings on $\FR^4$ arise, which are the higher analogues of monopoles and protected from many quantum corrections by supersymmetry. Just as for monopoles, such self-dual strings should be characterised by non-singular field configurations. Currently, the only known non-Abelian self-dual string solutions which come close \cite{Palmer:2013haa} still suffer from singularities in the potential forms.

Various approaches were followed to overcome this issue, and the most important one is certainly a generalisation of the gauge structure. Since the results of \cite{Saemann:2012uq} are developed using a categorical language, any choice of higher gauge structure directly and unambiguously leads to a corresponding candidate for the $(2,0)$-theory. Put differently, the search for a classical $(2,0)$-theory is thus reduced to a search for an appropriate higher gauge structure. 

Whilst the discussion in \cite{Saemann:2012uq} was based on non-Abelian principal 2-bundles with strict structure 2-groups, an additional categorification to principal 3-bundles with strict structure 3-groups was performed in \cite{Saemann:2013pca}. This involved a derivation of the full non-Abelian Deligne cohomology, which was later reproduced from a different perspective \cite{Wang:2013dwa}. Within this framework, the 3-form curvature is truly non-Abelian.  A full and very detailed description of principal 2-bundles with semistrict structure 2-groups and the connections they carry was then developed in \cite{Jurco:2014mva}, where also the candidate $(2,0)$-theory was derived. See also \cite{Zucchini:2011aa,Soncini:2014zra} and \cite{Schreiber:1303.4663} for different approaches to semistrict higher gauge theory.

In this paper, we aim to exhaustively address this issue once and for all by considering higher principal bundles over higher spaces whose gauge symmetries are encoded in higher Lie groupoids. The most general accessible geometric model for both higher spaces and higher groupoids seems to be that of Kan simplicial manifolds, and we will therefore use the language of Joyal's quasi-categories \cite{Joyal:2002:207-222,Lurie:0608040}. This framework is very general, and it contains, for example, the smooth 2-groups employed in the string 2-group model of \cite{Schommer-Pries:0911.2483}. These 2-groups, in turn, are equivalent to special Lie quasi-groupoids, as was shown in \cite{Zhu:0609420}. Moreover, this string group model is one of the most interesting higher gauge structure for our purposes, and an explicit formulation of the corresponding higher gauge theory has been worked out in \cite{Demessie:2016ieh}.

Higher principal bundles based on Kan simplicial sets have been discussed earlier, see e.g.\ \cite{Bakovic:2009aa,Bakovic:0902.3436} and \cite{Nikolaus:1207ab,nikolaus1207,Schreiber:2013pra}. Furthermore, higher principal bundles over a particular higher space were discussed in \cite{Ritter:2015zur}. This present paper generalises these results by allowing for arbitrary higher base spaces, arbitrary higher gauge groupoids, and higher connections. We present a rather straightforward approach to the computation of all the kinematical ingredients to corresponding higher gauge theories given by higher Deligne cocycles and coboundaries. In particular, we describe higher gauge potentials, their higher curvatures, finite gauge transformations, and the globalisation of this data by gluing.

Our approach will be substantially different from the one used e.g.\ in \cite{Sati:0801.3480,Fiorenza:2010mh}. Instead of considering a local description with infinitesimal gauge symmetries which then need to be integrated, we start from the global gauge symmetries. We then differentiate the Lie quasi-groupoid using a method proposed by \v Severa \cite{Severa:2006aa}. Within this approach, the underlying higher gauge algebra arises as moduli of descent data of certain higher principal bundles. Considering isomorphisms between such descent data, we readily glean the form of finite gauge transformations of corresponding higher connections. These finite gauge transformations can then be used to patch together local connection data to a global connection on higher principal bundles. We present thus a very direct approach to an explicit description of the higher Deligne cohomology classes describing principal Lie quasi-groupoid bundles with connection.

Finally, we demonstrate the usefulness of our formalism by constructing the field equations resulting from holomorphic principal Lie quasi-groupoid bundles over the twistor space appropriate for self-dual 3-forms in six dimensions. We find a significant generalisation of our previous results \cite{Saemann:2012uq,Saemann:2013pca,Jurco:2014mva}. In particular since we work with Lie quasi-groupoids instead of Lie quasi-groups, we have an additional matter field coupling to the usual tensor multiplet fields of a non-Abelian (2,0)-theory.

Our explicit and general treatment of higher gauge theory is now an ideal starting point for exploring physical questions arising in the context of string and M-theory. Particularly important is the search for examples of principal Lie quasi-groupoid bundles with connections that satisfy physically relevant field equations which differ significantly from the known, Abelian examples. The most important such field equations are certainly the self-dual string equation in four dimensional Euclidean space as well as the self-duality equation for a 3-form curvature in six-dimensional Minkowksi space. Such solutions would directly lead to a deeper understanding of M-theory. Moreover, their existence would have a similar effect as the discovery of non-Abelian monopole and instanton solutions in mathematics, and make accessible a whole new area within higher differential geometry.

\subsection*{Outline}

We start with a self-contained and detailed review of simplicial sets and how they can model higher Lie groupoids. We made a conscious attempt to keep our discussion accessible to interested theoretical and mathematical physicists with a very basic background knowledge in category theory. 

Once this framework is set up, we can readily introduce Lie quasi-groupoid bundles in Section \ref{sec:LieqgBundles} as simplicial maps between the \v Cech nerve of a cover of a base manifold and a Lie quasi-groupoid. We also define isomorphisms, pullbacks, and restrictions between such bundles as well as a notion of trivial Lie quasi-groupoid bundle. We work out the detailed description of such a bundle for Lie quasi-2-groupoids, before we allow for more general base spaces.

In Section \ref{sec:differentiation}, we review N$Q$-manifolds, their presheaves and relevant internal homomorphisms. These underlie a method to differentiate Lie quasi-groupoids to $L_\infty$-algebroids conceived by \v Severa \cite{Severa:2006aa}. We explain this method in much detail, presenting new proofs for the relevant theorems. In addition, we extend the differentiation procedure in two ways. Firstly, we study a notion of equivalence relations on the resulting $L_\infty$-algebroids, which can be used to derive finite gauge transformations for flat connections on Lie quasi-groupoid bundles. Secondly, we show how one can extend this method to non-flat connections using the d\'ecalage of a simplicial manifold.

Section \ref{sec:ConStr} deals with the explicit construction of connections on Lie quasi-groupoid bundles. We propose a new approach to defining local such connection data and deriving explicit finite gauge transformations, which is inspired by both \v Severa's differentiation method and the description of local kinetic data in terms of N$Q$-manifolds \cite{Bojowald:0406445,Kotov:2007nr,Gruetzmann:2014ica}, see also \cite{Sati:0801.3480,Fiorenza:2011jr}. We also describe how the local data can be glued together to a global connection on the Lie quasi-groupoid bundle and how our constructions extend to higher base spaces.

Finally, as an application of our constructions, we study the Penrose--Ward transform which maps certain holomorphic Lie quasi-groupoid bundles over a suitable twistor space to solutions of six-dimensional field equations in section \ref{sec:PenroseWard}. We find that the relevant constraint equations are superconformal and involve a non-Abelian self-dual 3-form curvature.

\subsection*{Remarks on notation and conventions}

Let us briefly comment on our notation and our conventions to make our discussion more accessible. We use capital Roman letters to denote sets, spaces, manifolds, and graded manifolds such as $X,Y,M$. Categories and groupoids are denoted by calligraphic letters such as $\CC$ and $\CG$. By $\shom_\CC(a,b)$, we mean morphisms in $\CC$ from $a$ to $b$, even though most of our notation will be right-to-left: $(f\circ g)(a)=f(g(a))$. Functors between categories as well as presheaves are denoted by capital Greek letters $\Phi$, $\Psi$, etc.\ with two exceptions: the special presheaves forming simplicial sets and simplicial manifolds are denoted by calligraphic letters $\CCX$, $\CCY$, $\ldots$ and $N$ will be the nerve functor, mapping a small category to a simplicial set. The functor category between two categories $\CC$ and $\CD$ is denoted by $\CatFun(\CC,\CD)$.

For the purpose of this paper, all our manifolds and supermanifolds are smooth (real or complex) and $\CatDiff$ and $\CatSMfd$ denote the categories of smooth manifolds and smooth supermanifolds.

\section{Simplicial manifolds and \texorpdfstring{$\infty$}{infinity}-Lie groupoids}\label{sec:SimpMan}

Let us start with a self-contained and rather detailed review of simplicial manifolds and so-called quasi-categories. The latter were introduced by Boardman and Vogt \cite{Boardman:1973:MHS} and Joyal \cite{Joyal:2002:207-222} suggested them as a suitable definition of $\infty$-categories, see also \cite{Lurie:0608040} for a comprehensive review. In modern terminology, they are geometric models for $(\infty,1)$-categories\footnote{Recall that in an $(m,n)$-category, all $k$-morphisms with $k>m$ are identities and all $k$-morphisms with $k>n$ are invertible.} and they are much more convenient to work with than many other notions of higher categories, such as bicategories, tricategories, etc. Restricting quasi-categories further, one can obtain quasi-groupoids or $(\infty,0)$-categories. These will serve as higher gauge groupoids in the higher bundles we shall construct later.

For a more extensive introduction to simplicial sets see \cite{Friedman:0809.4221} as well as the excellent text books \cite{0387984038,0226511812,Goerss:1999aa,Hovey:1999}. Further details on quasi-categories can be found e.g.\ in  \cite{Baez:9702014,Verity:0410412,Henriques:2006aa,Getzler:0404003,Zhu:0812.4150,Zhu:0801.2057,Li:2014}.

\subsection{Simplicial sets and simplicial manifolds}\label{ssec:simplicial_sets}

Let us recall the definition of simplicial sets in terms of the simplex category. Our discussion of higher Deligne cocycles later will benefit from a slight deviation of the usual convention in the literature of simplicial sets: we define the small category $[p]$ as the ordered set $\{0,1,\ldots, p\}$ together with morphisms $i\rightarrow j$ for $i\geq j$, as opposed to $i\leq j$, as more commonly used in the literature. Correspondingly, composition of morphisms will always be right-to-left, i.e.\ $(f\circ g)(x)=f(g(x))$. We keep, however, the notation $\shom(x,y)$ for the morphisms from $x$ to $y$. 

We begin with the following full subcategory of the category $\CatCat$ of small categories.
\begin{definition}
 The \uline{simplex category} or \uline{ordinal number category} $\mathbf{\Delta}$ is the category that has the categories $[p]$, $p\in \NN_0$, as its objects and functors between them (i.e.\ order-preserving maps between the underlying sets) as its morphisms. 
 \end{definition}
 
\noindent
The morphisms of $\mathbf{\Delta}$ are generated by the so-called \uline{coface maps}, denoted by $\phi^p_i$, and \uline{codegeneracy maps}, denoted by $\delta^p_i$, which are defined by
\begin{equation}\label{eq:CoFandCoD}
\begin{minipage}{11cm}
\tikzset{->-/.style={decoration={markings,mark=at position #1 with {\arrow{>}}},postaction={decorate}}}
\begin{tikzpicture}[scale=1,every node/.style={scale=1}]
   %\phi
   \draw (1,1) node {$\phi^p_i:[p-1]\ \to\ [p]$};
   \draw (-.3,0) node {$0$};
   \draw (-.3,-.5) node {$1$};
   \draw (0,-.9) node {$\vdots$};
   \draw (2,-.9) node {$\vdots$};
   \draw (-.6,-1.5) node {$i-1$};
   \draw (-.3,-2) node {$i$};
   \draw (0,-2.4) node {$\vdots$};
   \draw (2,-2.9) node {$\vdots$};
   \draw (-.6,-3) node {$p-1$};
   \draw (2.3,0) node {$0$};
   \draw (2.3,-.5) node {$1$};
   \draw (2.6,-1.5) node {$i-1$};
   \draw (2.28,-2) node {$i$};
   \draw (2.6,-2.5) node {$i+1$};
   \draw (2.28,-3.5) node {$p$};
   \draw[->-=.5] (0,0) -- (2,0);   
   \draw[->-=.5] (0,-.5) -- (2,-.5);   
   \draw[->-=.5] (0,-1.5) -- (2,-1.5);   
   \draw[->-=.5] (0,-2) -- (2,-2.5);   
   \draw[->-=.5] (0,-3) -- (2,-3.5);   
   \filldraw [black] (0,0) circle (1.5pt);
   \filldraw [black] (0,-.5) circle (1.5pt);
   \filldraw [black] (0,-1.5) circle (1.5pt);
   \filldraw [black] (0,-2) circle (1.5pt);
   \filldraw [black] (0,-3) circle (1.5pt);
   \filldraw [black] (2,0) circle (1.5pt);
   \filldraw [black] (2,-.5) circle (1.5pt);
   \filldraw [black] (2,-1.5) circle (1.5pt);
   \filldraw [black] (2,-2) circle (1.5pt);
   \filldraw [black] (2,-2.5) circle (1.5pt);
   \filldraw [black] (2,-3.5) circle (1.5pt);
   %delta
   \draw (7,1) node {$\delta^p_i:[p+1]\ \to\ [p]$};
   \draw (5.7,0) node {$0$};
   \draw (5.7,-.5) node {$1$};
   \draw (5.7,-1.5) node {$i$};
   \draw (5.4,-2) node {$i+1$};
   \draw (5.4,-2.5) node {$i+2$};
   \draw (5.4,-3.5) node {$p+1$};
   \draw (8.3,0) node {$0$};
   \draw (8.3,-.5) node {$1$};
   \draw (8.25,-1.5) node {$i$};
   \draw (8.6,-2) node {$i+1$};
   \draw (8.3,-3) node {$p$};
   \draw (6,-.9) node {$\vdots$};
   \draw (6,-2.9) node {$\vdots$};
   \draw (8,-.9) node {$\vdots$};
   \draw (8,-2.4) node {$\vdots$};
   \draw[->-=.5] (6,0) -- (8,0);   
   \draw[->-=.5] (6,-.5) -- (8,-.5);   
   \draw[->-=.5] (6,-1.5) -- (8,-1.5);   
   \draw[->-=.5] (6,-2) -- (8,-1.5);   
   \draw[->-=.5] (6,-2.5) -- (8,-2);  
   \draw[->-=.5] (6,-3.5) -- (8,-3);   
   \filldraw [black] (6,0) circle (1.5pt);
   \filldraw [black] (6,-.5) circle (1.5pt);
   \filldraw [black] (6,-1.5) circle (1.5pt);
   \filldraw [black] (6,-2) circle (1.5pt);
   \filldraw [black] (6,-2.5) circle (1.5pt);
   \filldraw [black] (6,-3.5) circle (1.5pt);
   \filldraw [black] (8,0) circle (1.5pt);
   \filldraw [black] (8,-.5) circle (1.5pt);
   \filldraw [black] (8,-1.5) circle (1.5pt);
   \filldraw [black] (8,-2) circle (1.5pt);
   \filldraw [black] (8,-3) circle (1.5pt);
  \end{tikzpicture}
  \end{minipage}
\end{equation}
Specifically, any order-preserving map $f:[p]\to [q]$ can be written as
\begin{equation}
 f\ =\ \phi_{i_m}\circ\cdots\circ\phi_{i_1}\circ\delta_{j_1}\circ\cdots\circ\delta_{j_n}
\end{equation}
with $p+m-n=q$, $0\leq i_1<\cdots<i_m\leq q$, and $0\leq j_1<\cdots<j_n< p$. See Figure \ref{fig:3morph} for a few examples or e.g.~\cite[Chapter 5]{0387984038} for more.

Next, let $\CatTop$ be the category of topological spaces. The objects in $\mathbf{\Delta}$ have a geometric realisation\footnote{cf.\ \cite[Chapter 1]{Goerss:1999aa} for details} in terms of the standard topological $p$-simplex 
\begin{equation}
 |\Delta^p|\ :=\ \left\{(t_0,\ldots,t_p)\in\FR^{p+1}\,|\, \sum_{i=0}^p t_i=1~\mbox{and}~ t_i\geq 0\right\}
\end{equation}
by means of the functor $\mathbf{\Delta}\to\CatTop$ defined by $[p]\mapsto|\Delta^p|$ and 
\begin{equation}
  \Big([p]\ \overset{f}{\longrightarrow}\ [q]\Big)\ \mapsto\  \begin{pmatrix}|\Delta^p|& \overset{}{\longrightarrow}& |\Delta^q|\\ (t_0,\ldots,t_p)& \mapsto& \big(\sum_{f(i)=0}t_i,\ldots,\sum_{f(i)=q}t_i\big)\end{pmatrix}.
\end{equation}
Thus, the coface map $\phi^p_i$ induces the injection $|\Delta^p|\hookrightarrow |\Delta^{p+1}|$ given by $(t_0,\ldots,t_p)\mapsto (t_0,\ldots,t_{i-1},0, t_{i},\ldots, t_p)$, sending $|\Delta^p|$ to the $i$-th face of $|\Delta^{p+1}|$. Likewise, the codegeneracy map $\delta^p_i$ induces the projection $|\Delta^p|\rightarrow |\Delta^{p-1}|$ by $(t_p,\ldots,t_0)\mapsto (t_0,\ldots,t_{i}+t_{i+1},\ldots, t_p)$, sending $|\Delta^p|$ to $|\Delta^{p-1}|$ by collapsing together the vertices $i$ and $i+1$.

Using the simplex category $\mathbf{\Delta}$ and its opposite, $\mathbf{\Delta}^{\rm op}$, we can now define simplices in the category of sets, $\CatSet$.
\begin{definition}
 A \uline{simplicial set} $\CCX$ is a $\CatSet$-valued presheaf on $\mathbf{\Delta}$. That is, $\CCX$ is a functor $\CCX:\mathbf{\Delta}^{\rm op}\rightarrow \CatSet$. 
\end{definition}

\noindent
Unwrapping this definition, we arrive at the following statements. A simplicial set is a non-negatively graded set $\CCX=\bigcup_{p\in\NN_0} \CCX_p$ for which $\CCX_p:=\CCX([p])$ is called the set of \uline{simplicial  $p$-simplices}; the elements of $\CCX_0$ are also called the \uline{vertices} of $\CCX$. The coface and codegeneracy maps translate into the \uline{face maps}, $\sff^p_i:=\CCX(\phi^p_i):\CCX_p\rightarrow \CCX_{p-1}$, and the \uline{degeneracy maps}, $\sfd^p_i:=\CCX(\delta^p_i):\CCX_p\rightarrow \CCX_{p+1}$, respectively. They satisfy 
\begin{equation}\label{eq:axioms_simplicial_set}
 \begin{gathered}
  \sff_i\circ\sff_j\ =\ \sff_{j-1}\circ\sff_i\efor i\ <\ j~,\quad\sfd_i\circ\sfd_j\ =\ \sfd_{j+1}\circ\sfd_i\efor i\ \leq\ j~,\\
  \sff_i\circ\sfd_j\ =\ \sfd_{j-1}\circ\sff_{i}\efor i\ <\ j~,\quad \sff_i\circ\sfd_j\ =\ \sfd_j\circ\sff_{i-1}\efor i\ >\ j+1~,\\
  \sff_i\circ\sfd_i\ =\ \id\ =\ \sff_{i+1}\circ\sfd_i~,
 \end{gathered}
\end{equation}
where the domains of all the maps have to be chosen appropriately. These relations are straightforwardly obtained from similar relations for the coface and codegeneracy maps. Note that the last line of \eqref{eq:axioms_simplicial_set} implies that the face maps are surjective and the degeneracy maps are injective. We usually depict simplicial sets by writing arrows for the face maps,
\begin{equation}
 \left\{ \cdots \quadarrow \CCX_2\triplearrow \CCX_1\doublearrow \CCX_0\right\}.
\end{equation}

\begin{rem}
Any ordinary set $X$ can be identified with the simplicial set\linebreak $\big\{ \cdots \quadarrow X\triplearrow X\doublearrow X\big\}$, where all the face and degeneracy maps are identities. Such a set is called a \uline{simplicially constant simplicial set}.
\end{rem}

\vspace{20pt}
\begin{figure}[h]
\begin{center}
\begin{minipage}{10cm}
\tikzset{->-/.style={decoration={markings,mark=at position #1 with {\arrow{>}}},postaction={decorate}}}
\begin{tikzpicture}[scale=.7,every node/.style={scale=.7}]
   \draw (6.5,-1.5) node {(i)};
   \draw (-.3,0) node {$0$};
   \draw (-.3,-.5) node {$1$};
   \draw (3.3,0) node {$0$};
   \draw (3.3,-.5) node {$1$};
   \draw[->-=.5] (0,0) -- (3,0);     
   \draw[->-=.5] (0,-.5) -- (3,0);     
   \filldraw [black] (0,0) circle (2pt);
   \filldraw [black] (0,-.5) circle (2pt);
   \filldraw [black] (3,0) circle (2pt);
   \filldraw [black] (3,-.5) circle (2pt);
   \draw (4.7,0) node {$0$};
   \draw (4.7,-.5) node {$1$};
   \draw (8.3,0) node {$0$};
   \draw (8.3,-.5) node {$1$};
   \draw[->-=.5] (5,0) -- (8,0);   
   \draw[->-=.5] (5,-.5) -- (8,-.5);   
   \filldraw [black] (5,0) circle (2pt);
   \filldraw [black] (5,-.5) circle (2pt);
   \filldraw [black] (8,0) circle (2pt);
   \filldraw [black] (8,-.5) circle (2pt);
   \draw (9.7,0) node {$0$};
   \draw (9.7,-.5) node {$1$};
   \draw (13.3,0) node {$0$};
   \draw (13.3,-.5) node {$1$};
   \draw[->-=.5] (10,0) -- (13,-.5);   
   \draw[->-=.5] (10,-.5) -- (13,-.5);   
   \filldraw [black] (10,0) circle (2pt);
   \filldraw [black] (10,-.5) circle (2pt);
   \filldraw [black] (13,0) circle (2pt);
   \filldraw [black] (13,-.5) circle (2pt);
  \end{tikzpicture}
  \end{minipage}
  
\vspace{20pt}
\hspace*{-100pt}
  \begin{minipage}{10cm}
\tikzset{->-/.style={decoration={markings,mark=at position #1 with {\arrow{>}}},postaction={decorate}}}
\begin{tikzpicture}[scale=.7,every node/.style={scale=.7}]
   \draw (9,-2) node {(ii)};
   \draw (-.3,0) node {$0$};
   \draw (-.3,-.5) node {$1$};
   \draw (-.3,-1) node {$2$};
   \draw (3.3,0) node {$0$};
   \draw (3.3,-.5) node {$1$};
   \draw[->-=.5] (0,0) -- (3,0);     
   \draw[->-=.5] (0,-.5) -- (3,0);     
   \draw[->-=.5] (0,-1) -- (3,0);     
   \filldraw [black] (0,0) circle (2pt);
   \filldraw [black] (0,-.5) circle (2pt);
   \filldraw [black] (0,-1) circle (2pt);
   \filldraw [black] (3,0) circle (2pt);
   \filldraw [black] (3,-.5) circle (2pt);
   \draw (4.7,0) node {$0$};
   \draw (4.7,-.5) node {$1$};
   \draw (4.7,-1) node {$2$};
   \draw (8.3,0) node {$0$};
   \draw (8.3,-.5) node {$1$};
   \draw[->-=.5] (5,0) -- (8,0);   
   \draw[->-=.5] (5,-.5) -- (8,0);   
   \draw[->-=.5] (5,-1) -- (8,-.5);   
   \filldraw [black] (5,0) circle (2pt);
   \filldraw [black] (5,-.5) circle (2pt);
    \filldraw [black] (5,-1) circle (2pt);
   \filldraw [black] (8,0) circle (2pt);
   \filldraw [black] (8,-.5) circle (2pt);
   \draw (9.7,0) node {$0$};
   \draw (9.7,-.5) node {$1$};
   \draw (9.7,-1) node {$2$};
   \draw (13.3,0) node {$0$};
   \draw (13.3,-.5) node {$1$};
   \draw[->-=.5] (10,0) -- (13,0);   
   \draw[->-=.5] (10,-.5) -- (13,-.5);   
   \draw[->-=.5] (10,-1) -- (13,-.5);   
   \filldraw [black] (10,0) circle (2pt);
   \filldraw [black] (10,-.5) circle (2pt);
   \filldraw [black] (10,-1) circle (2pt);
   \filldraw [black] (13,0) circle (2pt);
   \filldraw [black] (13,-.5) circle (2pt);
   \draw (14.7,0) node {$0$};
   \draw (14.7,-.5) node {$1$};
   \draw (14.7,-1) node {$2$};
   \draw (18.3,0) node {$0$};
   \draw (18.3,-.5) node {$1$};
   \draw[->-=.5] (15,0) -- (18,-.5);   
   \draw[->-=.5] (15,-.5) -- (18,-.5);   
    \draw[->-=.5] (15,-1) -- (18,-.5);   
   \filldraw [black] (15,0) circle (2pt);
   \filldraw [black] (15,-.5) circle (2pt);
   \filldraw [black] (15,-1) circle (2pt);
   \filldraw [black] (18,0) circle (2pt);
   \filldraw [black] (18,-.5) circle (2pt);
  \end{tikzpicture}
  \end{minipage}
  \begin{minipage}{14cm}
  \caption{\label{fig:3morph}\small The three morphisms $(0,0)=\phi^1_1\circ\delta^0_0$, $(0,1)=\id$, and $(1,1)=\phi^1_0\circ\delta^0_0$ of  $\shom_\mathbf{\Delta}([1],[1])$ in diagram (i) and the four morphisms $(0,0,0)=\phi^1_1\circ\delta^0_0\circ\delta^1_0$, $(0,0,1)=\delta^1_0$, $(0,1,1)=\delta^1_1$, and $(1,1,1)=\phi^1_0\circ\delta^0_0\circ\delta^1_0$ of  $\shom_\mathbf{\Delta}([2],[1])$ in diagram (ii).}
  \end{minipage}
  \end{center}
  \end{figure}
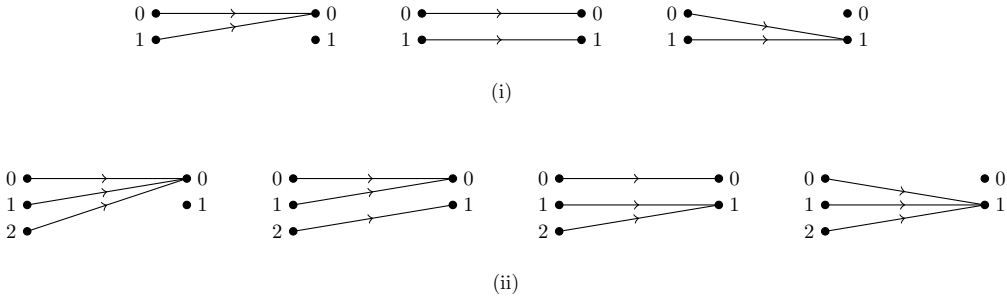

An important example of a simplicial set is given by the standard simplicial $n$-simplex which is defined as follows.

\begin{definition}
 The \uline{standard simplicial $n$-simplex}, denoted by $\Delta^n$, is the simplicial set $\shom_\mathbf{\Delta}(-,[n]):\mathbf{\Delta}^{\rm op}\rightarrow \CatSet$.
 \end{definition}
\noindent 

\begin{figure}[h]
\hspace*{60pt}
\begin{minipage}{10cm}
\tikzset{->-/.style={decoration={markings,mark=at position #1 with {\arrow{>}}},postaction={decorate}}}
\begin{tikzpicture}[scale=.7,every node/.style={scale=.7}]
   \draw (1,.5) node {$(0,0)\circ\phi^1_0$};
   \draw (-2.3,0) node {$0$};
   \draw (4.3,0) node {$0$};
   \draw (4.3,-.5) node {$1$};
   \draw[->-=.5] (-2,0) -- (1,-.5);     
   \draw[->-=.5] (1,-.5) -- (4,0);     
   \filldraw [black] (-2,0) circle (2pt);    
   \filldraw [black] (1,0) circle (2pt);
   \filldraw [black] (1,-.5) circle (2pt);
   \filldraw [black] (4,0) circle (2pt);
   \filldraw [black] (4,-.5) circle (2pt);
      \draw (9,.5) node {$(0,0)\circ\phi^1_1$};
    \draw (5.7,0) node {$0$};
   \draw (12.3,0) node {$0$};
   \draw (12.3,-.5) node {$1$};
   \draw[->-=.5] (6,0) -- (9,0);     
   \draw[->-=.5] (9,0) -- (12,0);     
   \filldraw [black] (6,0) circle (2pt);    
   \filldraw [black] (9,0) circle (2pt);
   \filldraw [black] (9,-.5) circle (2pt);
   \filldraw [black] (12,0) circle (2pt);
   \filldraw [black] (12,-.5) circle (2pt);
  \end{tikzpicture}
  \end{minipage}

\hspace*{60pt}
\begin{minipage}{10cm}
\tikzset{->-/.style={decoration={markings,mark=at position #1 with {\arrow{>}}},postaction={decorate}}}
\begin{tikzpicture}[scale=.7,every node/.style={scale=.7}]
   \draw (1,.5) node {$(0,1)\circ\phi^1_0$};
   \draw (-2.3,0) node {$0$};
   \draw (4.3,0) node {$0$};
   \draw (4.3,-.5) node {$1$};
   \draw[->-=.5] (-2,0) -- (1,-.5);     
   \draw[->-=.5] (1,-.5) -- (4,-.5);     
   \filldraw [black] (-2,0) circle (2pt);    
   \filldraw [black] (1,0) circle (2pt);
   \filldraw [black] (1,-.5) circle (2pt);
   \filldraw [black] (4,0) circle (2pt);
   \filldraw [black] (4,-.5) circle (2pt);
      \draw (9,.5) node {$(0,1)\circ\phi^1_1$};
    \draw (5.7,0) node {$0$};
   \draw (12.3,0) node {$0$};
   \draw (12.3,-.5) node {$1$};
   \draw[->-=.5] (6,0) -- (9,0);     
   \draw[->-=.5] (9,0) -- (12,0);     
   \filldraw [black] (6,0) circle (2pt);    
   \filldraw [black] (9,0) circle (2pt);
   \filldraw [black] (9,-.5) circle (2pt);
   \filldraw [black] (12,0) circle (2pt);
   \filldraw [black] (12,-.5) circle (2pt);
  \end{tikzpicture}
  \end{minipage}
 
 \hspace*{60pt} 
\begin{minipage}{10cm}
\tikzset{->-/.style={decoration={markings,mark=at position #1 with {\arrow{>}}},postaction={decorate}}}
\begin{tikzpicture}[scale=.7,every node/.style={scale=.7}]
   \draw (1,.5) node {$(1,1)\circ\phi^1_0$};
   \draw (-2.3,0) node {$0$};
   \draw (4.3,0) node {$0$};
   \draw (4.3,-.5) node {$1$};
   \draw[->-=.5] (-2,0) -- (1,-.5);     
   \draw[->-=.5] (1,-.5) -- (4,-.5);     
   \filldraw [black] (-2,0) circle (2pt);    
   \filldraw [black] (1,0) circle (2pt);
   \filldraw [black] (1,-.5) circle (2pt);
   \filldraw [black] (4,0) circle (2pt);
   \filldraw [black] (4,-.5) circle (2pt);
      \draw (9,.5) node {$(1,1)\circ\phi^1_1$};
    \draw (5.7,0) node {$0$};
   \draw (12.3,0) node {$0$};
   \draw (12.3,-.5) node {$1$};
   \draw[->-=.5] (6,0) -- (9,0);     
   \draw[->-=.5] (9,0) -- (12,-.5);     
   \filldraw [black] (6,0) circle (2pt);    
   \filldraw [black] (9,0) circle (2pt);
   \filldraw [black] (9,-.5) circle (2pt);
   \filldraw [black] (12,0) circle (2pt);
   \filldraw [black] (12,-.5) circle (2pt);
  \end{tikzpicture}
  \end{minipage}
  \begin{center}
  \begin{minipage}{14cm}
\caption{\label{fig:compphi}\small The composition of the morphisms $(0,0)$, $(0,1)$, and $(1,1)$ of $\shom_\mathbf{\Delta}([1],[1])$ with the coface maps $\phi^1_0$ and $\phi^1_1$.}
\end{minipage}
\end{center}
\end{figure}
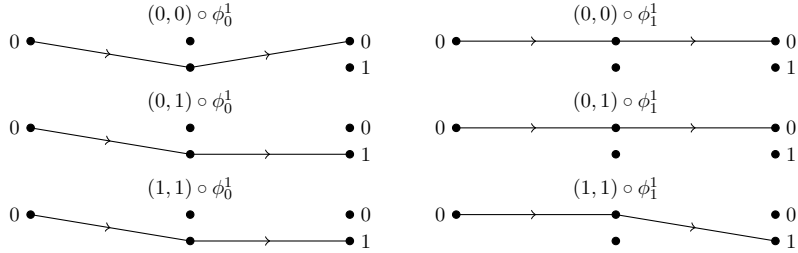

Thus, simplicial $p$-simplices in $\Delta^n$ are the elements of $\shom_\mathbf{\Delta}([p],[n])$, and we write $\Delta^n=\bigcup_{p\in\NN_0} \Delta^n_p$ with $\Delta^n_p:=\shom_\mathbf{\Delta}([p],[n])$. Furthermore, whenever such a map is injective, it is called a \uline{non-degenerate simplicial simplex}; otherwise, it is called degenerate. There is a unique non-degenerate simplicial $n$-simplex in $\Delta^n$ given by the identity at $[n]$ and all simplicial simplices for $p>n$ in $\Delta^n$ are degenerate. Note that the set $\shom_\mathbf{\Delta}([p],[n])$ has $\binom{p+n+1}{n}$ elements. We shall make use of the notation in which the elements of $\shom_\mathbf{\Delta}([p],[n])$ are labelled by their images.  For instance, $\shom_\mathbf{\Delta}([1],[1])$ contains the three morphisms $(0,0)$, $(0,1)$, and $(1,1)$ while $\shom_\mathbf{\Delta}([2],[1])$ contains the four morphisms $(0,0,0)$, $(0,0,1)$, $(0,1,1)$ and $(1,1,1)$ all of which are depicted in Figure \ref{fig:3morph}. Moreover, the face map $\sff^p_i:\Delta^n_p\to \Delta^n_{p-1}$ acts on a simplicial $p$-simplex $\sigma$ in the standard simplicial $n$-simplex $\Delta^n$ according to
\begin{subequations}
\begin{equation}
 \Big([p]\ \overset{\sigma}{\longrightarrow}\ [n]\Big)\ \mapsto\  \Big([p-1]\ \overset{\sigma\circ\phi^p_i}{\longrightarrow}\ [n]\Big)
\end{equation}
while the degeneracy maps $\sfd^p_i:\Delta^n_p\to\Delta^n_{p+1}$ act as
\begin{equation}
 \Big([p]\ \overset{\sigma}{\longrightarrow}\ [n]\Big)\ \mapsto\  \Big([p+1]\ \overset{\sigma\circ\delta^p_i}{\longrightarrow}\ [n]\Big)~.
\end{equation}
\end{subequations}
For example, when acting on the morphisms $(0,0)$, $(0,1)$ and $(1,1)$ the face maps yield $\sff^1_0(0,0)=0=\sff^1_1(0,0)$, $\sff^1_0(0,1)=1$ and $\sff^1_1(0,1)=0$, and $\sff^1_0(1,1)=1=\sff^1_1(1,1)$ as can be seen from Figure \ref{fig:compphi}.

\vspace{10pt}

Having defined simplicial sets, let us now move on to maps between them.
\begin{definition}
A \uline{simplicial map} between simplicial sets is a natural transformation between the functors  defining the simplicial sets as presheaves. 
\end{definition}
\noindent
Put differently, a simplicial map $g:\CCX\rightarrow \CCY$ between two simplicial sets $\CCX$ and $\CCY$ is a map of degree zero of graded sets $g=(g^p:\CCX_p\rightarrow \CCY_p)$, which commutes with the face and degeneracy maps,
\begin{subequations}
\begin{equation}
 g \circ\sff_i^\CCX\ =\ \sff_i^\CCY\circ g \eand g \circ \sfd_i^\CCX\ =\ \sfd_i^\CCY\circ g~,
\end{equation}
that is, the diagrams
\begin{equation}
     \myxymatrix{
    \cdots~ & \CCX_2 \ar@<.7ex>[r]\ar@<-.7ex>[r]\ar@<0ex>[r]\ar@{->}[d]_{g^2} & \CCX_1 \ar@<.3ex>[r]\ar@<-.3ex>[r]\ar@{->}[d]_{g^1} & \CCX_0\ar@{->}[d]_{g^0}\\
    \cdots & \CCY_2 \ar@<.7ex>[r]\ar@<-.7ex>[r]\ar@<0ex>[r] & \CCY_1 \ar@<.3ex>[r]\ar@<-.3ex>[r] & \CCY_0
    }~~~\eand~~~
     \myxymatrix{
    \cdots~ & \CCX_2 \ar@{->}[d]_{g^2} & \CCX_1 \ar@<.3ex>[l]\ar@<-.3ex>[l]\ar@{->}[d]_{g^1} & \CCX_0\ar@{->}[d]_{g^0} \ar@<0ex>[l]\\
    \cdots & \CCY_2  & \CCY_1 \ar@<.3ex>[l]\ar@<-.3ex>[l] & \CCY_0 \ar@<0ex>[l]
    }
\end{equation}
\end{subequations}
are commutative. If all the maps $g^p$ are embeddings (that is, injections), then $\CCX$ is called a \uline{simplicial subset} of $\CCY$. If no confusion arises, we shall suppress the dependence of the face and degeneracy maps on the simplicial sets to which they belong.

\begin{exam}\label{exam:Yoneda}
The Yoneda lemma (see e.g.~\cite[Chapter 4]{LeinsterBook} for a proof) asserts that for any $\CatSet$-valued presheaf on a category $\CC$, that is any functor $\Phi:\CC^{\rm op}\to\CatSet$, the set of natural transformations $\shom_\CC(-,c)\Rightarrow \Phi$ is bijective to the set $\Phi(c)$ for any object $c\in\CC$. By virtue of this lemma, any simplicial map $\Delta^m\to \Delta^n$, which is a natural transformation $\shom_{\mathbf{\Delta}}(-,[m])\Rightarrow \shom_{\mathbf{\Delta}}(-,[n])$ corresponds bijectively to a morphism $[m]\to[n]$ in the simplex category $\mathbf{\Delta}$.
\end{exam}

\vspace{10pt}
\begin{figure}[h]
\begin{center}
\tikzset{->-/.style={decoration={markings,mark=at position #1 with {\arrow{>}}},postaction={decorate}}}
\begin{tikzpicture}[scale=0.7,every node/.style={scale=0.7}]
  \begin{scope}[xshift = -5cm, yshift= 3cm]   
   \draw[->-=.5] (3,0) -- (0,0);
   \draw[->-=.5] (1.5,2.6) -- (0,0);   
   \draw[->-=.5] (3,0) -- (1.5,2.6);
   \filldraw [black] (0,0) circle (2pt);
   \filldraw [black] (1.5,2.6) circle (2pt);
   \filldraw [black] (3,0) circle (2pt);
   \draw (-.3,0) node {$0$};
   \draw (1.5,2.95) node {$1$};
   \draw (3.3,0) node {$2$};
   \draw (1.5,-.4) node {$(0,2)$};
   \draw (.1,1.3) node {$(0,1)$};
   \draw (2.9,1.3) node {$(1,2)$};
   \draw (1.5,.7) node {$(0,1,2)$};
  \end{scope}
   %tetrahedron
   \draw (-0.2,-0.2) node {$0$};
   \draw (4.7,3.2) node {$3$};
   \draw (10.3,-0.2) node {$2$};
   \draw (5,8.3) node {$1$};
   \draw (2.1,4.4) node {$(0,1)$};
   \draw (7.9,4.4) node {$(1,2)$};
   \draw (5.6,5.5) node {$(1,3)$};
   \draw (2.2,1.9) node {$(0,3)$};
   \draw (7.8,1.9) node {$(3,2)$};
   \draw (5,-.4) node {$(0,2)$};
   \draw (6.6,3.8) node {\rotatebox{-59}{$(1,3,2)$}};
   \draw (3.4,3.8) node {\rotatebox{59}{$(0,1,3)$}};
   \draw (5,1.3) node {$(0,3,2)$};
   \draw (5.0,2.3) node {$(0,1,2)$};
   \draw (5.0,4.5) node {$(0,1,3,2)$};
   \draw[->-=.5] (5,8) -- (0,0);   
   \draw[->-=.5] (10,0) -- (0,0);   
   \draw[dashed,->-=.5] (5,3) -- (0,0);   
   \draw[dashed,->-=.5] (10,0) -- (5,3);   
   \draw[->-=.5] (10,0) -- (5,8);   
   \draw[dashed,->-=.5] (5,8) -- (5,3);
   \filldraw [black] (0,0) circle (2pt);
   \filldraw [black] (5,8) circle (2pt);
   \filldraw [black] (10,0) circle (2pt);
   \filldraw [black] (5,3) circle (2pt);
\end{tikzpicture}
\begin{minipage}{14cm}
\caption{\label{fig:2_3_simplices}\small The simplicial 2- and 3-simplices $(0,1,2)$ and $(0,1,2,3)$ and their faces.}
\end{minipage}
\end{center}
\end{figure}
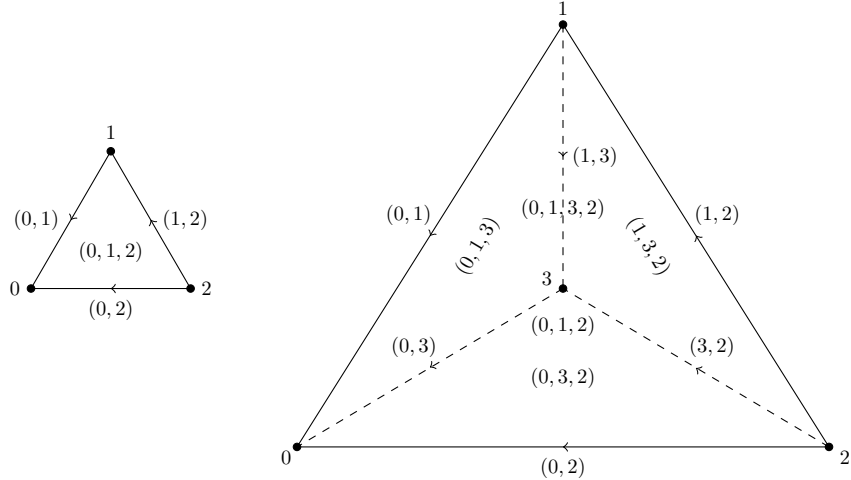

Furthermore, we shall need the following definition.

\begin{definition}
 Let $\CCX$ be a simplicial set and let $\{x_1,\ldots, x_i\}$ be a set of elements of $\CCX$. The simplicial subset of $\CCX$ \uline{generated by} $\{x_1,\ldots, x_i\}$ is the smallest simplicial subset of $\CCX$ which contains all these elements.
\end{definition}

\noindent
Particularly important examples of such simplicial subsets are the faces of $\Delta^n$.
\begin{definition}
For each $i$, the \uline{$i$-th face} of $\Delta^n$ is the simplicial subset of $\Delta^n$, denoted by $\sff_i\Delta^n$, that is generated by the image $\phi^n_i(0,1,\ldots,n)$ of the $i$-th coface map. 
\end{definition}

\noindent 
Explicitly, the simplicial map $\sff_i\Delta^n\hookrightarrow \Delta^n$ is given by the post-composition with $\phi^n_i$, that is, $\shom_\mathbf{\Delta}([p],[n-1])\ni g\mapsto\phi_i^n\circ g\in \shom_\mathbf{\Delta}([p],[n])$ so that $\sff_i\Delta^n=\bigcup_{p\in\mathbbm{N}_0}(\sff_i\Delta^n)_p$ with $(\sff_i\Delta^n)_p:=\{\phi_i^n\circ g\in \shom_\mathbf{\Delta}([p],[n])\,|\,g\in \shom_\mathbf{\Delta}([p],[n-1])\}$. It follows that $\sff_i\Delta^n\cong\Delta^{n-1}$. This is exemplified in Figure \ref{fig:2_3_simplices}, where the simplicial 2-simplex $(0,1,2)$ in $\Delta^2$ and its three faces $(0,1)$, $(0,2)$, and $(1,2)$ are depicted.

We can now combine simplicial sets and maps between them into a category.
\begin{definition} 
The \underline{category of simplicial sets} $\CatsSet$ is defined as the functor category $\CatFun(\mathbf{\Delta}^{\rm op},\CatSet)$.
\end{definition}
\noindent The Yoneda lemma, see Example \ref{exam:Yoneda}, immediately implies the following lemma.
\begin{lemma}
 Given a simplicial set $\CCX=\bigcup_{p\in\mathbbm{N}_0}\CCX_p$, we have $\CCX_p\cong\shom_\CatsSet(\Delta^p,\CCX)$.
\end{lemma}

We will usually work over simplicial sets with an underlying smooth structure. To describe these, we consider simplicial objects internal to the category $\CatDiff$, the category of smooth manifolds. 

\begin{definition}
A \uline{simplicial manifold} is a $\CatDiff$-valued presheaf on $\mathbf{\Delta}$ and we shall denote the \uline{category of simplicial manifolds} by $\CatsDiff:=\CatFun(\mathbf{\Delta}^{\rm op},\CatDiff)$.
\end{definition}

\noindent 
The simplicial simplices of a simplicial manifold are thus manifolds and face and degeneracy maps are smooth submersions and smooth immersions, respectively. 
 
The functor defining the simplicial manifold $\CCX\in \CatFun(\mathbf{\Delta}^{\rm op},\CatDiff)$ can be composed with the tangent functor $T$ to yield another simplicial manifold $T\CCX:\mathbf{\Delta}^{\rm op}\to\CatDiff$, the \uline{simplicial tangent bundle} $T\CCX$ of $\CCX$:
\begin{equation}
\mathbf{\Delta}^{\rm op}\ \overset{\CCX}{\longrightarrow}\ \CatDiff\ \overset{T}{\longrightarrow}\ \CatVect\ \longrightarrow\ \CatDiff~,
\end{equation}
where $\CatVect$ is the category of vector bundles and the last functor is the forgetful functor. 

Finally, note that by replacing $\CatDiff$ with $\CatSMfd$, the category of supermanifolds, we obtain the category of simplicial supermanifolds $\CatsSMfd$. We shall make use of this category when discussing the differentiation of Lie quasi-groupoids.

\subsection{Simplicial homotopies}

As we shall discuss now, simplicial maps between simplicial sets also carry a simplicial structure. This statement can then be iterated to simplicial homotopies, which are maps between simplicial maps.

Firstly, observe that $\CatsSet$ is a strict symmetric monoidal category. In particular, it comes with a product.
\begin{definition}
 Given two simplicial sets $\CCX$ and $\CCY$, the \uline{product} $\CCX\times \CCY$ is the simplicial set with simplicial $p$-simplices $(\CCX\times \CCY)_p:=\CCX_p\times \CCY_p$, and the face and degeneracy maps act as $\sff_i^{\CCX\times \CCY}(x_p,y_p):=(\sff_i^\CCX x_p,\sff_i^\CCY y_p)$ and $\sfd_i^{\CCX\times \CCY}(x_p,y_p):=(\sfd_i^\CCX x_p,\sfd_i^\CCY y_p)$ for all $(x_p,y_p)\in (\CCX\times \CCY)_p$.
\end{definition}

\noindent 
Given a product in a category, it is natural to try to construct its right-adjoint, known as the internal hom.
\begin{definition}
Let $\CCX$ and $\CCY$ be simplicial sets. The simplicial set $\inthom(\CCX,\CCY)$, called the \uline{internal hom}, has $\inthom_p(\CCX,\CCY):=\shom_\CatsSet(\Delta^p\times \CCX,\CCY)$ as its simplicial $p$-simplices and its  face and degeneracy maps are given by
\begin{equation}\label{eq:FaceDegHom}
\begin{gathered}
\sff^p_i\,:\, \Big(\Delta^p\times \CCX\  \overset{f}{\longrightarrow}\ \CCY\Big)\ \mapsto\ \Big(\Delta^{p-1}\times \CCX\ \overset{\phi^p_i\times\id_\CCX}{\longrightarrow}\ \Delta^p\times \CCX\  \overset{f}{\longrightarrow}\ \CCY\Big)~,\\
  \sfd^p_i\,:\,\Big(\Delta^p\times \CCX\  \overset{f}{\longrightarrow}\ \CCY\Big)\ \mapsto\ \Big(\Delta^{p+1}\times \CCX\ \overset{\delta^p_i\times\id_\CCX}{\longrightarrow}\ \Delta^p\times \CCX\  \overset{f}{\longrightarrow}\ \CCY\Big)~,
  \end{gathered}
\end{equation}
where $\phi^p_i$ and $\delta^p_i$ are the coface maps and codegeneracy maps introduced in \eqref{eq:CoFandCoD}.\footnote{Here,  we make use of the previous statement that any simplicial map $\Delta^m\to \Delta^n$ corresponds bijectively to a morphism $[m]\to[n]$ in the simplex category $\mathbf{\Delta}$ by virtue of the Yoneda lemma, cf.\ Example \ref{exam:Yoneda}.} 
\end{definition}
\noindent
We immediately see that $\inthom_0(\CCX,\CCY)$ consists of the simplicial maps between $\CCX$ and $\CCY$. Note also that by the Yoneda lemma, we have
\begin{equation}
 \shom_\CatsSet(\Delta^p\times \CCX,\CCY)\ =\ \inthom_p(\CCX,\CCY)\ \cong\ \shom_\CatsSet(\Delta^p,\inthom(\CCX,\CCY))~,
\end{equation}
which we can generalise further.

\begin{lemma}\label{lem:3SimpSets}
 Let $\CCX$, $\CCY$, and $\CCZ$ be simplicial sets. Then the internal hom is indeed the right adjoint of the product\footnote{and $\CatsSet$ is Cartesian-closed}. That is, 
 \begin{equation}\label{eq:desiredIso}
   \shom_\CatsSet(\CCX\times \CCY,\CCZ)\ \cong\ \shom_\CatsSet(\CCX,\inthom(\CCY,\CCZ))~.
 \end{equation}
\end{lemma}

\noindent
{\it Proof:} For any fixed $g\in\shom_\CatsSet(\CCX\times \CCY,\CCZ)$, the assignment
\begin{equation}\label{eq:defofsimpmapfromsimpmap}
\shom_\CatsSet(\Delta^p,\CCX)\cong \CCX_p\ \ni\ x_p\ \mapsto\ g\circ(x_p\times {\rm id}_\CCY)\ \in\ \inthom_p(\CCY,\CCZ)
\end{equation}
 yields maps $h^p:\CCX_p\to  \inthom_p(\CCY,\CCZ)$. Using \eqref{eq:FaceDegHom}, we see that these commute with the face maps
 \begin{equation}
 \begin{aligned}
  (\sff^p_i\circ h^p)(x_p)\ &=\ h^p(x_p)\circ(\phi^p_i\times\id_\CCY)\\
   &=\ g\circ(x_p\times {\rm id}_\CCY)\circ(\phi^p_i\times\id_\CCY)\\
   &=\ g\circ (\sff^p_i(x_p)\times\id_\CCY)\\
   &=\ (h^{p-1}\circ\sff^p_i)(x_p)~,
 \end{aligned}
 \end{equation}
and the relation $\sfd^p_i\circ h^p=h^{p+1}\circ\sfd^p_i$ follows similarly. The map $h^p$ is therefore a simplicial map and any element of $\shom_\CatsSet(\CCX\times \CCY,\CCZ)$  yields an element of $\shom_\CatsSet(\CCX,\inthom(\CCY,\CCZ))$ by the assignment \eqref{eq:defofsimpmapfromsimpmap}.  

The converse is also true, and it is rather straightforward to see that \eqref{eq:defofsimpmapfromsimpmap} gives the desired isomorphism \eqref{eq:desiredIso}. Indeed, given a simplicial map $h\in \shom_\CatsSet(\CCX,\inthom(\CCY,\CCZ))$, we have a simplicial map $h^p(x_p):\Delta^p\times \CCY\to \CCZ$ by evaluating $h^p$ at some $x_p\in \CCX_p$. In turn, $h^p(x_p)$ can be used to construct maps $g^p:\CCX_p\times \CCY_p\to \CCZ_p$ by setting 
\begin{equation}\label{eq:defofsimpmapfromsimpmap-inverse}
g^p(x_p,y_p)\ :=\ (h^p(x_p))^p(s_!{}^p_p,y_p)\efor x_p\ \in\ \CCX_p\eand y_p\ \in\ \CCY_p~,
\end{equation}
 where $s_!{}^p_p$ is the unique non-degenerate simplicial $p$-simplex in $\Delta^p$. These maps define a simplicial map $g\in\shom_\CatsSet(\CCX\times \CCY,\CCZ)$, since, for instance, from $\sff^p_i\circ h^p=h^{p-1}\circ\sff^p_i$ together with \eqref{eq:FaceDegHom}, we obtain
\begin{equation}
 (\sff^p_i\circ h^p)(x_p)\ =\ h^p(x_p)\circ(\phi^p_i\times{\rm id}_\CCY)\ =\ (h^{p-1}\circ\sff^p_i)(x_p)~.
\end{equation}
Upon applying this equation to the simplicial $(p-1)$-simplex $(\sff^p_p(s_!{}^p_p),\sff^p_i(y_p))$, we find
\begin{equation}
\begin{aligned}
(h^p(x_p))^{p-1}\circ\big((\phi^p_i\times{\rm id}_\CCY)(\sff^p_p(s_!{}^p_p),\sff^p_i(y_p))\big)\ &=\ (h^p(x_p))^{p-1}(\sff^p_i(s_!{}^p_p),\sff^p_i(y_p))\\
 &=\ ((h^p(x_p))^{p-1}\circ\sff^p_i)(s_!{}^p_p,y_p)\\
&=\ (\sff^p_i\circ (h^p(x_p))^p)(s_!{}^p_p,y_p)\\
&=\ (\sff^p_i\circ g^p)(x_p,y_p)~,\\
 ((h^{p-1}\circ\sff^p_i)(x_p))^{p-1}(\sff^p_p(s_!{}^p_p),\sff^p_i(y_p))\ &=\ (h^{p-1}(\sff^p_i(x_p)))^{p-1}(\sff^p_p(s_!{}^p_p),\sff^p_i(y_p))\\ 
 &=\ (g^{p-1}\circ\sff^p_i)(x_p,y_p)~,
 \end{aligned}
\end{equation}
that is, $\sff^p_i\circ g^p=g^{p-1}\circ\sff^p_i$. The relation $\sfd^p_i\circ g^p=g^{p+1}\circ\sfd^p_i$ follows similarly. As one may check, \eqref{eq:defofsimpmapfromsimpmap-inverse} constitutes the inverse construction to \eqref{eq:defofsimpmapfromsimpmap}. \hfill $\Box$

\begin{rem}\label{rem:ComLemHomSimp}
  Note that statements involving the internal hom $\inthom(\CCX,\CCY)$, as e.g.\ Lemma \ref{lem:3SimpSets}, are problematic in the category of simplicial manifolds, as mapping spaces are in general not manifolds. As usual, the way out is to use the category of diffeological spaces. We shall be able to ignore this issue in the subsequent discussion since we shall only be interested in the internal homs $\underline{\shom}(\Delta^k,\CCX)$, which are simplicial manifolds if $\CCX$ is.
\end{rem}

Let us now discuss simplicial homotopies in slightly more detail. They are maps between simplicial maps and therefore they should be given by 
$\inthom_1(\CCX,\CCY)$.

\begin{definition}\label{def:DefSimpHomo}
 Let $\CCX$ and $\CCY$ be two simplicial sets. A \uline{simplicial homotopy} between two simplicial maps $g,\tilde g:\CCX\rightarrow \CCY$ is a simplicial map $h:\Delta^1\times \CCX\to \CCY$ that renders
 \begin{equation}\label{eq:DefSimpHomo}
    \xymatrixcolsep{5pc}
    \myxymatrix{
    \Delta^0\times \CCX\cong \CCX\ar@{->}[rd]^{g} \ar@{->}[d]_{\id\times \phi_1^1} & \\
    \Delta^1\times \CCX\ar@{->}[r]^{h} & \CCY\\
    \Delta^0\times \CCX\cong \CCX\ar@{->}[ru]^{\tilde g} \ar@{->}[u]^{\id\times \phi_0^1} & 
    }
 \end{equation}
commutative. Here, $\phi_0^1$ and $\phi_1^1$ are the coface maps introduced in \eqref{eq:CoFandCoD}.
\end{definition}
\noindent Note that this definition makes sense since simplicial $p$-simplices in $\Delta^n$ are given by elements in $\shom_\mathbf{\Delta}([p],[n])$ which may be post-composed with $\phi^{n+1}_i$ to obtain elements in $\shom_\mathbf{\Delta}([p],[n+1])$.\footnote{Effectively, this is a version of the previous statement  that any simplicial map $\Delta^m\to \Delta^n$ corresponds bijectively to a morphism $[m]\to[n]$ in the simplex category $\mathbf{\Delta}$. See Example \ref{exam:Yoneda}.} For $n=0$, this leaves the two possibilities $\phi_0^1$ and $\phi_1^1$ so that the morphism $[p]\to [0]$ yields morphisms $[p]\to [1]$ by either post-composing with $\phi_0^1$ or  $\phi_1^1$, respectively.\footnote{Note that the simplicial $p$-simplex $\Delta^0_p$ in $\Delta^0=\bigcup_{p\in \mathbbm{N}_0}\Delta^0_p$ contains only one single morphism $[p]\to[0]$ given by $0\mapsto 0$, $1\to0$, $\ldots$, $p\mapsto 0$.}  Furthermore, the commutativity of the diagram ensures that $h^p(x_p,(0,\ldots,0))=g^p(x_p)$ and $h^p(x_p,(1,\ldots,1))=\tilde g^p(x_p)$ for all simplicial $p$-simplices $x_p\in \CCX_p$ for $p\geq0$. For more details, we refer to \cite[Chapter 1]{Goerss:1999aa}. 

We can now generalise this to higher homotopies.

\begin{definition}\label{def:simplicial_k_homotopy}
A \uline{simplicial $k$-homotopy} is an element of $\inthom_k(\CCX,\CCY)$.
\end{definition}

An equivalent description of simplicial homotopies that will be useful to us later is the following.

\pagebreak
\begin{lemma}\label{lem:HomMap}
 Let $\CCX$ and $\CCY$ be simplicial sets. A simplicial map $h:\CCX\to\inthom(\Delta^1,\CCY)$ is of the form $h^p:\CCX_p\to \inthom_p(\Delta^1,\CCY)$ with  $h^p=(h^p_0,\ldots,h^p_p)$ so that $h^p_i:\CCX_p\to \CCY_{p+1}$ for $0\leq i\leq p$ and
 \begin{subequations}\label{eq:HomMap}
 \begin{equation}\label{eq:HomMap-2}
 \begin{gathered}
  \sff_i\circ h_j\ =\ h_{j-1}\circ \sff_i\efor i\ <\ j~,\quad
  \sfd_i\circ h_j\ =\ h_{j+1}\circ \sfd_i\efor i\ \leq\ j~,\\
  \sfd_i\circ h_j\ =\ h_j\circ \sfd_{i-1}\efor i\ >\ j~,\quad
  \sff_i\circ h_j\ =\ h_j\circ\sff_{i-1}\efor i\ >\ j+1~,\\
  \sff_i \circ h_i\ =\ \sff_i\circ h_{i-1}~,
 \end{gathered}
 \end{equation}
 where the domains are appropriately chosen. The simplicial map $h$ describes a simplicial homotopy between the simplicial maps 
 \begin{equation}\label{eq:HomMap-1}
  \sff_0\circ h_0\,:\,\CCX\ \to\  \CCY\eand \sff_{p+1}\circ h_p\,:\,\CCX\ \to\ \CCY~.
 \end{equation}
  \end{subequations}
\end{lemma}

\noindent
{\it Proof:} Firstly, for $p=0$, we have $h^0:\CCX_0\to \inthom_0(\Delta^1,\CCY)$. Since $\inthom_0(\Delta^1,\CCY)=\shom_{\CatsSet}(\Delta^0\times\Delta^1,\CCY)\cong\shom_{\CatsSet}(\Delta^1,\CCY)\cong \CCY_1$, we thus obtain a map $h^0_0:\CCX_0\to \CCY_1$ and so $\sff^1_0\circ h^0_0:\CCX_0\to \CCY_0$. Next, we note that the simplicial set $\Delta^p\times\Delta^1$ is characterised by its $p+1$ non-degenerate simplicial $(p+1)$-simplices\footnote{Note that  all simplicial $q$-simplices for $q>p+1$ in $\Delta^p\times\Delta^1$ are degenerate and thus can be obtained by applying appropriate degeneracy maps to the non-degenerate simplicial $(p+1)$-simplices.} 
\begin{equation}\label{eq:NonDegSimDeltap1}
\begin{aligned}
  s_{0}^p\ &:=\ ((0,0,1,2,\ldots,p),(0,1,1,\ldots,1))~,\\
  s_{1}^p\ &:=\ ((0,1,1,2,\ldots,p),(0,0,1,\ldots,1))~,\\
    &~~ \vdots\\
  s_{p}^p\ &:=\ ((0,1,2,\ldots,p,p),(0,0,\ldots,0,1))~.\\
\end{aligned}
\end{equation}
 Thus, any simplicial map $\Delta^p\times\Delta^1\to \CCY$ is fixed by assigning these simplicial simplices to simplicial $(p+1)$-simplices in $\CCY$. Consequently, a map $h^p:\CCX_p\to \inthom_p(\Delta^1,\CCY)=\shom_{\CatsSet}(\Delta^p\times\Delta^1,\CCY)$ is determined by $p+1$ maps $h^p_i:\CCX_p\to \CCY_{p+1}$ which are given by 
 \begin{equation}
 h^p_i(-)\ :=\ h^p(-)(s_{i}^p)~.
 \end{equation}
  Hence, $\sff^{p+1}_0\circ h^p_0:\CCX_p\to \CCY_p$ and $\sff^{p+1}_{p+1}\circ h^p_p:\CCX_p\to \CCY_p$ and so we have obtained the desired maps \eqref{eq:HomMap-1}. 
 
Now since $h:\CCX\to \inthom(\Delta^1,\CCY)$ is a simplicial map, it commutes with the face and degeneracy maps on $\inthom(\Delta^1,\CCY)$ as given in \eqref{eq:FaceDegHom}:
 \begin{equation}
 \begin{gathered}
   (\sff^p_i\circ h^p)(-)\ =\ h^{p}(-)\circ(\phi^p_i\times\id)\ =\ (h^{p-1}\circ \sff^p_i)(-)~,\\
   (\sfd^p_i\circ h^p)(-)\ =\ h^{p}(-)\circ(\delta^p_i\times\id)\ =\ (h^{p+1}\circ \sfd^p_i)(-)~.
   \end{gathered}
 \end{equation}
The first of these implies that
\begin{equation}
  h^{p}(-)\big((\phi^p_i\times\id)(s^{p-1}_j)\big)\ =\ h^{p-1}\big(\sff^p_i(-)\big)(s^{p-1}_j)\ =\ (h^{p-1}_j\circ\sff^p_i)(-)~.
\end{equation}
As a short calculation reveals, for $i<j$ the left-hand-side of this equation is given by
\begin{equation}
\begin{aligned}
 h^{p}(-)\big((\phi^p_i\times\id)(s^{p-1}_j)\big)\ &=\ h^{p}(-)\big(\sff_i^{p+1}(s^p_{j+1})\big)\\
 &=\ \sff^{p+1}_i\circ h^p(-)(s^p_{j+1})\ =\ (\sff^{p+1}_i\circ h^p_{j+1})(-)
\end{aligned}
\end{equation}
so that $\sff^{p+1}_i\circ h^p_{j+1}=h^{p-1}_j\circ\sff^p_i$. This establishes the first relation in \eqref{eq:HomMap-2}. The remaining relations in \eqref{eq:HomMap-2} follow in a similar fashion.

Finally, we verify that  \eqref{eq:HomMap-1} are simplicial maps, that is, they do indeed commute with the face and degeneracy maps on $\CCX$ and $\CCY$. We only check the relation $\sff^p_i\circ\sff^{p+1}_0\circ h^p_0=\sff^p_0\circ h^{p-1}_0\circ\sff^p_i$ as the others are more of the same: for $i=0$, we have
\begin{subequations}
\begin{equation}
\sff^p_0\circ h^{p-1}_0\circ\sff^p_0\ =\ \sff^p_0\circ \sff^{p+1}_0\circ h^p_1\ =\ \sff^p_0\circ\sff^{p+1}_0\circ h^p_0
\end{equation}
while for $i>0$ we find
\begin{equation}
\sff^p_0\circ h^{p-1}_0\circ\sff^p_i\ =\ \sff^p_0\circ \sff^{p+1}_{i+1}\circ h^p_0\ =\ \sff^p_i\circ\sff^{p+1}_0\circ h^p_0~,
\end{equation}
\end{subequations}
where we have used \eqref{eq:axioms_simplicial_set} and \eqref{eq:HomMap-2}. This concludes the proof. \hfill $\Box$

\subsection{Kan simplicial sets}

Particularly important simplicial sets are the horns of the standard simplicial $n$-simplex $\Delta^n$, which lead to the definition of horns of general simplicial sets.

\begin{definition}
For each $i$, the \uline{$(n,i)$-horn $\Lambda^n_i$} of $\Delta^n$ is the simplicial subset of $\Delta^n$ generated by the union of all faces of $\Delta^n$ except for the $i$-th one. The horns $\Lambda^n_i$ with $0<i<n$ are called the \uline{inner horns} of $\Delta^n$ while the horns $\Lambda^n_0$ and $\Lambda^n_n$ are called the \uline{outer horns} of $\Delta^n$. 
 \end{definition}

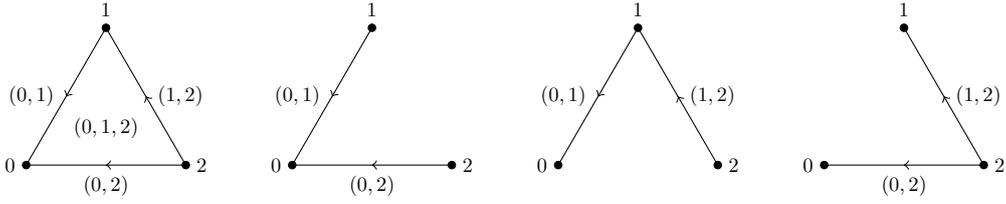
\begin{figure}[h]
\begin{center}
\tikzset{->-/.style={decoration={markings,mark=at position #1 with {\arrow{>}}},postaction={decorate}}}
\begin{tikzpicture}[scale=.7,every node/.style={scale=.7}]
   \draw[->-=.5] (3,0) -- (0,0);
   \draw[->-=.5] (1.5,2.6) -- (0,0);   
   \draw[->-=.5] (3,0) -- (1.5,2.6);
   \filldraw [black] (0,0) circle (2pt);
   \filldraw [black] (1.5,2.6) circle (2pt);
   \filldraw [black] (3,0) circle (2pt);
   \draw (-.3,0) node {$0$};
   \draw (1.5,2.95) node {$1$};
   \draw (3.3,0) node {$2$};
   \draw (1.5,-.4) node {$(0,2)$};
   \draw (.1,1.3) node {$(0,1)$};
   \draw (2.9,1.3) node {$(1,2)$};
   \draw (1.5,.7) node {$(0,1,2)$};
   \draw[->-=.5] (8,0) -- (5,0);  
   \draw[->-=.5] (6.5,2.6) -- (5,0);   
   \filldraw [black] (5,0) circle (2pt);
   \filldraw [black] (6.5,2.6) circle (2pt);
   \filldraw [black] (8,0) circle (2pt);
   \draw (4.7,0) node {$0$};
   \draw (6.5,2.95) node {$1$};
   \draw (8.3,0) node {$2$};
   \draw (6.5,-.4) node {$(0,2)$};
   \draw (5.1,1.3) node {$(0,1)$};
   \draw[->-=.5] (11.5,2.6) -- (10,0);   
   \draw[->-=.5] (13,0) -- (11.5,2.6);  
   \filldraw [black] (10,0) circle (2pt);
   \filldraw [black] (11.5,2.6) circle (2pt);
   \filldraw [black] (13,0) circle (2pt);
   \draw (9.7,0) node {$0$};
   \draw (11.5,2.95) node {$1$};
   \draw (13.3,0) node {$2$};
   \draw (10.1,1.3) node {$(0,1)$};
   \draw (12.9,1.3) node {$(1,2)$};
   \draw[->-=.5] (18,0) -- (15,0);  
   \draw[->-=.5] (18,0) -- (16.5,2.6);   
   \filldraw [black] (15,0) circle (2pt);
   \filldraw [black] (16.5,2.6) circle (2pt);
   \filldraw [black] (18,0) circle (2pt);
   \draw (14.7,0) node {$0$};
   \draw (16.5,2.95) node {$1$};
   \draw (18.3,0) node {$2$};
   \draw (16.5,-.4) node {$(0,2)$};
   \draw (17.9,1.3) node {$(1,2)$};
\end{tikzpicture}

\begin{minipage}{14cm}
\caption{\label{fig:horns}\small The simplicial 2-simplex and its horns. The simplicial 2-simplex  $(0,1,2)$ of $\Delta^2$ is depicted as a double arrow $(0,1,2)$ with its faces $(0,1)=\sff^2_2(0,1,2)$, $(1,2)=\sff^2_0(0,1,2)$, and $(0,2)=\sff^2_1(0,1,2)$. Its horns in $\Lambda^2_0$, $\Lambda^2_1$, and $\Lambda^2_2$ are depicted next to it.}
\end{minipage}
\end{center}
\vspace{10pt}
\end{figure}

\noindent
Put differently, we may say that the horn $\Lambda^n_i$ is the simplicial subset of $\Delta^n$ that is generated by the images of $\{\phi^n_0,\ldots,\phi^n_{i-1},\phi^n_{i+1},\ldots,\phi^n_n\}$. This is exemplified in Figure \ref{fig:horns} by considering the simplicial 2-simplex $(0,1,2)$ in $\Delta^2$ together with its three horns in $\Lambda_0^2$, $\Lambda_1^2$, and $\Lambda_2^2$. We generalise the notion of horns to arbitrary simplicial sets as follows.

\begin{definition}
The \uline{$(n,i)$-horns} of a simplicial set $\CCX$ are the simplicial maps $\Lambda^n_i\rightarrow \CCX$.
\end{definition}

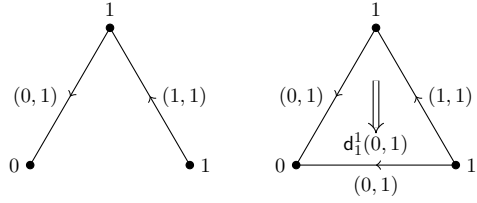
\begin{figure}[h]
\begin{center}
\tikzset{->-/.style={decoration={markings,mark=at position #1 with {\arrow{>}}},postaction={decorate}}}
\begin{tikzpicture}[scale=.7,every node/.style={scale=.7}]
   \draw[->-=.5] (1.5,2.6) -- (0,0);   
   \draw[->-=.5] (3,0) -- (1.5,2.6);
   \filldraw [black] (0,0) circle (2pt);
   \filldraw [black] (1.5,2.6) circle (2pt);
   \filldraw [black] (3,0) circle (2pt);
   \draw (-.3,0) node {$0$};
   \draw (1.5,2.95) node {$1$};
   \draw (3.3,0) node {$1$};
   \draw (.1,1.3) node {$(0,1)$};
   \draw (2.9,1.3) node {$(1,1)$};
   \draw[->-=.5] (8,0) -- (5,0);  
   \draw[->-=.5] (6.5,2.6) -- (5,0);
   \draw[->-=.5] (8,0) -- (6.5,2.6);  
   \draw [-implies,double equal sign distance] (6.5,1.6) -- (6.5,0.6);
   \filldraw [black] (5,0) circle (2pt);
   \filldraw [black] (6.5,2.6) circle (2pt);
   \filldraw [black] (8,0) circle (2pt);
   \draw (4.7,0) node {$0$};
   \draw (6.5,2.95) node {$1$};
   \draw (8.3,0) node {$1$};
   \draw (5.1,1.3) node {$(0,1)$};
   \draw (6.5,-.4) node {$(0,1)$};
   \draw (7.9,1.3) node {$(1,1)$};
   \draw (6.5,.4) node {$\sfd^1_1(0,1)$};
  \end{tikzpicture}
\begin{minipage}{14cm}
\caption{\label{fig:hornfill}\small A horn that can trivially be filled with $\sfd^1_1(0,1)=(0,1,1)$. We use a double arrow pointing at the face of the filler which was added.}
\end{minipage}
\end{center}
\end{figure}

Since the horns $\Lambda^n_i$ of $\Delta^n$ arise by removing the unique non-degenerate simplicial $n$-simplex from $\Delta^n$ as well the $i$-th non-degenerate simplicial $(n-1)$-simplex, they can again be completed to simplices. If the horns of a simplicial set can be filled, we say that certain Kan conditions are satisfied. Note that certain horns can trivially be filled, cf.\ Figure \ref{fig:hornfill}.

\pagebreak

\begin{definition}
 Let $i:\Lambda^p_i\hookrightarrow\Delta^p$ be the inclusion. A simplicial map $g:\CCX\rightarrow \CCY$ satisfies the \uline{Kan condition $\Kan(p,i)$} if for any horn $\lambda:\Lambda^p_i\to \CCX$ and any compatible simplicial simplex $\delta:\Delta^p\to \CCY$ with $\delta\circ i=g\circ \lambda$, there exists a lift $\tilde\delta :\Delta^p\to \CCX$ of $\delta:\Delta^p\to \CCY$ such that 
  \begin{equation}
    \myxymatrix{
    \Lambda^p_i\ar@{->}[r]^{\lambda} \ar@{^{(}->}[d] & \CCX \ar@{->}[d]^{g}\\
    \Delta^p \ar@{-->}[ru]^{\tilde{\delta}} \ar@{->}[r]^{\delta} & \CCY
    }
 \end{equation}
is commutative. Whenever the lifts $\tilde\delta :\Delta^p\to \CCX$ are unique, we say that $g$ satisfies the \uline{unique Kan condition $\Kan!(p,i)$}.
\end{definition}

\noindent If the Kan conditions are satisfied for a large number of horns, we obtain Kan fibrations and Kan simplicial sets.

\begin{definition}
A \uline{Kan fibration} is a simplicial map which satisfies the Kan conditions $\Kan(p,i)$ for all $p\geq 1$ and $0\leq i\leq p$. An \uline{inner Kan fibration}\footnote{In the nomenclature of Boardman \& Voigt \cite{boardman:1973aa}, these are called weak Kan fibrations.} is a simplicial map which satisfies the Kan conditions $\Kan(p,i)$ for all $p\geq 1$ and $0< i< p$ (i.e.~for inner horns).
\end{definition}

\noindent
Let $*$ be the simplicially constant simplicial set of the one-point set. 
 
\pagebreak

\begin{definition}
A \uline{(inner) Kan simplicial set}\footnote{Note that (inner) Kan simplicial sets are also known as (inner) Kan complexes.} is a simplicial set $\CCX$ such that $\CCX\rightarrow *$ is a (inner) Kan fibration.  
\end{definition}

\noindent
In particular, if $\CCX$ is a Kan simplicial set, then any horn $\lambda:\Lambda^p_i\rightarrow \CCX$ can be filled. That is, there is a simplicial map $\tilde\delta:\Delta^p\to \CCX$ such that
 \begin{equation}
    \myxymatrix{
    \Lambda^p_i\ar@{->}[r]^{\lambda} \ar@{^{(}->}[d] & \CCX \\
    \Delta^p \ar@{->}[ru]^{\tilde\delta} & 
    }
 \end{equation}
is commutative. We may rephrase this horn filling property by saying that the natural restrictions 
\begin{equation}\label{eq:HornFillProp}
\CCX_p\ \cong\ \shom_\CatsSet(\Delta^p,\CCX)\ \to\ \shom_\CatsSet(\Lambda^p_i,\CCX)
\end{equation}
are surjective for all $p\geq1$ and $0\leq i\leq p$.

\begin{definition}
A \uline{Kan simplicial manifold} is a simplicial manifold so that all the restrictions \eqref{eq:HornFillProp} are surjective submersions.
\end{definition}

\begin{exam}
The simplicial set $\Delta^1$ is not a Kan simplicial set (in fact none of the $\Delta^n$ for $n\geq 1$ is) since it does not satisfy the Kan condition ${\rm Kan}(2,0)$, for instance. Specifically, define a horn  $\lambda:\Lambda^2_2\to \Delta^1$ (see also Figure \ref{fig:horns}) by setting $\lambda:(1,2)\mapsto (1,2)$ and $\lambda:(0,2)\mapsto (2,2)$. This fixes $\lambda$ completely since from $\lambda\circ \sff^1_i=\sff^1_i\circ\lambda$ we deduce that $\lambda:0\mapsto 2$, $\lambda:1\mapsto 1$, and $\lambda:2\mapsto 2 $, respectively. There is, however, no $\delta:\Delta^2\to\Delta^1$ since $\lambda:1\mapsto 1$ and $\lambda:0\mapsto 2$ would imply $\delta: (0,1)\mapsto(2,1)$ and $(2,1)\not\in\Delta^1$.
\end{exam}

\begin{exam}
Let $S:\CatTop\to\CatsSet$ be the singular functor that sends a topological space $X$ to the simplicial set $S(X)=\bigcup_{p\in\mathbbm{N}_0} \shom_\CatTop(|\Delta^p|,X)$, called its singular set. The singular set is a Kan simplicial set (see {e.g.}~\cite[Chapter 1]{Goerss:1999aa} for details).
\end{exam}

\begin{exam}\label{exam:SimpGroup}
Let $\CatGrp$ be the category of groups. A \uline{simplicial group} is a functor $\mathbf{\Delta}^{\rm op}\to\CatGrp$. Any simplicial group is a Kan simplicial set due to a result of Moore \cite{Moore1954}. Note that any ordinary group $\sG$ can be regarded as a simplicially constant simplicial group, that is, the simplicial group which has a copy of $\sG$ in each dimension and all the face and degeneracy maps are identities.
\end{exam}

Simplicial homotopies will give rise to gauge symmetries later, and we are therefore interested in the cases in which simplicial homotopies yield an equivalence relation.

\begin{prop}\label{prop:EquivHom}
Let $\CCX$ and $\CCY$ be two simplicial sets. Consider the relation $g\sim \tilde g$ on the set of all simplicial maps between $\CCX$ and $\CCY$ defined by saying that $g$ is related to $\tilde g$ whenever there exists a simplicial homotopy from $g$ to $\tilde g$. If $\CCY$ is a Kan simplicial set then this is an equivalence relation.
\end{prop}

\noindent
{\it Proof:}
Reflexivity is automatic but symmetry and transitivity follow from filling horns. Suppose there is a homotopy $h_1$ between two simplicial maps $g_1:\CCX\to \CCY$ and $g_2:\CCX\to \CCY$. We may then define a simplicial map $\CCX\times\Lambda^2_0\to \CCY$ by letting $\CCX\times\{(1,2)\}\to \CCY$ be the homotopy $h_1$ and $\CCX\times\{(0,2)\}\to \CCY$ be the simplicial map $g_1$. By the Kan extension property, this map extends to a map  $\CCX\times\Delta^2\to \CCY$, and the restriction of this map to $\CCX\times\{(0,1)\}\to \CCY$ gives the desired homotopy between $g_2$ and $g_1$. Transitivity works similarly. Indeed, if $h_2$ is another homotopy between $g_2$ and another simplicial map $g_3:\CCX\to \CCY$, we define a simplicial map $\CCX\times\Lambda^2_1\to \CCY$ with $h_1:\CCX\times\{(1,2)\}\to \CCY$ and $h_2:\CCX\times\{(0,1)\}\to \CCY$. Again, this map extends to a simplicial map $\CCX\times\Delta^2\to \CCY$, and the restriction of this map to $\CCX\times\{(0,2)\}\to \CCY$ gives the desired homotopy between $g_1$ and $g_3$.
\hfill $\Box$

\vspace{10pt}
Note that Proposition \ref{prop:EquivHom} generalises to higher homotopies as well as to homotopies between simplicial manifolds. This will be of particular importance in the context of defining \v Cech equivalence for principal bundles later on. 

\subsection{Nerves of categories}\label{sec:CatNerSimpMan}

Each small category comes with a defining simplicial set called its nerve.
 
\begin{definition}
 The \uline{nerve}  $N(\CC)$ of a small category $\CC$ is the simplicial set $N(\CC):\mathbf{\Delta}^{\rm op}\rightarrow \CatSet$ whose simplicial $p$-simplices are $N_{p}(\CC):=\CatFun([p],\CC)$.
\end{definition}

\noindent 
Put differently, the simplicial $p$-simplices are given by the functor category consisting of functors $[p]\rightarrow \CC$ and natural transformations between them. We usually suppress the latter when talking about nerves. Functors $[p]\rightarrow \CC$ yield strings of $p$ composable morphisms in $\CC$ together with all possible compositions between these morphisms. Therefore, the nerve of a category $\CC=(\CC_0,\CC_1)$ has the objects $\CC_0$ as its vertices, the morphisms $\CC_1$ as its simplicial 1-simplices, and strings of $p$ composable morphisms as its simplicial $p$-simplices. The face maps $\sff_i^p:N_p(\CC)\rightarrow N_{p-1}(\CC)$ are given by composition of the $i$-th and $(i+1)$-th morphism for $0< i< p$. For $i=0$ and $i=p$, $\sff_i^p$ removes the $i$-th object. The degeneracy maps $\sfd_i^p:N_p(\CC)\rightarrow N_{p+1}(\CC)$ insert an identity morphism before the $i$-th morphism. Explicitly, the lowest face and degeneracy maps are
\begin{equation}
 \begin{gathered}
  \sfd_0^0(x)\ =\ \id_x~,\\
  \sff_0^1(f)\ =\ \sfs(f)~,~~~\sff_1^1(f)\ =\ \sft(f)~,\\
  \sfd^1_0(f)\ =\ \id_{\sft(f)}\circ f~,~~~\sfd^1_1(f)\ =\ f\circ\id_{\sfs(f)}~,\\
  \sff_0^2(f_2,f_1)\ =\ f_1~,~~~\sff_1^2(f_2,f_1)\ =\ f_2\circ f_1~,~~~ \sff_2^2(f_2,f_1)\ =\ f_2~,
 \end{gathered}
\end{equation}
for $x\in \CC_0\equiv N_0(\CC)$, $f\in \CC_1\equiv N_1(\CC)$, and $(f_2,f_1)\in N_2(\CC)$. Here, $\id$, $\sfs$, and $\sft$ are the usual structure maps of the category $\CC$. The simplicial identities guarantee the usual relations between the structure equations such as $\sfs(\id_x)=\sft(\id_x)=x$. 

Two simple examples of nerves are the nerve of the category $[p]$, which is $N([p])=\Delta^p$ and the nerve of a discrete category $(X\rightrightarrows X)$, which is the simplicially constant simplicial set $\big\{ \cdots \quadarrow X\triplearrow X\doublearrow X\big\}$.

From the above description, it is clear that every category is fully determined by its nerve. The following characterises a simplicial set which is the nerve of a category (see e.g.\ \cite[Proposition 1.1.2.2]{Lurie:0608040}):
\begin{prop}\label{prop:ss_is_category}
 A simplicial set is the nerve of a category if and only if it is an inner Kan simplicial set for which all the inner Kan conditions are uniquely satisfied.
\end{prop}

\noindent 
Instead of giving a full explanation of this proposition, let us merely motivate it. Consider again the simplicial 2-simplex in $\Delta^2$ and its horns (see also Figure \ref{fig:horns}). The inner horn is $\Lambda_1^2$, and the fact that it can be filled uniquely means that the composition of morphisms $(0,1)$ and $(1,2)$ is well defined. If we were also able to fill the horns $\Lambda_0^2$ and $\Lambda_2^2$, then we would have a left-inverse to the morphism $(0,1)$ as well as a right-inverse to the morphism $(1,2)$, respectively. We shall return to this point when discussing groupoids. Note that the inner horns $\Lambda_i^3$, $i=1,2$, have a unique horn filler, which amounts to associativity of the composition of morphisms.

\begin{prop}\label{prop:nerve_functor}
 The nerve construction induces a functor $N:\CatCat\to\CatsSet$, called the \uline{nerve functor}.\footnote{In fact, this functor is the right-adjoint of the geometric realisation functor, whose image on $\mathbf{\Delta}$ was briefly mentioned in Section \ref{ssec:simplicial_sets}.}
\end{prop}
In particular, any functor $\Phi:\CC\rightarrow \CD$ between two categories $\CC$ and $\CD$ induces a simplicial map $N(\Phi):N(\CC)\to N(\CD)$,
\begin{equation}
 \myxymatrix{
  \mathbf{\Delta}^{\rm op}   \ar@/_4ex/[rr]_{N(\CD)}="g1"
  \ar@/^4ex/[rr]^{N(\CC)}="g2"
  \ar@{=>}^{N(\Phi)} "g2"+<0ex,-2.5ex>;"g1"+<0ex,2.5ex>
&& \CatsSet
}
\end{equation} 
between the associated nerves $N(\CC)=\bigcup_{p\in\mathbbm{N}_0}\CatFun([p],\CC)$ and $N(\CD)=\bigcup_{p\in\mathbbm{N}_0}\CatFun([p],\CD)$. Explicitly, a simplicial $p$-simplex $F\in\CatFun([p],\CC)$ is mapped to the simplicial $p$-simplex $N(\Phi)(F)=\Phi\circ F\in\CatFun([p],\CD)$. Likewise, a morphism $\alpha$ in $\CatFun([p],\CC)$ is mapped to the morphism $\Phi(\alpha)\in \CatFun([p],\CD)$. Notice also that the nerve functor factors through the product of categories, and we have a canonical isomorphism $N(\CC\times \CD)\cong N(\CC)\times N(\CD)$. 
 
The next obvious step is to look for an interpretation of natural transformations in the context of nerves. Recall that a natural transformation between two functors $\Phi_{0,1}:\CC\rightarrow \CD$ can alternatively be regarded as a functor $\Psi:\CC\times [1] \rightarrow\CD$ rendering the diagram
 \begin{equation}\label{diag:simplicial_homotopy}
    \xymatrixcolsep{5pc}
    \myxymatrix{
    \CC\times[0]\cong \CC\ar@{->}[rd]^{\Phi_0} \ar@{->}[d]_{1\times \phi^1_1} & \\
    \CC\times[1]\ar@{->}[r]^{\Psi} & \CD\\
    \CC\times[0]\cong \CC\ar@{->}[ru]^{\Phi_1} \ar@{->}[u]^{1\times \phi^1_0} & 
    }
 \end{equation}
commutative. Here, $\phi^1_0$ and $\phi^1_1$ are the coface maps defined in \eqref{eq:CoFandCoD}. To see this more explicitly, consider a morphism $f\in \CC(x,y)$. Commutativity of the diagram \eqref{diag:simplicial_homotopy} amounts to $\Phi_0(x)=\Psi(x,0)$, $\Phi_1(x)=\Psi(x,1)$, $\Phi_0(f)=\Psi(f\times (0,0))$, and $\Phi_1(f)=\Psi(f\times(1,1))$. Consequently, upon applying the functor $\Psi$ to the commutative diagram 
 \begin{equation}
    \xymatrixcolsep{5pc}
    \myxymatrix{
    (x,0)\ar@{->}[d]_{f\times(0,0)}   & (x,1) \ar@{->}[l]_{\id_x\times(0,1) } \ar@{->}[d]^{f\times(1,1)}\\
    (y,0)& (y,1)\ar@{->}[l]^{\id_y\times(0,1)} 
    }
 \end{equation}
in $\CC\times[1],$ we obtain the commutative diagram 
\begin{equation}\label{eq:NatTrans}
    \xymatrixcolsep{5pc}
    \myxymatrix{
    \Phi_0(x)\ar@{->}[d]_{\Phi_0(f)}   & \Phi_1(x) \ar@{->}[d]^{\Phi_1(f)}\ar@{->}[l]_{\alpha_x } \\
    \Phi_0(y) & \Phi_1(y)\ar@{->}[l]^{\alpha_y}
    }
 \end{equation}
 in $\CD$, where we have made use of the identifications  $\Phi_0(x)=\Psi(x,0)$, $\Phi_1(x)=\Psi(x,1)$, $\Phi_0(f)=\Psi(f\times(0,0))$, $\Phi_1(f)=\Psi(f\times(1,1))$, and set $\alpha_x:=\Psi(\id_x\times(0,1))$. The diagram \eqref{eq:NatTrans} is, however, the standard definition of a natural transformation $\alpha:\Phi_0\Rightarrow\Phi_1$ between the functors $\Phi_0$ and $\Phi_1$. 
 
Upon applying the nerve functor $N:\CatCat\to\CatsSet$  to the diagram \eqref{diag:simplicial_homotopy} and making use of the factorisation property $N(\CC\times \CD)\cong N(\CC)\times N(\CD)$ together with $N([p])=\Delta^p$, we recover the definition of a simplicial homotopy \eqref{eq:DefSimpHomo} between the simplicial maps $N(\Phi_0):N(\CC)\to N(\CD)$ and $N(\Phi_1):N(\CC)\to N(\CD)$ that are induced by the functors $\Phi_0$ and $\Phi_1$. In summary, natural transformations between functors translate to simplicial homotopies between the corresponding simplicial maps.

\subsection{Quasi-categories}\label{sec:quasiCategory}

Characterising categories by Kan conditions suggests a definition of higher categories as (inner) Kan simplicial sets, see e.g.~\cite{Joyal:2002:207-222,Lurie:0608040}. The resulting formulations of higher categories are much simpler than other common definitions, as for example those in terms of weak 2- or 3-categories.

\begin{definition}
A \uline{quasi-category} or \uline{$\infty$-category} is an inner Kan simplicial set.
\end{definition}

\noindent
Here, the vertices are to be understood as the objects and the simplicial $1$-simplices are essentially $1$-morphisms in a higher category. Recall that the composition of 1-morphisms in a category corresponds to filling a horn in the category's nerve. This was unique, as the nerve of a category is uniquely Kan. In quasi-categories, this uniqueness is removed, and we therefore do no longer have a unique composition of 1-morphisms. We rather have a set of possible compositions given by the set of all horn fillers. This is very much in line with the general philosophy of categorification, in which entities are replaced by sets of equivalent entities. For further clarification of this, see the discussion of the Duskin nerve of a weak 2-category in Appendix \ref{app:Duskin_nerve}.

One of the important benefits of working with quasi-categories is the fact that all transfors (functors, natural transformations, modifications, etc.) become very simple, and this will dramatically simplify our formulation of categorified principal bundles.

\begin{definition}
A \uline{quasi-functor} or \uline{0-quasi-transfor} between two quasi-categories is a simplicial map between them. A \uline{quasi-natural transformation} or \uline{1-quasi-transfor} between two quasi-functors is a simplicial homotopy between the corresponding simplicial maps. In general, a \uline{$k$-quasi-transfor}, for $k\geq2$, is simply a simplicial $k$-homotopy, see Definition \ref{def:simplicial_k_homotopy}.
\end{definition}

A common specialisation we will use is to require the Kan conditions to be satisfied uniquely from some order upwards.
\begin{definition}\label{def:nQuasi}
An \uline{$n$-quasi-category} is a quasi-category for which the Kan conditions $\Kan(p,i)$ are satisfied uniquely for all $p>n$ and $i\in \{1,\ldots,p-1\}$ .
\end{definition}

\noindent
Notice that quasi-categories and $n$-quasi-categories form geometric models for $(\infty,1)$- and $(n,1)$-categories, respectively. Notice also that we can internalise quasi-categories in other categories, e.g.\ in $\CatDiff$. In case of the latter, the result will be a simplicial manifold satisfying the evident inner Kan conditions. 

\subsection{Lie quasi-groupoids}\label{ssec:Groupoids}

To describe the gauge structure of the simplicial bundles later on, we need a higher group structure. Such structures are captured by Lie quasi-groupoids, which are higher versions of groupoids. Let us start off by briefly recalling some basic notions about groupoids.

\begin{definition}
 A \uline{groupoid} is a small category in which every morphism is invertible. 
\end{definition}

\noindent
In particular, a groupoid $\CG$ consists of two sets $\CG_0$ (the objects) and $\CG_1$ (the morphisms) with source and target maps $\sfs,\sft:\CG_1\doublearrow \CG_0$ as well as an identity or inclusion map $\id:\CG_0\embd \CG_1$. We have $\sfs(\id_x)=\sft(\id_x)=x$ for all $x\in \CG_0$. If two morphisms $f_1$ and $f_2$ have matching source and target, $\sft(f_1)=\sfs(f_2)$, we can compose them into a new morphism $f_2\circ f_1$ with $\sfs(f_2\circ f_1)=\sfs(f_1)$ and $\sft(f_2\circ f_1)=\sft(f_2)$. The composition is associative and every morphism $f\in \CG_1$ has an inverse $f^{-1}\in \CG_1$, that is, $f^{-1}\circ f=\id_{\sfs(f)}$ and $f\circ f^{-1}=\id_{\sft(f)}$. 

All the groupoids and higher groupoids we are interested in will be smooth, and we directly internalise in $\CatDiff$.\footnote{Recall that internalisation in $\CatDiff$ is non-trivial as this category does not have all pullbacks. However, this is a technicality which we can fix by requiring source and target maps to be surjective submersions.} 

\begin{definition}
A \uline{Lie groupoid} is a groupoid internal to $\CatDiff$. 
\end{definition}

\noindent
This implies that both the set of objects and and the set of morphisms of a Lie groupoid are manifolds and all its structure maps are smooth. 

\begin{exam}\label{exam:fibgroupoid} 
Let  $f:Y\to X$ be a surjective submersion between two manifolds $Y$ and $X$ and denote the fibre product of $Y$ with itself over $X$ by $Y^{[2]}:=Y\times_X Y:=\{(y_1,y_2)\in Y\times Y\,|\,f(y_1)=f(y_2)\}$. The \uline{\v Cech groupoid} $\check{\CC}(Y\rightarrow X)$ of $f$ is the natural Lie groupoid $Y\times_X Y\doublearrow Y$ with pairs $(y_1,y_2)$ for $y_1,y_2\in Y$ satisfying $f(y_1)=f(y_2)$ as its morphisms. It has the source, target, composition, and identity maps given by  $\sfs(y_1,y_2):=y_2$, $\sft(y_1,y_2):=y_1$, $\id_x:=(x,x)$, and $(y_1,y_2)\circ(y_2,y_3):=(y_1,y_3)$. Note that when $X$ is just the one-point set, the \v Cech groupoid is also known as the \uline{pair groupoid} of $Y$.	
\end{exam}

\begin{exam}
A simple example of a Lie groupoid is the \uline{delooping} $\sB\sG=(\sG\doublearrow *)$ of a Lie group $\sG$, where the source and target maps are trivial, $\id_*=\unit_\sG$, and the composition is group multiplication in $\sG$. Note that $\sB\sG$ differs from $\check \CC(\sG\rightarrow *)$.
\end{exam}

Because a groupoid $\CG=(\CG_0,\CG_1)$ is a category, we can consider its nerve. As before, the sets of simplicial 0-, 1-, and 2-simplices of the nerve $N(\CG)$ are identified with $\CG_0$, $\CG_1$, and the set of composable morphisms, respectively. In general, the simplicial $p$-simplices are given by $p$-tuples of composable morphisms. In particular, the \uline{\v Cech nerve} of the \v Cech groupoid $\check \CC (Y\rightarrow X)$ is 
\begin{equation}
N(Y\times_X Y\doublearrow Y)\ =\ \left\{ \cdots \quadarrow Y\times_X Y\times_X Y\triplearrow Y\times_X Y\doublearrow Y\right\},
\end{equation}
with face and degeneracy maps given by
\begin{equation}
\begin{aligned}
 \sff^p_i(x_0,\ldots,x_{p})\ &:=\ (x_0,\ldots,x_{i-1},x_{i+1},\ldots,x_{p})~,\\
  \sfd^p_i(x_0,\ldots,x_{p})\ &:=\ (x_0,\ldots,x_{i-1},x_i,x_{i},\ldots,x_{p})~.
 \end{aligned}
\end{equation}
Also, the nerve $N(\sB\sG)$ of the delooping $\sB\sG$ is the simplicial manifold
\begin{equation}
 N(\sB\sG)\ =\ \left\{ \cdots \quadarrow \sG\times \sG\triplearrow \sG\doublearrow *\right\}
\end{equation}
with the obvious face and degeneracy maps. 

Proposition \ref{prop:ss_is_category} characterises simplicial sets which are nerves of categories. The analogue statement for groupoids is given by the following proposition.

\begin{prop}\label{prop:ss_is_groupoid}
 A simplicial set is the nerve of a groupoid if and only if it is a Kan simplicial set in which all the Kan conditions (i.e.\ inner and outer) are uniquely satisfied.
\end{prop}

\noindent 
As a corollary to Proposition \ref{prop:ss_is_groupoid}, it follows that a simplicial manifold is the nerve of a groupoid if it is a Kan simplicial manifold, and the Kan conditions are uniquely satisfied.

It is now completely clear how to generalise from groupoids to quasi-groupoids, Lie quasi-groupoids and Lie $n$-quasi-groupoids, cf.\ \cite{Henriques:2006aa,Getzler:0404003,Zhu:0812.4150} or \cite{Mehta:1012.4103}. 
\begin{definition}\label{def:QuasiGroupoid}
A \uline{quasi-groupoid} is a Kan simplicial set. A \uline{Lie quasi-groupoid} is a Kan simplicial manifold. A \uline{Lie $n$-quasi-groupoid} is a Lie quasi-groupoid for which the Kan conditions $\Kan(p,i)$ are satisfied uniquely for $p>n$, $i\in \{0,\ldots,p\}$.
\end{definition}

\noindent
Note that in the case of Lie $n$-quasi-groupoids, this means that  the restrictions \eqref{eq:HornFillProp} are diffeomorphisms for $p>n$.

\begin{rem}\label{rem:InftGrp}
A \uline{reduced} (Lie) quasi-groupoid is a (Lie) quasi-groupoid with a single simplicial 0-simplex or object. We shall follow the delooping hypothesis and identify reduced (Lie) quasi-groupoids with \uline{(Lie) quasi-groups}. Likewise, whenever the Kan conditions $\Kan(p,i)$ are satisfied uniquely for $p>n$, we shall speak of \uline{(Lie) $n$-quasi-groups}. The categories of quasi-groups and simplicial groups (see Example \ref{exam:SimpGroup}) are equivalent due to a classical result of Quillen \cite{quillen1969}. This also holds true in the context of Lie quasi-groups and Lie simplicial groups as was shown in Nikolaus {\it et al.}~\cite[Proposition 3.35]{nikolaus1207}.
\end{rem}

\begin{lemma}\label{lem:KanHomXY}
 Let $\CCX$ and $\CCY$ be simplicial sets. If $\CCY$ is a quasi-groupoid then so is  $\inthom(\CCX,\CCY)$.
\end{lemma}

\noindent
{\it Proof:}  Let us sketch the proof. By  Definition \ref{def:QuasiGroupoid}, a quasi-groupoid is a Kan simplicial set, that is, a Kan fibration over the one-point simplicial set (see Section \ref{sec:SimpMan})
\begin{equation}
    \myxymatrix{
    \Lambda^p_i\ar@{->}[r] \ar@{^{(}->}[d] & \CCY \\
    \Delta^p \ar@{->}[ru] & 
    }
 \end{equation}
This diagram can be extended to a commutative diagram
\begin{equation}
    \myxymatrix{
    \Lambda^p_i\times \CCX\ar@{->}[r] \ar@{^{(}->}[d] & \CCY \\
    \Delta^p\times \CCX \ar@{->}[ru] & 
    }
 \end{equation}
 for any simplicial manifold $\CCX$ (see e.g.~\cite[Lemma 1.5]{Curtis:1971:107-209}). By virtue of Lemma \ref{lem:3SimpSets}, the latter diagram is equivalent to
 \begin{equation}
    \myxymatrix{
    \Lambda^p_i\ar@{->}[r] \ar@{^{(}->}[d] & \inthom(\CCX,\CCY) \\
    \Delta^p \ar@{->}[ru] & 
    }
 \end{equation}
 and this proves the Lemma.  \hfill $\Box$

\section{Lie quasi-groupoid bundles}\label{sec:LieqgBundles}

Having introduced the notion of Lie $n$-quasi-groups, we now wish to define and discuss principal bundles that have such quasi-groups as their structure groups. Our constructions naturally extend to Lie quasi-groupoids, and by a slight abuse of language we shall call the result a Lie quasi-groupoid bundle. These bundles form the underlying geometrical structure of higher gauged sigma models. All our definitions are given for manifolds but they extend to supermanifolds. In Section \ref{ssec:higher_base_1}, we shall give generalisations for higher base spaces. 
Higher principal bundles based on Kan simplicial sets have been discussed earlier, see e.g.\ \cite{Bakovic:2009aa,Bakovic:0902.3436} and \cite{Nikolaus:1207ab,nikolaus1207,Schreiber:2013pra}. Furthermore, higher principal bundles over a particular higher base space were discussed in \cite{Ritter:2015zur}. We shall start off with the simplical definition of principal $\sG$-bundles for $\sG$ a Lie group.

\subsection{Principal bundles and Lie groupoid bundles from simplicial maps}

Recall that given a Lie group $\sG$ and a surjective submersion $Y\rightarrow X$ between two manifolds $X$ and $Y$, a principal $\sG$-bundle over $X$ subordinate to the cover $Y\rightarrow X$  is a functor from the \v Cech groupoid $\check \CC(Y\rightarrow X)=(Y\times_X Y\doublearrow Y)$ to the delooping $\sB\sG=(\sG\rightrightarrows *)$. Moreover, two such principal $\sG$-bundles are called equivalent if and only if there is a natural isomorphism between the underlying functors. 

Recall from Section \ref{ssec:Groupoids} that the nerves of the \v Cech groupoid and $\sB\sG$ read as
\begin{equation}
\begin{aligned}
 N(\check \CC(Y\rightarrow X))\ &=\ \left\{ \cdots \quadarrow Y\times_X Y\times_X Y\triplearrow Y\times_X Y\doublearrow Y\right\},\\
 N(\sB\sG)\ &=\ \left\{ \cdots \quadarrow \sG\times \sG\triplearrow \sG\doublearrow *\right\},
\end{aligned}
\end{equation}
respectively. With Propositions \ref{prop:nerve_functor} and \ref{prop:EquivHom}, the definitions of principal $\sG$-bundles and their isomorphisms are recast in the simplicial language as follows.
\begin{definition}\label{def:principal_G_bundle}
 For $\sG$ a Lie group, a \uline{principal $\sG$-bundle} over a manifold $X$ subordinate to a surjective submersion $Y\rightarrow X$  is a simplicial map $g:N(\check \CC(Y\rightarrow X))\to N(\sB\sG)$. Two such principal $\sG$-bundles $g,\tilde g:N(\check \CC(Y\rightarrow X))\rightarrow N(\sB\sG)$ are called \uline{equivalent} if and only if there is a simplicial homotopy between $g$ and $\tilde g$.
\end{definition}

In the following, we shall specialise to the usual setting of surjective submersions given by an ordinary cover $\frU=\dot\bigcup_{a\in A}U_a\to X$, where $\dot\cup$ denotes the disjoint union. The corresponding \v Cech groupoid $\check \CC(\frU\rightarrow X)$  has the pairs $(x,a)$ with $x\in U_a$ as objects and the triples $(x,a,b)$ with $x\in U_a\cap U_b\neq \varnothing$ as morphisms, and the structure maps are 
 \begin{equation}
 \begin{gathered}
  \sfs(x,a,b)\ :=\ (x,b)~,~~~\sft(x,a,b)\ :=\ (x,a)~,~~~\id_{(x,a)}\ :=\ (x,a,a)~,\\
  (x,a,b)\circ (x,b,c)\ :=\ (x,a,c)~.
 \end{gathered}
 \end{equation}
Furthermore, the nerve $N(\check \CC(\frU\rightarrow X))$ of $\check \CC(\frU\rightarrow X)$ is given by
\begin{equation}
 N(\check \CC(\frU\rightarrow X))\ =\ \left\{ \cdots \quadarrow \dot\bigcup_{a,b,c\in A}U_a\cap U_b\cap U_c\triplearrow \dot\bigcup_{a,b\in A}U_a\cap U_b \doublearrow \dot\bigcup_{a\in A}U_a\right\}.
\end{equation}
The degeneracy maps on a simplicial 0-simplex $(x,a)$ and on a simplicial 2-simplex $(x,a,b)$ are $\sfd^0_0(x,a)=(x,a,a)$, $\sfd^1_0(x,a,b)=(x,a,a,b)$, and $\sfd^1_1(x,a,b)=(x,a,b,b)$, and likewise for the higher simplices. The face maps on a simplicial $1$-simplex $(x,a,b)$ are $\sff^1_0(x,a,b)=(x,b)$ and $\sff^1_1(x,a,b)=(x,a)$ and on a simplicial 2-simplex $(x,a,b,c)$ we have $\sff^2_0(x,a,b,c)=(x,b,c)$, $\sff^2_1(x,a,b,c)=(x,a,c)$, and $\sff^2_2(x,a,b,c)=(x,a,b)$, respectively.

To recover the usual description of a principal $\sG$-bundle in terms of \v Cech cocycles from Definition \ref{def:principal_G_bundle}, we note that the simplicial map $g:N(\check \CC(\frU\rightarrow X))\rightarrow N(\sB\sG)$ consists of individual maps, the lowest of which are
\begin{equation}
\begin{gathered}
 g_a(x)\ :=\ g^0(x,a)\ =\ *~,~~~g_{ab}(x)\ :=\ g^1(x,a,b)\ \in\ \sG~,\\
 g_{abc}(x)\ :=\ g^2(x,a,b,c)\ =\ (g_{abc}^1(x),g_{abc}^2(x))\ \in\ \sG\times \sG~,~~~\mbox{etc.}
\end{gathered}
\end{equation}
The fact that $g$ commutes with the face maps gives the trivial relation $g\circ \sff^1_i=\sff^1_i\circ g$ together with the non-trivial relation $g\circ \sff^2_i=\sff^2_i\circ g$. The latter implies
\begin{equation}
 g_{abc}^1(x)\ =\ g_{ab}(x)~,~~~g^1_{abc}(x) g^2_{abc}(x)\ =\ g_{ac}(x)~,~~~g_{abc}^2(x)\ =\ g_{bc}(x)~,
\end{equation}
which yields the usual cocycle conditions for the \v{C}ech cocycles describing a principal $\sG$-bundle. For brevity, we shall often suppress the $x$-dependence in the following provided no confusion arises. Because of $g\circ \sfd^0_0=\sfd^0_0\circ g$, we have $g_{aa}=\unit_{\sG}$ and $g\circ \sfd^1_i=\sfd^1_i\circ g$ implies $g_{aab}=(\unit_{\sG},g_{ab})$ as well as $g_{abb}=(g_{ab},\unit_{\sG})$. Recall that all information about the structure of a category is encoded in its nerve's simplicial 0-, 1-, and 2-simplices, respectively, and all higher-dimensional simplices are completely fixed by the lower-dimensional ones. Since we extracted all non-trivial conditions for $g$ being a simplicial map from the relevant low-dimensional simplices, all higher conditions will be satisfied automatically.

 \vspace{10pt}
 \begin{figure}[h]
\begin{center}
\tikzset{->-/.style={decoration={markings,mark=at position #1 with {\arrow{>}}},postaction={decorate}}}
\begin{tikzpicture}[scale=0.7,every node/.style={scale=0.7}]
     \draw (2.3,10.8) node {$*$};
   \draw (2.3,16.3) node {$*$};
   \draw (7.8,10.8) node {$*$};
   \draw (7.8,16.3) node {$*$};
   \draw (5,10.7) node {$h_{aa,01}$};
   \draw (5,16.3) node {$h_{bb,01}$};
   \draw (5.6,13.5) node {$h_{ab,01}$};
   \draw (2.1,13.5) node {$\tilde g_{ab}$};
   \draw (7.9,13.5) node {$g_{ab}$};
   \draw[->-=.5] (2.5,11) -- (7.5,11); 
   \draw[->-=.5] (2.5,16) -- (2.5,11); 
   \draw[->-=.5] (2.5,16) -- (7.5,16); 
   \draw[->-=.5] (2.5,16) -- (7.5,11); 
   \draw[->-=.5] (7.5,16) -- (7.5,11); 
   \filldraw [black] (2.5,11) circle (2pt);
   \filldraw [black] (7.5,11) circle (2pt);
   \filldraw [black] (7.5,16) circle (2pt);
   \filldraw [black] (2.5,16) circle (2pt);
\end{tikzpicture}
\begin{minipage}{14cm}
\caption{\small \label{fig:SimpHomOrdPrinBun} Coboundary transformation for two principal $\sG$-bundles $g,\tilde g:N(\check \CC(\frU\rightarrow X))\rightarrow N(\sB\sG)$ for $\sG$ a Lie group.}
\end{minipage}
\end{center}
\end{figure}
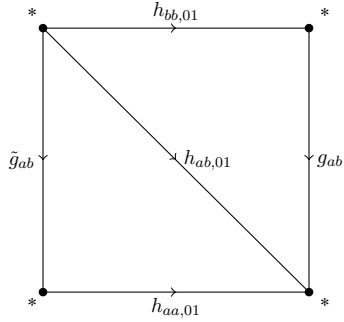

Let us now recover the \v Cech coboundary relations. According to Definition \ref{def:DefSimpHomo}, a simplicial homotopy between two principal $\sG$-bundles $g$ and $\tilde g$ subordinate to the cover $\frU\rightarrow X$ gives rise to a simplicial map $h: N(\check \CC(\frU\rightarrow X))\times \Delta^1\to N(\sB\sG)$ such that  
\begin{equation}
 \begin{aligned}
  h^p((x,a_0,\ldots,a_p),(0,\ldots,0))\ &=\ g^p(x,a_0,\ldots,a_p)\ =:\ g_{a_0\cdots a_p}(x)~, \\
  h^p((x,a_0,\ldots,a_p),(1,\ldots,1))\ &=\ \tilde g^p(x,a_0,\ldots,a_p)\ =:\ \tilde g_{a_0\cdots a_p}(x)
 \end{aligned}
\end{equation}
for all simplicial $p$-simplices $(x,a_0,\ldots,a_p)\in N_p(\check \CC(\frU\rightarrow X))$ and $p\geq0$. Using
\begin{equation}
\begin{aligned}
&N(\check \CC(\frU\rightarrow X))\times \Delta^1\ =\ \bigcup_{p\in\mathbbm{N}_0}\Big( N_p(\check \CC(\frU\rightarrow X))\times \Delta^1_p\Big)\ =\ \\
 &\kern1cm=\ \left\{ \cdots \triplearrow \dot\bigcup_{a,b\in A}U_a\cap U_b \times\{((0,0),(0,1),(1,1)\} \doublearrow \dot\bigcup_{a\in A}U_a\times\{0,1\}\right\},
 \end{aligned}
\end{equation}
the lowest components of $h$ are
\begin{equation}
\begin{gathered}
h^0((x,a), 0)\ =\ *\ =\ h^0((x,a),1)~,\\
g_{ab}(x)\ =\ h^1((x,a,b),(0,0))\eand \tilde g_{ab}(x)\ =\ h^1((x,a,b),(1,1))~,\\
h_{ab,01}(x)\ :=\ h^1((x,a,b),(0,1))~.
\end{gathered}
\end{equation}
Since $h$ is a simplicial map, it commutes with the face maps and coface maps, and, consequently, it follows that the diagram in Figure \ref{fig:SimpHomOrdPrinBun} is commutative. This, in turn yields the usual coboundary relations
\begin{equation}
g_{ab}h_{bb,01}\ =\ h_{aa,01} \tilde g_{ab} 
\end{equation}
for \v Cech 1-cocycles.
Just as in the case of the cocycle conditions, there are no further constraints arising from considering higher-dimensional simplices.
 
Our Definition \ref{def:principal_G_bundle} of principal $\sG$-bundles generalises to nerves of Lie groupoids.
\begin{definition}
 For $\CG$ a Lie groupoid, a \uline{Lie groupoid bundle} or \uline{principal $\CG$-bundle} over $X$ is a simplicial map $g:N(\check \CC(\frU\rightarrow X))\to N(\CG)$. Two such principal $\CG$-bundles $g,\tilde g:N(\check \CC(\frU\to X))\rightarrow N(\CG)$ over $X$ are called \uline{equivalent} if and only if there is a simplicial homotopy between $g$ and $\tilde g$.
\end{definition}
In terms of \v Cech cocycles and coboundaries, we have the following picture. A simplicial map $g:N(\check \CC(\frU\to X))\rightarrow N(\CG)$, where $\CG=(\CG_1\rightrightarrows \CG_0)$ has lowest components
\begin{equation}
\begin{gathered}
 g_a(x)\ :=\ g^0(x,a)\ \in\ \CG_0~,~~~g_{ab}(x)\ :=\ g^1(x,a,b)\ \in\ \CG_1~,\\
 g_{abc}(x)\ :=\ g^2(x,a,b,c)\ =\ (g_{abc}^1(x),g_{abc}^2(x))\ \in\ \CG_1\times \CG_1~,~~~\mbox{etc.}
\end{gathered}
\end{equation}
The fact that $g$ is a simplicial map gives rise to the following conditions:
\begin{equation}
\begin{gathered}
 \sfs(g_{ab})\ =\ g_b~,~~~\sft(g_{ab})\ =\ g_a~,~~~g_{aa}\ =\ \id_{g_a}~,\\
 g_{abc}^1\ =\ g_{ab}~,~~~g^1_{abc}\circ g^2_{abc}\ =\ g_{ac}~,~~~g_{abc}^2\ =\ g_{bc}~.
\end{gathered}
\end{equation}
Obviously, this then yields $g_{ac}=g_{ab}\circ g_{bc}$.

A simplicial homotopy $h:g\rightarrow \tilde g$ is given by 
\begin{equation}
 h_{aa,01}\,:\,\tilde g_a\ \to\  g_a\eand g_{ab}\circ h_{bb,01}\ =\  h_{aa,01}\circ \tilde g_{ab}~,
\end{equation}
where now $g_a(x)=h^0((x,a), 0)$ and $\tilde g_a(x)=h^0((x,a), 1)$ and the rest as presented above. 

Note that Lie groupoid bundles are the geometric structure underlying gauged sigma models. Here, $\CG$ is the action groupoid of the gauge group acting on the space $\CG_0$. We shall discuss this point in more detail in Section \ref{ssec:local_finite_gauge}.

\subsection{Generalisation to Lie quasi-groupoid bundles}

After the preceding discussion, the generalisation to Lie quasi-groupoid bundles is now rather obvious.

\begin{definition}\label{def:EquivLQuasGrpBun}
 For $\CCG$ a Lie quasi-groupoid, a \uline{Lie quasi-groupoid bundle} or \uline{principal} \uline{$\CCG$-bundle} over $X$ is a simplicial map $g:N(\check \CC(\frU\to X))\to \CCG$. Two such principal $\CCG$-bundles $g,\tilde g:N(\check \CC(\frU\to X))\rightarrow \CCG$ over $X$ are called \uline{equivalent} if and only if there is a simplicial homotopy between $g$ and $\tilde g$.
\end{definition}

\noindent 
Again, our notion of equivalence is sensible since $\CCG$ is a Kan simplical manifold and with Proposition \ref{prop:EquivHom} it follows that simplicial homotopies give rise to equivalence relations.

To illustrate this definition, let us work through the details for a Lie 2-quasi-groupoid $\CCG$. In this case, a simplicial map $g:N(\check \CC(\frU\to X))\rightarrow \CCG$ has the lowest components
\begin{equation}\label{eq:SimpMapsBund}
\begin{gathered}
 g_a(x)\ :=\ g^0(x,a)\ \in\ \CCG_0~,~~~g_{ab}(x)\ :=\ g^1(x,a,b)\ \in\ \CCG_1~,\\
 g_{abc}(x)\ :=\ g^2(x,a,b,c)\ \in\ \CCG_2~.
 \end{gathered}
\end{equation}
Beyond this, the uniqueness of the horn fillers fully determines the map $g$. The fact that $g$ is a simplicial map implies that the face maps acting on $g_{ab}$ and $g_{abc}$ yield
\begin{equation}
\begin{gathered}
 (\sff^1_0\circ g^1)(x,a,b)\ =\ g_b(x)~,~~~
 (\sff^1_1\circ g^1)(x,a,b)\ =\ g_a(x)~,\\
 (\sff^2_0\circ g^2)(x,a,b,c)\ =\ g_{bc}(x)~,\quad
  (\sff^2_1\circ g^2)(x,a,b,c)\ =\ g_{ac}(x)~,\\
   (\sff^2_2\circ g^2)(x,a,b,c)\ =\ g_{ab}(x)~.
  \end{gathered}
\end{equation}

\begin{figure}[t]
\begin{center}
\tikzset{->-/.style={decoration={markings,mark=at position #1 with {\arrow{>}}},postaction={decorate}}}
\begin{tikzpicture}[scale=0.7,every node/.style={scale=0.7}]
   \filldraw[pattern=north west lines, pattern color=green!20,draw=white] (0,0) -- (5,3) -- (5,8) -- cycle;
   \draw (5,-1.5) node {(iii)};
   \draw (5,9.5) node {(ii)};
   \draw (5,15) node {(i)};
   \draw (6.3,16.3) node {$g_a$};
   \draw (3.8,16.3) node {$g_b$};
   \draw (5,16.2) node {$g_{ab}$};
   \draw (2.3,10.8) node {$g_c$};
   \draw (5,13.8) node {$g_b$};
   \draw (7.8,10.8) node {$g_a$};
   \draw (3.5,12.5) node {$g_{bc}$};
   \draw (6.6,12.5) node {$g_{ab}$};
   \draw (5,10.7) node {$g_{ac}$};
    \draw [-implies,double equal sign distance,thick] (5,12.5) -- (5,11.5);
       \draw (5.5,12) node {$g_{abc}$};

   \draw (-0.2,-0.2) node {$g_c$};
   \draw (4.7,3.2) node {$g_b$};
   \draw (10.3,-0.2) node {$g_a$};
   \draw (5,8.3) node {$g_d$};
   \draw [dashed,-implies,double equal sign distance,thick] (5.8,3.33) -- (6.7,3.68);
   \draw (6,3.8) node {$g_{abd}$};
   \draw [dashed,-implies,double equal sign distance,thick] (3.5,3.85) -- (4.2,4.6);
   \draw (3.4,4.4) node {$g_{bcd}$};
   \draw [dashed,-implies,double equal sign distance,thick] (5,1.8) -- (5,0.8);
   \draw (4.5,1.3) node {$g_{abc}$};
   \draw (2.2,4.4) node {\color{green!50!black} $g_{cd}$};
   \draw (7.8,4.4) node { $g_{ad}$};
   \draw (5.4,5.5) node {$g_{bd}$};
   \draw (2.2,1.8) node {\color{green!50!black} $g_{bc}$};
   \draw (7.8,1.8) node {\color{green!50!black} $g_{ab}$};
   \draw (5,-.3) node {$g_{ac}$};
   %tetrahedron
   \draw[draw=green!50!black,very thick,->-=.5] (5,8) -- (0,0);   
   \draw[->-=.5] (0,0) -- (10,0);   
   \draw[draw=green!50!black,very thick,dashed,->-=.5] (0,0) -- (5,3);   
   \draw[draw=green!50!black,very thick,dashed,->-=.5] (5,3) -- (10,0);   
   \draw[->-=.5] (5,8) -- (10,0);   
   \draw[dashed,->-=.5] (5,8) -- (5,3);
   %triangle 
   \draw[->-=.5] (2.5,11) -- (7.5,11); 
   \draw[->-=.5] (2.5,11) -- (5,13.5); 
   \draw[->-=.5] (5,13.5) -- (7.5,11);  
   %line
    \draw[->-=.5] (4,16.5) -- (6,16.5); 
   \filldraw [black] (0,0) circle (2pt);
   \filldraw [black] (5,8) circle (2pt);
   \filldraw [black] (10,0) circle (2pt);
   \filldraw [black] (5,3) circle (2pt);
   \filldraw [black] (2.5,11) circle (2pt);
   \filldraw [black] (7.5,11) circle (2pt);
   \filldraw [black] (5,13.5) circle (2pt);
   \filldraw [black] (4,16.5) circle (2pt);
   \filldraw [black] (6,16.5) circle (2pt);
      \draw [-implies,double distance=2pt,draw=green!50!black,thick] (5.5,3.5) -- (4.4,4);
      \draw [-implies,draw=green!50!black, thick]  (5.5,3.5) -- (4.4,4);
   \draw (5.2,4.1) node {\color{green!50!black}$g_{abcd}$};
   \draw [-implies,double equal sign distance,thick] (4.5,2.3) -- (5.5,2.9);
   \draw (5.3,2.3) node {$g_{acd}$};
\end{tikzpicture}
\begin{minipage}{14cm}
\caption{\small \label{fig:TransCocPrin} Transition functions and cocycle conditions of a principal $\CCG$-bundle $g:N(\check \CC(\frU\to X))\rightarrow \CCG$ for $\CCG$ a Lie 2-quasi-groupoid. As before, the arrows indicate the interpretation of the 2- and 3-simplices as horn fillers.}
\end{minipage}
\end{center}
\end{figure}

\noindent
Note that we also have
\begin{equation}
\begin{gathered}
  g_{aa}(x)\ =\ (\sfd_0^0\circ g^0)(x,a)~,\\ g_{aab}(x)\ =\ (\sfd^1_0\circ g^1)(x,a,b)~,~~~g_{abb}(x)\ =\ (\sfd^1_1\circ g^1)(x,a,b)~,\\g_{aaa}(x)\ =\ (\sfd^1_0\circ(\sfd^0_0\circ g^0))(x,a)\ =\ (\sfd^1_1\circ( \sfd^0_0\circ g^0))(x,a)~.
  \end{gathered}
\end{equation}

\begin{figure}[h!]
\begin{center}
\tikzset{->-/.style={decoration={markings,mark=at position #1 with {\arrow{>}}},postaction={decorate}}}
\begin{tikzpicture}[scale=0.7,every node/.style={scale=0.7}]
   \filldraw[pattern=north west lines, pattern color=green!20,draw=white] (0,0) -- (0,5) -- (5,8) -- cycle;
   \filldraw[pattern=north west lines, pattern color=blue!20,draw=white] (5,3) -- (10,5) -- (5,8) -- cycle;
   \filldraw[pattern=north west lines, pattern color=red!20,draw=white] (0,0) -- (5,3) -- (10,0) -- cycle;
   \draw (5,-1.5) node {(ii)};
   \draw (5,9.5) node {(i)};
   \draw (2.3,10.8) node {$\tilde g_a$};
   \draw (2.3,16.3) node {$\tilde g_b$};
   \draw (7.8,10.8) node {$g_a$};
   \draw (7.8,16.3) node {$g_b$};
   \draw (5,10.7) node {$h_{aa,01}$};
   \draw (5,16.3) node {$h_{bb,01}$};
   \draw (5.6,13.5) node {$h_{ab,01}$};
   \draw (2.1,13.5) node {$\tilde g_{ab}$};
   \draw (7.9,13.5) node {$g_{ab}$};
    \draw [-implies,double equal sign distance,thick] (6.5,15) -- (5.5,14);
   \draw [-implies,double equal sign distance,thick] (4.5,13) -- (3.5,12);
     \draw (5.4,14.8) node {$h_{abb,001}$};
   \draw (3.5,13) node {$h^{-1}_{aab,011}$};
   \draw (-0.2,-0.2) node {$\tilde g_c$};
   \draw (4.7,3.2) node {$\tilde g_b$};
   \draw (10.3,-0.2) node {$\tilde g_a$};
   \draw (-0.2,5.2) node {$g_c$};
   \draw (5,8.3) node {$g_b$};
   \draw (10.3,5.2) node {$g_a$};
   \draw [dashed,-implies,double equal sign distance,thick] (5.94,6.5) -- (6.5,5.6);
   \draw [dashed,-implies,double equal sign distance,thick] (8.14,2.98) -- (8.7,2.08);
   \draw (7.1,6) node {$h_{abb,001}$};
   \draw (7.6,2.6) node {$h^{-1}_{aab,011}$};
   \draw [dashed,-implies,double equal sign distance,thick] (1,4.6) -- (2,4.2);
   \draw [dashed,-implies,double equal sign distance,thick] (3.2,3.72) -- (4.2,3.32);
   \draw (1,4.1) node {$h_{bcc,001}$};
   \draw (4,4) node {$h^{-1}_{bbc,001}$};
   \draw [-implies,double equal sign distance,thick] (2,4) -- (3,3.5);
   \draw [-implies,double equal sign distance,thick] (6.5,1.75) -- (7.5,1.25);
   \draw (2.2,3.3) node {$h_{acc,001}$};
   \draw (6.5,1.3) node {$h^{-1}_{aac,011}$};
   \draw [-implies,double equal sign distance,thick] (4.7,6.8) -- (4.7,5.8);
   \draw (5.2,6.3) node {$g_{abc}$};
   \draw [dashed,-implies,double equal sign distance,thick] (5,1.8) -- (5,0.8);
   \draw (4.5,1.3) node {$\tilde g_{abc}$};
   \draw (-0.6,2.5) node {$h_{cc,01}$};
   \draw (5.5,5.5) node {$h_{bb,01}$};
   \draw (10.6,2.5) node {$h_{aa,01}$};
   \draw (5.5,2.3) node {$h_{ac,01}$};
   \draw (7.3,4.3) node {$h_{ab,01}$};
   \draw (3.3,4.4) node {$h_{bc,01}$};
   \draw (2.4,6.8) node {$g_{bc}$};
   \draw (7.8,6.8) node { $g_{ab}$};
   \draw (5.4,4.7) node {$g_{ac}$};
   \draw (2.2,1.8) node {$\tilde g_{bc}$};
   \draw (7.8,1.8) node {$\tilde g_{ab}$};
   \draw (5,-.3) node {$\tilde g_{ac}$};
   %prism
   \draw[->-=.5] (0,0) -- (0,5); 
   \draw[->-=.5] (0,5) -- (5,8); 
   \draw[->-=.5]  (5,8) -- (10,5);
   \draw[->-=.5] (0,5) -- (10,5);
   \draw[->-=.5] (0,0) -- (10,5);
   \draw[dashed,->-=.5] (0,0) -- (5,8);   
   \draw[->-=.5] (0,0) -- (10,0);   
   \draw[dashed,->-=.5] (0,0) -- (5,3);   
   \draw[dashed,->-=.5] (5,3) -- (10,0);   
   \draw[->-=.5] (10,0) -- (10,5);   
   \draw[dashed,->-=.5] (5,3) -- (10,5);   
   \draw[dashed,->-=.5] (5,3) -- (5,8);
   %square 
   \draw[->-=.5] (2.5,11) -- (7.5,11); 
   \draw[->-=.5] (2.5,16) -- (2.5,11); 
   \draw[->-=.5] (2.5,16) -- (7.5,16); 
   \draw[->-=.5] (2.5,16) -- (7.5,11); 
   \draw[->-=.5] (7.5,16) -- (7.5,11); 
   \filldraw [black] (0,0) circle (2pt);
   \filldraw [black] (0,5) circle (2pt);
   \filldraw [black] (5,8) circle (2pt);
   \filldraw [black] (10,5) circle (2pt);
   \filldraw [black] (10,0) circle (2pt);
   \filldraw [black] (5,3) circle (2pt);
   \filldraw [black] (2.5,11) circle (2pt);
   \filldraw [black] (7.5,11) circle (2pt);
   \filldraw [black] (7.5,16) circle (2pt);
   \filldraw [black] (2.5,16) circle (2pt);
\end{tikzpicture}
\begin{minipage}{14cm}
\caption{\small \label{fig:SimpHomBun} Coboundary transformation relating the dimension-0 maps and dimension-1 maps in diagram (i) and the dimension-2 maps in diagram (ii) of two principal $\CCG$-bundles $g,\tilde g:N(\check \CC(\frU))\rightarrow \CCG$ for $\CCG$ a Lie 2-quasi-groupoid. The notation $h_{aab,011}^{-1}$ is suggestive; it denotes the horn filler $h_{aab,011}$ with an arrow between the filled face and its opposite vertex.}
\end{minipage}
\end{center}
\end{figure}

\noindent
We thus regard $g_{ab}$ as the simplicial 1-simplex connecting the simplicial 0-simplices $g_b$ and $g_a$ as is depicted in diagram (i) of Figure \ref{fig:TransCocPrin}. Similarly, $g_{abc}$ is the simplicial 2-simplex connecting the simplicial 1-simplices $g_{bc}$ and $g_{ab}$ to the 1-simplex $g_{ac}$ as is depicted in diagram (ii) of Figure \ref{fig:TransCocPrin}.  Moreover, we have a unique horn filler $g_{abcd}$ for the 3-horn $(-,g_{acd},g_{abd},g_{abc})$. This can be seen as connecting the simplicial 2-simplices $g_{abc}$, $g_{abd}$, and $g_{acd}$ to the simplicial 2-simplex $g_{bcd}$, which, pictorially is given by a tetrahedron as is depicted in diagram (iii) of Figure \ref{fig:TransCocPrin}.  Filling the 3-horn $(-,g_{acd},g_{abd},g_{abc})$ then amounts to regarding these three 2-simplices as faces of the tetrahedron $g_{abcd}$ with the fourth face $g_{bcd}$. Altogether, the diagrams (i)--(iii) of Figure \ref{fig:TransCocPrin} are then to be understood as the \v Cech cocycle conditions of a quasi-groupoid bundle with structure Lie 2-quasi-groupoid $\CCG$. 

\begin{figure}[h]
\begin{center}
\tikzset{->-/.style={decoration={markings,mark=at position #1 with {\arrow{>}}},postaction={decorate}}}
\begin{tikzpicture}[scale=0.7,every node/.style={scale=0.7}]
   %\filldraw[fill=green!20,draw=white] (0,0) -- (0,5) -- (5,8) -- cycle;
   \filldraw[pattern=north west lines, pattern color=green!20,draw=white] (0,0) -- (0,5) -- (5,8) -- cycle;
   %\filldraw[fill=blue!20,draw=white] (7,1) -- (7,6) -- (12,3) -- cycle; 
   \filldraw[pattern=north west lines, pattern color=blue!20,draw=white] (7,1) -- (7,6) -- (12,3) -- cycle;    
   %\filldraw[fill=red!20,draw=white] (4,-4) -- (9,-1) -- (14,-4) -- cycle;  
   \filldraw[pattern=north west lines, pattern color=red!20,draw=white] (4,-4) -- (9,-1) -- (14,-4) -- cycle; 
   \draw (-0.2,-0.2) node {$\tilde g_c$};
   \draw (1.8,-2.2) node {$\tilde g_c$};
   \draw (3.8,-4.2) node {$\tilde g_c$};
   \draw (6.7,1.2) node {$\tilde g_b$};
   \draw (8.7,-0.8) node {$\tilde g_b$};
   \draw (14.3,-4.2) node {$\tilde g_a$};
   \draw (-0.2,5.2) node {$g_c$};
   \draw (5,8.3) node {$g_b$};
   \draw (7,6.3) node {$g_b$};
   \draw (10.3,5.2) node {$g_a$};
   \draw (12.3,3.2) node {$g_a$};
   \draw (14.3,1.2) node {$g_a$};
   \draw [-implies,double equal sign distance,thick] (10,1.3) -- (11,0.3);
   \draw (10.3,.5) node {$h_{ab}$};
   \draw [-implies,double equal sign distance,thick] (1.5,1.3) -- (2.5,0.3);
   \draw (1.8,.5) node {$h_{bc}$};
   \draw [-implies,double equal sign distance,thick] (5,6.8) -- (5,5.8);
   \draw (5.5,6.3) node {$g_{abc}$};
   \draw [dashed,-implies,double equal sign distance,thick] (9,-2.2) -- (9,-3.2);
   \draw (9.5,-2.7) node {$\tilde g_{abc}$};
   \draw (-0.6,2.5) node {\color{green!50!black}$h_{cc,0}$};
   \draw (6.5,3.8) node {\color{blue!50!black}$h_{bb,0}$};
   \draw (14.55,-1.5) node {\color{red!50!black}$h_{aa,0}$};
   \draw (2.4,6.8) node {\color{green!50!black} $g_{bc}$};
   \draw (7.8,6.8) node {\color{green!50!black} $g_{ab}$};
   \draw (9.9,4.641) node {\color{blue!50!black}$g_{ab}$};
   \draw (5,4.7) node {$g_{ac}$};
   \draw (6.5,-2.1) node {\color{red!50!black}$\tilde g_{bc}$};
   \draw (4.5,0) node {\color{blue!50!black}$\tilde g_{bc}$};
   \draw (11.5,-2.1) node {\color{red!50!black}$\tilde g_{ab}$};
   \draw (9,-4.3) node {$\tilde g_{ac}$};
   %green tetrahedron
   \draw[draw=green!50!black,very thick,->-=.5] (0,0) -- (0,5); 
   \draw[draw=green!50!black,very thick,->-=.5] (0,5) -- (5,8); 
   \draw[draw=green!50!black,very thick,->-=.5]  (5,8) -- (10,5);
   \draw[->-=.5] (0,5) -- (10,5);
   \draw[->-=.5] (0,0) -- (10,5);
   \draw[dashed,->-=.5] (0,0) -- (5,8);   
   \draw [-implies,double equal sign distance,draw=green!50!black,thick] (3.2,4) arc (-40:-180:20pt);
   \draw [-implies,draw=green!50!black,thick] (3.2,4) arc (-40:-180:20pt);
   \draw (3.5,3.4) node {\color{green!50!black}$\sfa_{g_{ab},g_{bc},h_{cc,0}}$};
   %blue tetrahedron
   \draw[->-=.5] (2,-2) -- (12,3);
   \draw[dashed,-] (4.69,2.3) -- (7,6);
   \draw[->-=1] (2,-2) -- (4.69,2.3);
   \draw[draw=blue!50!black,dashed,very thick,->-=.5] (7,1) -- (7,6);
   \draw[draw=blue!50!black,dashed,very thick,->-=.5] (2,-2) -- (7,1);
   \draw[dashed,->-=.5] (7,1) -- (12,3);
   \draw[draw=blue!50!black,very thick,-] (9.268,4.641) -- (12,3);
   \draw[draw=blue!50!black,dashed,very thick,->-=1] (7,6) -- (9.268,4.641);
   \draw [-implies,double equal sign distance,thick] (5,1.5) -- (7.5,-1);
       \draw (6.5,-.5) node {$h_{ac}$};
   \draw [-implies,double equal sign distance,draw=blue!50!black,thick] (6.3,2.8) arc (-140:0:20pt);
   \draw [-implies,draw=blue!50!black,thick] (6.3,2.8) arc (-140:0:20pt);
   \draw (8.3,2.4) node {\color{blue!50!black}$\sfa^{-1}_{g_{ab},h_{bb,0},\tilde g_{bc}}$};
   %red tetrahedron  
   \draw[->-=.5] (4,-4) -- (14,1);
   \draw[->-=.5] (4,-4) -- (14,-4);
   \draw[draw=red!50!black,very thick,->-=.5] (14,-4) -- (14,1);
   \draw[draw=red!50!black,very thick,->-=.5] (4,-4) -- (9,-1);
   \draw[->-=.5] (9,-1) -- (14,1);
   \draw[draw=red!50!black,dashed,very thick,->-=.5] (9,-1) -- (14,-4);
   \draw [-implies,double equal sign distance,draw=red!50!black,thick] (11.2,-1.7) arc (50:170:20pt);
   \draw [-implies,draw=red!50!black,thick] (11.2,-1.7) arc (50:170:20pt);
   \draw (11.5,-1.2) node {\color{red!50!black}$\sfa_{h_{aa,0},\tilde g_{ab},\tilde g_{bc}}$};
   \draw[draw=white!80!black,-] (-.05,0) -- (3.95,-4);
   \draw[draw=white!80!black,-] (.05,0) -- (4.05,-4);
   \draw[draw=white!80!black,-] (9.95,5) -- (13.95,1);
   \draw[draw=white!80!black,-] (10.05,5) -- (14.05,1);
   \draw[draw=white!80!black,-] (6.95,1) -- (8.95,-1);
   \draw[draw=white!80!black,-] (7.05,1) -- (9.05,-1);
   \draw[draw=white!80!black,-] (4.95,8) -- (6.95,6);
   \draw[draw=white!80!black,-] (5.05,8) -- (7.05,6);
   \filldraw [black] (0,0) circle (2pt);
   \filldraw [black] (0,5) circle (2pt);
   \filldraw [black] (5,8) circle (2pt);
   \filldraw [black] (10,5) circle (2pt);
   \filldraw [black] (2,-2) circle (2pt);
   \filldraw [black] (12,3) circle (2pt);
   \filldraw [black] (7,6) circle (2pt);
   \filldraw [black] (7,1) circle (2pt);
   \filldraw [black] (4,-4) circle (2pt);
   \filldraw [black] (14,1) circle (2pt);
   \filldraw [black] (14,-4) circle (2pt);
   \filldraw [black] (9,-1) circle (2pt);
\end{tikzpicture}
\begin{minipage}{14cm}
\caption{\small  \label{fig:SimpHomBunDecom} Decomposition of the coboundary transformation relating the dimension-2 maps of two principal $\CCG$-bundles $g,\tilde g:N(\check \CC(\frU))\rightarrow \CCG$ for $\CCG$ a Lie 2-quasi-groupoid. The labels $h_{ab}$, $h_{bc}$ and $h_{ac}$ denote squares of two triangles as in diagram (i) of Figure \ref{fig:SimpHomBun}.}
\end{minipage}
\end{center}
\end{figure}
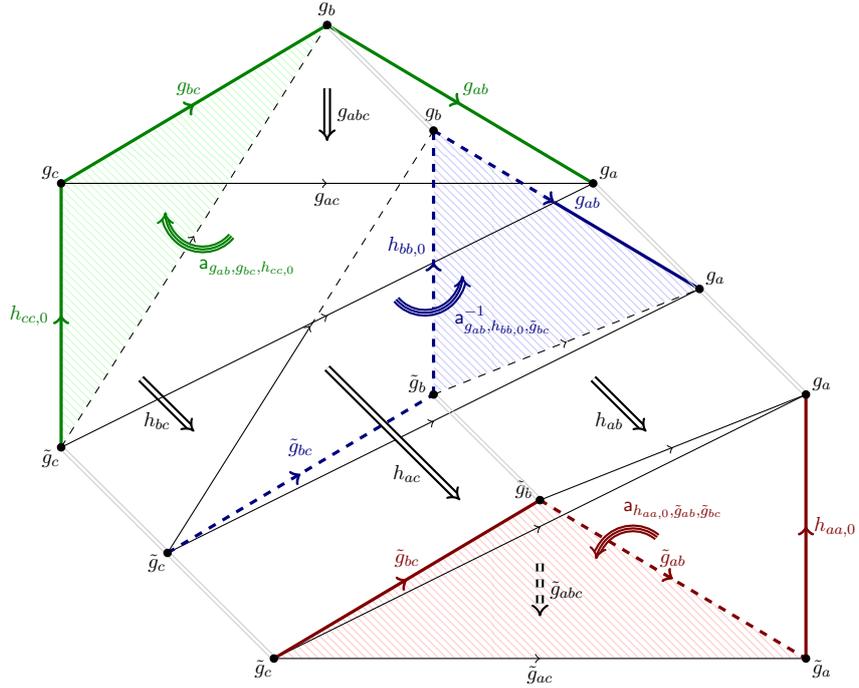

A simplicial homotopy $h$ between $g$ and $\tilde g$ in the case of a Lie 2-quasi-groupoid $\CCG$ has as lowest components
\begin{equation}\label{eq:SimpMapsBundEquiv}
\begin{gathered}
g_a(x)\ =\ h^0((x,a), 0)\eand  \tilde g_a(x)\ =\ h^0((x,a), 1)~,\\
g_{ab}(x)\ =\ h^1((x,a,b),(0,0))\eand \tilde g_{ab}(x)\ =\ h^1((x,a,b),(1,1))~,\\
h_{ab,01}(x)\ :=\ h^1((x,a,b),(0,1))~,\\
g_{abc}(x)\ =\ h^2((x,a,b,c),(0,0,0))\eand \tilde g_{abc}(x)\ =\ h^2((x,a,b,c),(1,1,1))~,\\
h_{abc,011}(x)\ :=\ h^2((x,a,b,c),(0,1,1))~,\\
h_{abc,001}(x)\ :=\ h^2((x,a,b,c),(0,0,1))~.
\end{gathered}
\end{equation}
The fact that $h$ is a simplicial map is reflected in the two diagrams displayed of Figure \ref{fig:SimpHomBun}. Since $\CCG$ is a Kan simplicial manifold, the horn fillers $h_{abb,001}$ and $h^{-1}_{aab,011}$ exist and they connect the pairs of 1-simplices $(g_{ab},h_{bb,01})$ and $(h_{aa,01},\tilde g_{ab})$. Furthermore, diagram (ii) of Figure \ref{fig:SimpHomBun}  can be understood as the composition of the three tetrahedra with vertices $(\tilde g_c,g_c,g_b,g_a)$, $(\tilde g_c,\tilde g_b,g_b,g_a)$, and $(\tilde g_c,\tilde g_b,\tilde g_a,g_a)$, as is displayed in Figure \ref{fig:SimpHomBunDecom}. In this figure,
$h_{ab}$ indicates applying the horn filler $h_{abb,001}$ followed by $h^{-1}_{aab,011}$, respectively.

\begin{rem}\label{rem:W2PB-1}
Let us comment on the connection with the weak principal 2-bundles of Jur\v co {\it et al.}\ \cite{Jurco:2014mva}. For this, we need to specialise the above discussion to a Lie 2-quasi-group, or a Lie 2-quasi-groupoid with a single 0-simplex, $\CCG$. According to Proposition \ref{prop:Duskin}, $\CCG$ is the Duskin nerve of a weak Lie 2-group $\sN\rightrightarrows \sM\rightrightarrows *$ and the simplicial map \eqref{eq:SimpMapsBund} can be identified with the cocycle data $(n_{abc},m_{ab})\in \CC^\infty(U_a\cap U_b\cap U_c,\sN)\times \CC^\infty(U_a\cap U_b,\sM)$ of a weak principal 2-bundle as follows:
\begin{subequations}
\begin{equation}\label{eq:CoCycleWeak2Group-A}
\begin{gathered}
  g_{ab}\ =\ m_{ab} \ewith m_{ab}\,:\,*\ \to\ *\eand m_{aa}\ =\ \id_{*}~,\\
  g_{abc}\ =\ (m_{bc},m_{ac},m_{ab};n_{abc})\ewith n_{abc}\,:\,m_{ab}\otimes m_{bc}\ \Rightarrow\ m_{ac}~,\\
   g_{aab}\ =\ (m_{ab},m_{ab},\id_{*};\sfr_{m_{ab}})\ewith 
   \sfr_{m_{ab}}\,:\,\id_{*}\otimes m_{ab}\ \Rightarrow\ m_{ab}~,\\
     g_{abb}\ =\ (\id_{*},m_{ab},m_{ab};\sfl_{m_{ab}})\ewith \sfl_{m_{ab}}\,:\,m_{ab}\otimes \id_{*}\ \Rightarrow\ m_{ab}~,\\
 g_{abcd}\ =\ (n_{bcd},n_{acd},n_{abd},n_{abc};\sfa_{m_{ab},m_{bc},m_{cd}})\\
 \mbox{with}\quad
 \sfa_{m_{ab},m_{bc},m_{cd}}\,:\,(m_{ab}\otimes m_{bc})\otimes m_{cd}\ \Rightarrow\ m_{ab}\otimes (m_{bc}\otimes m_{cd})
 \end{gathered}
\end{equation}
The existence of the 3-simplex depicted in Figure \ref{fig:TransCocPrin}, (iii), reflects the relation
\begin{equation}
 n_{acd}\circ (n_{abc}\otimes \id_{m_{cd}})\ =\ n_{abd}\circ (\id_{m_{ab}}\otimes n_{bcd})\circ\sfa_{m_{ab},m_{bc},m_{cd}}~.
\end{equation}
\end{subequations}
Likewise, \eqref{eq:SimpMapsBundEquiv} are related to a coboundary $(n_{ab},m_{a})\in \CC^\infty(U_a\cap U_b,\sN)\times \CC^\infty(U_{a},\sM)$ as follows:
\begin{equation}
\begin{gathered}
 n_{ab}\,:\,m_{ab}\otimes m_b\ \Rightarrow\ m_a\otimes\tilde m_{ab}~,\\
 n_{ac}\circ  (n_{abc}\otimes \id_{m_c})\ =\ (\id_{m_a}\otimes \tilde n_{abc})\circ \sfa_{m_a,\tilde m_{ab},\tilde m_{bc}}\circ (n_{ab}\otimes \id_{\tilde m_{bc}})\,\circ  \\
  \kern5cm \circ \, \sfa^{-1}_{m_{ab},m_b,\tilde m_{bc}}\circ (\id_{m_{ab}}\otimes  n_{bc})\circ \sfa_{m_{ab},m_{bc},m_c}~,
\end{gathered}
\end{equation}
where $n_{ab}$ is the composition of the simplices $h^{-1}_{aab,0}$ and $h_{abb,1}$, $m_a:=h_{aa,0}$, and $m_{ab}$ and $n_{abc}$ and their tilded versions are given by \eqref{eq:CoCycleWeak2Group-A}. Notice that the above relations can directly be read off the diagrams displayed in Figures \ref{fig:SimpHomBun} and \ref{fig:SimpHomBunDecom}. Altogether, we found clear correspondences reproducing the coboundary relations between normalised cocycles of weak principal 2-bundles as given in \cite{Jurco:2014mva}. 
\end{rem}

\begin{definition}\label{def:TrivialBundle}
 For $\CCG$ a Lie quasi-groupoid, a principal $\CCG$-bundle $g:N(\check \CC(\frU\rightarrow X))\to \CCG$ is called \uline{trivial} if and only if it is equivalent to a principal $\CCG$-bundle $\tilde g:N(\check \CC(\frU\rightarrow X))\to \CCG$ for which $\tilde g^p_{a_0\cdots a_{p}}(x)=(\sfd^{p-1}_{0}\circ\cdots\circ\sfd^1_0\circ\sfd^0_0\circ\tilde g^0)(x,a_0)\in\CCG_p$ for $p\geq1$.
\end{definition}

Consider a morphism of manifolds $f:X\to Y$. Let $\frU_Y=\dot\bigcup_{a\in A}U_a\rightarrow Y$ be a cover of $Y$ and  $\frU_X:=f^*\dot\bigcup_{a\in A}U_a\to X$ be the cover of  $X$ induced by $f$. Explicitly, $f^*\dot\bigcup_{a\in A}U_a=\bigcup_{a\in A}\{(a,x)\,|\,f(x)\in U_a\}=\dot\bigcup_{a\in A}f^{-1}(U_a)$. Consequently, we have a functor  $F:\check \CC(\frU_X\to X)\to \check \CC(\frU_Y\to Y)$ between the corresponding \v Cech groupoids  $\check\CC(\frU_X\to X)$  and  $\check\CC(\frU_Y\to Y)$  that sends objects $(x,a)\in\check\CC(\frU_X\to Y)$ to objects  $(f(x),a)\in\check\CC(\frU_Y)$ and morphisms $(x,a,b)\in\check\CC(\frU_X\to Y)$ to morphisms  $(f(x),a,b)\in\check\CC(\frU_Y)$, respectively. This enables us to define the pullback bundle and the restriction.

\begin{definition}\label{def:PullBackBundle}
Let $f:X\to Y$ be a morphism of manifolds $X$ and $Y$. Let $\frU_Y=\dot\bigcup_{a\in A}U_a\rightarrow Y$ be a cover of $Y$ and  $\frU_X:=f^*\dot\bigcup_{a\in A}U_a\to X$ be the cover of  $X$ induced by $f$. Furthermore, let $F:\check \CC(\frU_X\to X)\to \check \CC(\frU_Y\to Y)$ be the induced functor between the corresponding \v Cech groupoids  and  $N(F):N(\check \CC(\frU_X\to X))\to N(\check \CC(\frU_Y\to Y))$ the associated simplicial map between their nerves. For $\CCG$ a Lie quasi-groupoid, the \uline{pullback} of a principal $\CCG$-bundle $g:N( \CCC(\frU_Y\to Y))\to\CCG$ is the principal $\CCG$-bundle $g\circ N(F):N( \CCC(\frU_X\to X))\to\CCG$. We shall also write $f^*g$ for $g\circ N(F)$.
\end{definition}

\begin{definition}
Let $f:X\to Y$ be an embedding of a manifold $X$ into a manifold $Y$. For $\CCG$ a Lie quasi-groupoid, the \uline{restriction} of a principal $\CCG$-bundle over $Y$ to $X$ is its pullback along $f$.
\end{definition}

\subsection{Higher base spaces}\label{ssec:higher_base_1}

An important advantage of our construction of Lie quasi-groupoid bundles is its categorical nature. This allows us to generalise the base spaces of such bundles to higher or categorified spaces. Recall that a 2-space in the sense of \cite{Baez:2004in} is a category internal to $\CatDiff$, and in particular Lie groupoids are examples of 2-spaces. 

Notice that already in conventional gauge theory, there is ample motivation for considering Lie groupoids as base spaces. Consider, for example, a manifold $X$ with a group action by a group $\sG$ and suppose we wish to define gauge theory on the orbit space $X/\sG$. This space can be singular, e.g.\ in the case of orbifolds, which complicates a description of field theories. Instead, we can switch to gauge theory on the corresponding action Lie groupoid $\sG\times X\rightrightarrows X$, which is a smooth 2-space, were $\sfs(g,x):=x$ and $\sff(g,x):=gx$ for all $g\in\sG$ and $x\in X$. Moreover, Lie groupoids are presentations of differentiable stacks and there is growing interest on field theories on such spaces. Our construction might not yet be sufficient for an approach to higher gauge theory on differentiable stacks, as it is hard to see how it would respect Morita equivalence. 

The most general model of a higher space fitting our framework is an \uline{inner Kan}  \uline{simplicial manifold}, modelling an $(\infty,1)$-category internal to $\CatDiff$, as we have already discussed in Section \ref{sec:quasiCategory}. To cover such manifolds, we need the notion of surjective submersions.

\begin{definition}
 A \uline{surjective submersion of inner Kan simplicial manifolds} is a map of simplicial manifolds given by surjective submersions.
\end{definition}

\noindent 
Hence, for surjective submersions of inner Kan simplicial manifolds $f:\CCY\rightarrow \CCX$, we have surjective submersions $f_p:\CCY_p\rightarrow \CCX_p$ for their corresponding $p$-simplicial simplices with $p\geq0$. As before, we shall refer to a surjective submersion of inner Kan simplicial manifolds $\CCY\rightarrow \CCX$ as a \uline{cover} of $\CCX$. Note that sometimes, as, for instance, in the context Segal--Mitchison cohomology, it is useful to impose further conditions on a cover, cf.\ \cite{Brylinski:2001aa,Demessie:2014ewa}. Furthermore, we shall need the corresponding notion of a \v Cech nerve for which we employ bisimplicial manifolds.

\begin{definition}
 A \uline{bisimplicial manifold} is a functor $\CCB:\mathbf{\Delta}^{\rm op}\times \mathbf{\Delta}^{\rm op}\rightarrow \CatDiff$. 
\end{definition}

\noindent 
In particular, a bisimplicial manifold consists of $(p,q)$-simplical simplices with horizontal and vertical face maps yielding obvious commutative diagrams. Correspondingly, they can be horizontally or vertically constant and satisfy Kan conditions with respect to horizontal and vertical horns. We shall refer to such bisimplicial manifolds as \uline{horizontally} or {\uline{vertically Kan}.

\begin{definition}
The \uline{\v Cech nerve}  $\check \CCC(\CCY\rightarrow \CCX)$ of a surjective submersion $f:\CCY\to\CCX$ between two inner Kan simplicial manifolds $\CCY$ and $\CCX$ is the bisimplicial manifold whose simplicial $(p,q)$-simplices are 
\begin{subequations}
\begin{equation}
 \check \CCC_{p,q}(\CCY\rightarrow \CCX)\ :=\ \CCY^{[p]}_q
 \end{equation} 
 with
 \begin{equation}
  \CCY^{[p]}_q\ :=\ \{(y^1_q,\ldots,y^p_q)\in \CCY_q\times\cdots\times\CCY_q\,|\,f_q(y^1_q)=\ldots=f_q(y^p_q)\}~.
  \end{equation}
  \end{subequations}
\end{definition}

\noindent 
Note that $\check \CCC_{(\bullet,q)}(\CCY\rightarrow \CCX)=N(\check \CC(\CCY_q\to\CCX_q))$, that is, the restriction of $\check \CCC(\CCY\rightarrow \CCX)$ to the (vertical) simplicial manifolds $\check \CCC_{(\bullet,q)}(\CCU\rightarrow \CCX)$ for fixed $q\geq0$ is the nerve of the \v Cech groupoid $\check\CC(\CCY_q\to\CCX_q)$ for the surjective submersion $f_q:\CCY_q\rightarrow\CCX_q$. Since the nerve of any groupoid is always Kan (see Proposition \ref{prop:ss_is_groupoid}), it follows that the \v Cech nerve $\check \CCC(\CCY\rightarrow \CCX)$ is always vertically Kan. Hence, all vertical horns have horn fillers. However, since the simplicial manifold $\CCX$ was not necessarily Kan to begin with, $\check \CCC(\CCY\rightarrow \CCX)$ will not be horizontally Kan in general (but it will always be inner Kan).

As higher gauge groupoid, we can now either lift the simplicial manifold underlying a Lie quasi-groupoid to a horizontally constant bisimplicial manifold or we allow for a more general, bisimplicial manifold which is vertically Kan.\footnote{As an example of the latter possibility we could consider a bisimplicial manifold associated with a 2-nerve of a (Lie) 2-groupoid \cite{Tamsamani:1999aa, Simpson:9704006, Lack:0607271}.} We generalise our previous definitions as follows.

\begin{definition}
Let $\CCX$ be an inner Kan simplicial manifold and $\CCY\rightarrow \CCX$ a fixed cover of $\CCX$. For $\CCG$ a vertically Kan bisimplicial manifold, a \uline{principal $\CCG$-bundle} over $\CCX$ subordinate to the cover is a bisimplicial map 
 \begin{equation}
  g\,:\, \check \CCC(\CCY\rightarrow \CCX)~\rightarrow~\CCG~.
 \end{equation}
 Two such principal $\CCG$-bundles $g,\tilde g$ are called \uline{equivalent} if and only if there is a bisimplicial homotopy between $g$ and $\tilde g$.
\end{definition}

Notice that the special case of an inner Kan simplical manifold $N(T^*M\rightrightarrows M)$ was discussed in \cite{Ritter:2015zur} using double categories. Our discussion in this section generalises this picture.

\section{Differentiation of Lie quasi-groupoids}\label{sec:differentiation}

In order to endow our Lie quasi-groupoid bundles with connections, we first have to differentiate the structure groupoids to obtain gauge Lie quasi-algebroids. Here, we shall follow ideas presented by \v Severa \cite{Severa:2006aa}; see also Li \cite{Li:2014} for a more detailed picture. The presentation in \cite{Severa:2006aa} is very concise and somewhat abstract, and we give a detailed and constructive derivation that is readily applied to any explicit Lie quasi-groupoid. We then extend this approach, discussing equivalence relations between Lie quasi-algebroids, from which gauge transformations of connections on Lie quasi-groupoid bundles can be gleaned. 

\subsection{Presheaves on the category of supermanifolds}\label{ssec:presheaves_of_supermanifolds}

Recall the following canonical definitions, cf.\ \cite{Manin:1988ds} or \cite{Deligne:1999qp}. All our definitions involving a grading apply to $\RZ$-, $\RZ_2$, or $\NN$-gradings.

\begin{definition}
 A \uline{graded manifold} $X$ is a locally ringed space $X=(X_{\rm red},\CS_X)$ where $X_{\rm red}$ is a manifold, called the \uline{body} of $X$, and $\CS_X$ a sheaf of graded rings, called the \uline{structure sheaf}. A \uline{function on a graded manifold} is a (global) section of the structure sheaf, and the ring of global sections of the structure sheaf is denoted by $\CC^\infty(X):=\Gamma(X,\CS_X)$.
\end{definition}

\noindent
A typical example is the supermanifold $\FR^{m|n}:=(\FR^m,\CS_{\FR^m}\otimes\Lambda^\bullet\FR^n)$. In general, we can associate with any vector bundle $E$ over a manifold $X$ the supermanifold $\Pi E:=(X, \CS_X\otimes \Lambda^\bullet E)$. Here, $\Pi$ denotes the parity changing functor on linear spaces and $\Pi E$ is $E$ with the Gra{\ss}mann-parity of the fibres reversed. One can show that any real supermanifold must be of this form \cite{JSTOR:1998201}.  

Analogously, we have the $\NN$-graded manifold $E[1]$, where the square brackets $[k]$ indicate a shift of the degree of the elements of the relevant linear space by $k$. In particular, $E[1]$ is $E$ with degrees of the fibres shifted by $1$. As such, it is an $\NN$-graded manifold concentrated in degrees $0$ and $1$. We can apply a forgetful functor to $\NN$- or $\RZ$-graded manifolds, mapping them to supermanifolds.

\begin{definition}
Let $X=(X_{\rm red},\CS_X)$ and  $Y=(Y_{\rm red},\CS_Y)$ be two graded manifolds.
 A \uline{morphism of graded manifolds} $\varphi:X\to Y$ is a pair $\varphi=(\phi,\psi)$, where $\phi:X_{\rm red}\rightarrow Y_{\rm red}$ is a morphism of manifolds and $\psi:\CS_Y\rightarrow \phi_*\CS_X$ is a local morphism of graded rings between $\CS_Y$ and the zeroth direct image of $\CS_X$ under $\phi$.\footnote{Let $\phi:X\to Y$ be a morphism of manifolds and $\CS$ a sheaf on $X$. Then, the $q$-th direct image of $\CS$ under $\phi$ is the sheaf defined by the presheaf $V\mapsto H^q(\phi^{-1}(V),\CS)$ for $V\subseteq Y$.}
\end{definition}

\noindent
The category of supermanifolds $\CatSMfd$ is the category of $\RZ_2$-graded manifolds together with morphisms of $\RZ_2$-graded manifolds. Furthermore, by virtue of the above definition, $\shom_{\CatSMfd}(X,\FR)\subseteq \CC^\infty(X)$ for any supermanifold $X=(X_{\rm red},\CS_X)$ as  $\shom_{\CatSMfd}(X,\FR)\cong\Gamma(X,\CS_{\rm ev})$, where $\CS_{\rm ev}$ is the subsheaf of $\CS$ generated by even elements of $\CS$.

Also recall that a presheaf on a category $\CC$ is a functor $\CC^{\rm op}\rightarrow \CatSet$ and the set of presheaves $\hat \CC$ on $\CC$ generalises the objects in $\CC$ via the embedding $\CC_0\embd \hat \CC:X\mapsto \shom_\CC(-,X)$. We are here interested in certain presheaves arising as parameterised maps. For any $X,Y\in \CatSMfd$, we have the presheaf $\inthom(X,Y)$ on $\CatSMfd$, which is defined as
\begin{equation}
 \inthom(X,Y)(Z)\ :=\ \shom_{\CatSMfd}(X\times Z,Y)
\end{equation}
for $Z\in\CatSMfd$. If this presheaf is representable, i.e.\ $\inthom(X,Y)=\shom_{\CatSMfd}(-,Z)$ for some $Z\in \CatSMfd$, then it is called the \uline{internal hom functor}.

Following \cite{Kochan:0307303,Severa:2006aa}, we note that we may identify $\inthom(\FR^{0|1},X)$ with the grade-shifted tangent bundle $T[1]X$. To see this, let $X$ and $Y$ be two supermanifolds and choose local coordinates  $x=(x^i)$ on $X$, let $y\in Y$, and denote the  Gra{\ss}mann-odd generator of $\CS_{\FR^{0|1}}$ by $\theta$. Then,
\begin{equation}
 x^i(y,\theta)\ =\ \mathring{x}^i(y)+\xi^i(y)\theta
\end{equation}
describes an element of $\inthom(\FR^{0|1},X)(Y)$. Here, $\mathring{x}^i$ describes a morphism $f:Y\to X$ and $\xi^i$ a section  $Y\to f^* T[1]X$. This thus establishes the identification $\inthom(\FR^{0|1},X)\cong T[1]X$. Contrarily, $\shom_{\CatSMfd}(\FR^{0|1},X)\cong X$ since such maps are fully characterised by a morphism between the bodies of the supermanifolds, $*\rightarrow X$. Generally, we have the following proposition \cite{Severa:2006aa}.

\begin{prop}\label{lem:R0_k_homom}
  Let $X\in\CatSMfd$. Then, 
  \begin{equation}
    \inthom(\FR^{0|n},X)\ \cong\ (T[1])^nX~,
  \end{equation}
where $(T[1])^nX$ is the $n$-th iterated Gra{\ss}mann-odd tangent bundle of $X$.
\end{prop}

\noindent 
Explicitly, an element $f\in \inthom(\FR^{0|n},X)(Y)$ is given by the formal series expansion
\begin{equation}\label{eq:expansion_inner_hom}
 f(\theta_1,\ldots,\theta_n)\ =\ \mathring{f}+\sum_{j=1}^n\sum_{1\leq i_1<\cdots<i_j\leq n} \alpha_{i_1\cdots i_j}\theta_{i_1}\theta_{i_2}\cdots\theta_{i_j}~,
\end{equation}
where the $\theta_1,\ldots,\theta_n$ are the generators of $\CS_{\FR^{0|n}}$, $\mathring{f}\in \shom_{\CatSMfd}(Y,X)$, and the $\alpha_{i_1\cdots i_j}$s are sections of $\mathring{f}^*(T[1])^{j}X$ of degree $j$ in the category of $\NN$-graded manifolds and of degree $j\!\mod 2$ in the category of supermanifolds.

Note that there is a diagonal action of $\inthom(\FR^{0|1},\FR^{0|1})$ on $\FR^{0|n}$ and therefore on $f$ which lifts the $\RZ_2$-grading of the individual maps $\mathring{f}$ and $\alpha_{i_1\cdots i_j}$ to a $\RZ$-grading and endows the resulting $\RZ$-graded manifold with a differential. We therefore have the following.

\begin{cor}\label{cor:moduli_internal_hom}
For $X\in\CatSMfd$, there is an action of $\inthom(\FR^{0|1},\FR^{0|1})$ on $ \inthom(\FR^{0|n},X)\cong(T[1])^nX$ which induces a differential of degree 1 on $\CS_{(T[1])^nX}$.
\end{cor}

\noindent 
Specifically, we have a differential $Q$ acting on $f$ as in \eqref{eq:expansion_inner_hom} according to \cite{Severa:2006aa,Jurco:2014mva}
\begin{equation}\label{eq:differential_on_moduli}
 (Q f)(\theta_1,\ldots,\theta_n)\ :=\ \dder{\eps}f(\theta_1+\eps,\ldots,\theta_n+\eps)
\end{equation}
with $\varepsilon$ being Gra{\ss}mann-odd. This then induces the action on $\CS_{(T[1])^nX}$.

\subsection{\texorpdfstring{$L_\infty$}{L-infinity}-algebroids}\label{ssec:NQ_manifolds}

Recall the following definition.

\begin{definition}
  A \uline{Lie algebroid} $(X,E,[-,-],\rho)$ over a manifold $X$ is a vector bundle $E\rightarrow X$ together with a Lie bracket $[-,-]$ on its sheaf of sections $\CE_X$ and a morphism of vector bundles, called the \uline{anchor map} $\rho:E\rightarrow TX$, which is a Lie algebra homomorphism and satisfies $[s_1,fs_2]=(\rho(s_1)f)s_2+f[s_1,s_2]$ for all $f\in \CS_X(U)$ and $s_{1,2}\in \CE_X(U)$ and $U\subseteq X$.
\end{definition}
\noindent Note that for $X=*$, a Lie algebroid is simply a Lie algebra. Just as a Lie algebra arises from differentiating a Lie group, we can differentiate a given Lie groupoid $\CG_1\rightrightarrows\CG_0$ to a Lie algebroid 
\begin{equation}
  \sLie(\CG_1\rightrightarrows\CG_0)\ :=\ \ker \sfs_*\ =\ \dot\bigcup_{g_0\in \CG_0}T_{\id_{g_0}}\sft^{-1}(g_0)\ \subseteq\ T\CG_1~,
\end{equation}
where $\sfs_*$ and $\sft_*$ are the linear maps induced by $\sfs$ and $\sft$, the anchor map is given by the restriction of $\sft_*$ to $\sLie(\CG_1\rightrightarrows\CG_0)$, and the Lie bracket arises from the Lie bracket on vector fields. For example, given a manifold $Y$, we have a trivial surjective submersion $Y\rightarrow *$. The corresponding \v Cech groupoid $\check\CC(Y\rightarrow *)$ is the pair groupoid $Y\times Y\rightrightarrows Y$ of $Y$  and its Lie algebroid is simply $TY$. Note, however, that contrary to the case of Lie algebras, there are Lie algebroids which do not arise from differentiating a Lie groupoid \cite{Crainic:0105033}.
 
The description of Lie algebroids that allows for a straightforward categorification and that we shall use in the following is based on N$Q$-manifolds \cite{Severa:2001aa,Roytenberg:0203110}.\footnote{In physics terminology, $Q$ would play the role of a BRST operator and the grading corresponds to the ghost number.}
 
\begin{definition}
An \uline{N-manifold} is an $\NN_0$-graded manifold. A \uline{homological vector field} $Q$ on an N-manifold is a nil\-quadratic vector field of degree 1. An N-manifold endowed with a homological vector field $Q$ is called an \uline{N$Q$-manifold}. \uline{Morphisms of N$Q$-manifolds} are morphisms of graded manifolds which respect the action of the respective homological vector fields.
\end{definition} 

\noindent 
A simple example of an N$Q$-manifold is given by the usual Chevalley--Eilenberg description of a Lie algebra $\frg$: the N-manifold is $\frg[1]$ and $Q$ is the Chevalley--Eilenberg differential.

This is readily generalised to Lie algebroids, which are captured by N$Q$-manifolds concentrated in degrees $0$ and $1$. Such an N$Q$-manifold is necessarily a vector bundle\footnote{This is equivalent to saying that such an N$Q$-manifold is a split supermanifold, which already follows from results of \cite{JSTOR:1998201}.} $E[1]\rightarrow X$ \cite{Roytenberg:0203110} and in terms of some local coordinates $(x^i)=(x^1,\ldots,x^m)$ and $(\xi^\alpha)=(\xi^1,\ldots,\xi^n)$ on the base and on the fibres, respectively, the vector field $Q$ is of the form
\begin{equation}
 Q\ :=\ -\xi^\alpha\rho_{\alpha}^i(x) \der{x^i}-\tfrac12\xi^\alpha\xi^\beta {f_{\alpha\beta}}^\gamma(x)\der{\xi^\gamma}~.
\end{equation}
Here, $\rho^i_\alpha(x)$ encodes the anchor map $\rho$ and ${f_{\alpha\beta}}^\gamma(x)$ are the structure functions of a pointwise Lie bracket on $\Gamma(E)$. The condition $Q^2=0$ ensures that the ${f_{\alpha\beta}}^\gamma(x)$ satisfy the Jacobi identity, that $\rho$ is a Lie algebra homomorphism, and that the Lie bracket and $\rho$ are connected via a Leibniz rule. 

This motivates the following generalisation, see \cite{Severa:2001aa}.

\begin{definition}
 An \uline{$n$-term $L_\infty$-algebroid} is an N$Q$-manifold that is concentrated in degrees $0\leq p\leq n$. If $n$ is arbitrarily large, we shall simply speak of an \uline{$L_\infty$-algebroid}. Furthermore, we shall refer to the degree-0 component of the N$Q$-manifold as the \uline{base} of the $L_\infty$-algebroid. If the base is trivial, i.e.\ it consists of a single element, then we have an \uline{$L_\infty$-algebra}.
\end{definition}

\noindent 
Note that the vector field $Q$ acts on the algebra of functions on an $L_\infty$-algebroid $\frg$, which we write as
\begin{equation}
 \CC^\infty(\frg)\ =\ \mathsf{Sym}^\bullet_{\CC^\infty(\frg_0)}(\frg^\vee)\ =\ \CC^\infty(\frg_0)\otimes \big(\FC~\oplus~\frg_1^\vee~\oplus~(\frg_2^\vee\,\oplus\,\frg_1^\vee\odot \frg_1^\vee)~\oplus~\cdots\big)~,
\end{equation}
where $\frg^\vee$ is the dual of $\frg$. All of the above straightforwardly generalises to Lie $\infty$-superalgebroids or $L_\infty$-superalge\-broids using N$Q$-supermanifolds.

An important example of an N$Q$-manifold is the iterated grade-shifted tangent bundle\linebreak $\inthom(\FR^{0|n},X)\cong(T[1])^nX$ from Corollary \ref{cor:moduli_internal_hom}, where the differential of degree 1 plays the role of the vector field $Q$, hence the notation in \eqref{eq:differential_on_moduli}.

\subsection{1-jets of Lie quasi-groupoids}\label{ssec:1jets}

Consider the category $\CatSurSub$ of surjective submersions which has surjective submersions $Y\to X$ between supermanifolds $Y$ and $X$ as its objects and maps  as its morphisms such that ($Y_{1,2}\to X_{1,2}$ are surjective submersions)
\begin{equation}
    \myxymatrix{
    Y_1\ar@{->}[r] \ar@{->}[d] & Y_2\ar@{->}[d]\\
    X_1 \ar@{->}[r] & X_2
    }
\end{equation}
are commutative. As before, let $\CatsSMfd:=\CatFun(\mathbf{\Delta}^{\rm op},\CatSMfd)$ be the category of simplicial supermanifolds. 

Since the nerve $N$ of the \v Cech groupoid of an object in $\CatSurSub$ is an object of $\CatsSMfd$, an object $\CCX\in \CatsSMfd$ gives rise to a $\CatSet$-valued presheaf
\begin{equation}
 \shom_{\CatsSMfd}(N(-),\CCX)\,:\, \CatSurSub^{\rm op}\ \to\ \CatSet
\end{equation}
on $\CatSurSub$. We are now interested in linearisations of this presheaf, which we shall call $k$-jets in analogy with the well-known construction in differential geometry. In particular, taking the 1-jet of a quasi-groupoid turns out to be the appropriate higher analogue of the Lie functor, differentiating a Lie group to a Lie algebra.

The linearisation of the presheaf $ \shom_{\CatsSMfd}(N(-),\CCX)$ is performed by restricting to the full subcategory $\CatSurSub_k$ of $\CatSurSub$, whose objects are surjective submersions of the form $X\times \FR^{0|k}\rightarrow X$. We have 
\begin{equation}\label{eq:SurSubIso}
\begin{gathered}
 \shom_{\CatSurSub_k}(X_1\times \FR^{0|k}\rightarrow X_1,X_2\times \FR^{0|k}\rightarrow X_2)\ \cong\ 
 \\ \cong\ \shom_{\CatSMfd}(X_1,X_2)\times \shom_{\CatSMfd}(X_1\times \FR^{0|k},\FR^{0|k})~,
\end{gathered}
\end{equation}
which follows from the fact that for any two trivial fibrations $X_{1,2}\times E_{1,2}\to X_{1,2}$, a fibre-preserving map $\phi:X_1\times E_1\to X_2\times E_2$ is of the form
\begin{equation}
\phi\,:\,(x_1,f_1)\ \mapsto\ (x_2,f_2)\ :=\ (\phi_1(x_1),\phi_2(x_1,f_1))
\end{equation}
for $\phi_1\in \shom_{\CatSMfd}(X_1,X_2)$ and $\phi_2\in\shom_{\CatSMfd}(X_1\times E_1,E_2)$.

Because of the identification \eqref{eq:SurSubIso}, a presheaf on $\CatSurSub_k$ can equivalently be described by a presheaf on $\CatSMfd$ together with an action of  $\inthom(\FR^{0|k},\FR^{0|k})$. Let us denote $\CatSMfd$ together with an action by $\inthom(\FR^{0|k},\FR^{0|k})$ by $\CatSMfd_k$. For instance, $\CatSMfd_1$ is the category of N$Q$-supermanifolds since the action of  $\inthom(\FR^{0|1},\FR^{0|1})$ corresponds to the action of the vector field $Q$.  We then give the following definition \cite{Severa:2006aa}.
\begin{definition}
Given a presheaf on $\CatSurSub$, its restriction to $\CatSurSub_k$ yields a presheaf on $\CatSMfd_k$ which we call the  \uline{$k$-jet} of the presheaf on $\CatSurSub$. The \uline{$k$-jet of a simplicial} \uline{supermanifold} $\CCX$ is the $k$-jet of the presheaf $\shom_{\CatsSMfd}(N(-),\CCX)$.
\end{definition}

As stated above, we are particularly interested in the $k$-jets of a Lie $n$-quasi-groupoid $\CCG$, and \v Severa showed in \cite{Severa:2006aa}, that  the $k$-jets of $\CCG$ are representable as presheaves on $\CatSMfd$ for every $k\in \NN$, that is,  they are of the form $\shom_{\CatSMfd}(-,Z)$ for some $Z\in \CatSMfd$. Our constructive proof of this is given by an explicit construction of the supermanifold $Z$, which we explain in great detail for $k=1$. As we shall see, $Z$ carries an action of $\inthom(\FR^{0|1},\FR^{0|1})$, which induces the structure of a differential graded manifold. The resulting N$Q$-manifold is then the $L_\infty$-algebroid of the Lie $n$-quasi-groupoid $\CCG$. 

Let $\CCG$ be a Lie $n$-quasi-groupoid. The 1-jet of $\CCG$ corresponds to a functor from $\CatSMfd^{\rm op}$ to $\CCG$-valued descent data with respect to surjective submersions $X\times \FR^{0|1}\rightarrow X$ for $X\in \CatSMfd$. Equivalently, we have a principal $\CCG$-bundle over $X$ subordinate to the cover $X\times \FR^{0|1}\rightarrow X$ given in terms of a simplicial map $g:N(\check\CC(X\times\FR^{0|1}\to X))\to\CCG$, cf.~Definition \ref{def:TrivialBundle}. Note that the fibred products of $X\times\FR^{0|1}\to X$ are given by
\begin{equation}
 \underbrace{(X\times\FR^{0|1})\times_X(X\times\FR^{0|1})\times_X\cdots\times_X(X\times\FR^{0|1})}_{n{\rm -times}}\ \cong\ X\times \FR^{0|n}~,
\end{equation}
and therefore we have $N_p(\check\CC(X\times\FR^{0|1}\to X))\cong X\times \FR^{0|p+1}$ with face and degeneracy maps
\begin{equation}
\begin{aligned}
 \sff^p_i(x,\theta_0,\theta_1,\ldots,\theta_p)\ &:=\ (x,\theta_0,\ldots,\theta_{i-1},\theta_{i+1},\ldots,\theta_p)~,\\
 \sfd^p_i(x,\theta_0,\theta_1,\ldots,\theta_p)\ &:=\ (x,\theta_0,\ldots,\theta_{i-1},\theta_i,\theta_i,\ldots,\theta_p)
\end{aligned}
\end{equation}
with $x\in X$ and $\theta_i\in\FR^{0|1}$ for $i\in\{0,\ldots,p\}$. Thus, the simplicial map $g$ consists of maps $g^p\in \inthom(\FR^{0|p+1},\CCG_p)$ which, when evaluated on a supermanifold $X$, have a formal expansion
\begin{equation}\label{eq:gFormExp}
 g^p(x,\theta_0,\ldots,\theta_p)\ =\ \mathring{g}^p(x)+\sum_{j=0}^p\sum_{0\leq i_0<i_1<\cdots<i_j\leq p} \gamma^p_{i_0i_1\cdots i_j}(x)\theta_{i_0}\theta_{i_1}\cdots\theta_{i_j}~.
\end{equation}
Recall from our discussion in Section \ref{ssec:presheaves_of_supermanifolds} that the coefficients $\gamma^p_{i_0\cdots i_j}(x)\in (T[1])^{p+1}\CCG_p$ are of degree $j+1$; for instance, for $p=0$, we have $g^0(x,\theta_0)=\mathring{g}^0(x)+\gamma^0_0(x)\theta_0$ with $\mathring{g}^0(x)\in\CCG_0$ and  $\gamma^0_0(x)\in T_{\mathring{g}^0(x)}[1]\CCG_0$. In the following, we shall refer to $\gamma^p_{012\cdots p}(x)$  as the top component of  $g^p$.

We are now ready to construct $g:N(\check\CC(X\times\FR^{0|1}\to X)\to\CCG$ recursively, following \v Severa \cite{Severa:2006aa}. We start by defining the truncation $\CCZ^{(k)}$ of $N(\check\CC(X\times\FR^{0|1}\to X))$ at the simplicial $k$-simplices as follows:
\begin{equation}\label{eq:examples_Y_k}
 \begin{aligned}
  \CCZ^{(0)}\ &=\ \{(x,\underbrace{0,\ldots,0}_{r_0{\rm-times}})\,|\,r_0\in\NN_0\}~,\\
  \CCZ^{(1)}\ &=\ \{(x,\underbrace{\theta_0,\ldots,\theta_0}_{r_0{\rm-times}},\underbrace{0,\ldots,0}_{r_1\,{\rm-times}})\,|\,r_0,r_1\in\NN_0\}~,\\
  \CCZ^{(2)}\ &=\ \{(x,\underbrace{\theta_0,\ldots,\theta_0}_{r_0{\rm-times}},\underbrace{\theta_1,\ldots,\theta_1}_{r_1{\rm-times}},\underbrace{0,\ldots,0}_{r_2\,{\rm-times}})\,|\,r_0,r_1,r_2\in\NN_0\}~,\\
  &~\,\vdots
 \end{aligned}
\end{equation}
This yields a filtration of simplicial sets $\CCZ^{(0)}\subseteq \CCZ^{(1)}\subseteq\cdots\subseteq N(\check\CC(X\times\FR^{0|1}\to X))$ which we can use to iteratively construct simplicial maps $g^{(k)}:\CCZ^{(k)}\rightarrow \CCG$ with 
\begin{equation}\label{eq:reccon}
  g^{(k)}\big|_{\CCZ^{(k-1)}}\ =\ g^{(k-1)}\big|_{\CCZ^{(k-1)}}
\end{equation}
that converge towards $g$. The form of the $\CCZ^{(k)}$ implies that the $g^{(k)}$ are completely fixed by\footnote{Here, we have suppressed the $x$-dependence and we shall continue doing so in the following.}
\begin{equation}\label{eq:gFormExp-3}
g^{(k)}(\theta_0,\theta_1,\ldots,\theta_{k-1},0)\ =\ \mathring{g}^{(k)}+\sum_{j=0}^{k-1}\sum_{0\leq i_0<i_1<\cdots<i_j\leq k-1} \gamma^{(k)}_{i_0i_1\cdots i_j}\theta_{i_0}\theta_{i_1}\cdots\theta_{i_j}~,
\end{equation}
since the maps on all higher-dimensional and lower-dimensional simplicial simplices follow from the application of appropriate degeneracy and face maps. Because of \eqref{eq:reccon}, all the coefficients in \eqref{eq:gFormExp-3} apart from the top component are fixed by the lower level maps $g^{(0)}$, $g^{(1)}$, $\ldots$, and $g^{(k-1)}$. 

Given the maps $g^{(k)}$, we readily derive the component maps of the simplicial map $g$ as
\begin{equation}
 g^p(\theta_0,\ldots,\theta_p)\ =\ (\sff^{p+1}_{p+1}\circ g^{(p+1)})(\theta_0,\ldots,\theta_p,0)~.
\end{equation}

Recall from Corollary \ref{cor:moduli_internal_hom} that $(T[1])^{p+1}\CCG_p$ is endowed with the homological vector field $Q$,
\begin{equation}\label{eq:QonG}
 Qg^{(k)}(x,\theta_0,\ldots,\theta_j)\ :=\ \dder{\eps}g^{(k)}(x,\theta_0+\eps,\ldots,\theta_j+\eps)~.
\end{equation}
It is therefore convenient to change coordinates on $\CCZ^{(k)}$ according to
\begin{equation}\label{eq:ChangeTheta}
 \hat\theta_0\ :=\ \theta_0-\theta_1~,\quad\hat\theta_1\ :=\ \theta_1-\theta_2~,\quad\ldots~,\quad
 \hat\theta_{k-1}\ :=\ \theta_{k-1}-\theta_k~,\quad\hat\theta_k\ :=\ \theta_k~,
\end{equation}
since then the $\varepsilon$-shift will occur only in $\hat\theta_k$. Correspondingly, \eqref{eq:gFormExp} becomes
\begin{equation}\label{eq:gFormExp-2}
\begin{aligned}
 &g^{(k)}(\theta_0,\theta_1,\ldots,\theta_{k-1},0)\ =\ g^{(k)}\big(\theta_0(\hat\theta_0,\ldots,\hat\theta_k),\ldots,\theta_p(\hat\theta_0,\ldots,\hat\theta_k)\big)\ =\\ 
 &\kern1cm=\ \mathring{g}^{(k)}+\sum_{j=0}^{k-1}\sum_{0\leq i_0<i_1<\cdots<i_j\leq k-1} \hat\gamma^{(k)}_{i_0i_1\cdots i_j}\hat\theta_{i_0}\hat\theta_{i_1}\cdots\hat\theta_{i_j}~.
 \end{aligned}
\end{equation}
 Explicitly, the expansions of the $g^{(k)}$ for $k=0,1,2$ read as
\begin{equation}
\begin{aligned}
 g^{(0)}(\theta_0)\ &=\ \mathring{g}^{(0)}+\hat\gamma^{(0)}_0\theta_0~,\\
 g^{(1)}(\theta_0,\theta_1)\ &=\  \mathring{g}^{(1)}+\hat\gamma^{(1)}_0(\theta_0-\theta_1)+\hat\gamma^{(1)}_1\theta_1+\hat\gamma^{(1)}_{01}\theta_0\theta_1~,\\
  g^{(2)}(\theta_0,\theta_1,\theta_2)\ &=\  \mathring{g}^{(2)}+\hat\gamma^{(2)}_0(\theta_0-\theta_1)+\hat\gamma^{(2)}_1(\theta_1-\theta_2)+\hat\gamma^{(2)}_2\theta_2\,+\\
  &\kern1cm+\hat\gamma^{(2)}_{01}(\theta_0\theta_1-\theta_0\theta_2+\theta_1\theta_2)+\hat\gamma^{(1)}_{02}(\theta_0\theta_2-\theta_1\theta_2)+\hat\gamma^{(2)}_{012}\theta_0\theta_1\theta_2~.
\end{aligned}
\end{equation}

Let us now discuss the constructing of the low $g^{(k)}$ in detail, also in view of later applications, and then come to the general case.

At zeroth level, $g^{(0)}:\CCZ^{(0)}\to\CCG_0$ is fixed by
\begin{equation}
 g^{(0)}(0)\ :=\ \mathring{g}\ \in\ \CCG_0
\end{equation}
Hence, $g^{(0)}$ is parametrised by $\CCG_0$.

\vspace{15pt}
\begin{figure}[h]
\begin{center}
\tikzset{->-/.style={decoration={markings,mark=at position #1 with {\arrow{>}}},postaction={decorate}}}
\begin{tikzpicture}[scale=.8,every node/.style={scale=.8}]
   \filldraw [black] (0,0) circle (2pt);
   \draw (-.8,0) node {$\begin{smallmatrix} g^{(1)}(0)\\=\,g^{(0)}(0)\end{smallmatrix}$};
   \draw[->-=.5] (4,0) -- (7,0);
   \filldraw [black] (4,0) circle (2pt);
   \filldraw [black] (7,0) circle (2pt);
   \draw (3.2,0) node {$\begin{smallmatrix} g^{(1)}(0)\\=\,g^{(0)}(0)\end{smallmatrix}$};
   \draw (7.8,0) node {$\scriptstyle g^{(1)}(\theta_0)$};
   \draw (5.5,-.5) node {$\scriptstyle g^{(1)}(\theta_0,0)$};
\end{tikzpicture}

\begin{minipage}{14cm}
\caption{\label{fig:HornFillerL1}\small Construction of $g^{(1)}(\theta_0,0)$. The horn to the left, given by data constructed at level zero, can be filled as shown on the right due to the Kan property.}
\end{minipage}
\end{center}
\end{figure}

At first level, the simplicial set $\CCZ^{(1)}$ consists of simplicial simplices with tuples of the form $(\theta_0,\ldots,\theta_0,0,\ldots,0)$. Here, \eqref{eq:reccon} amounts to
\begin{equation}
 g^{(1)}(0,\ldots,0)\ =\ g^{(0)}(0,\ldots,0)\ =\ \mathring{g}~.
\end{equation} 
Since $\CCG$ is a Kan simplicial manifold, we can fill all horns and, in particular, we can fill the horn $g^{(1)}(0)$ by a simplicial 1-simplex, 
\begin{equation}\label{eq:L1fix}
 g^{(1)}(\theta_0,0)\ =\ \sfd^0_0(g^{(0)}(0))+\alpha_1\theta_0\ =\ \sfd^0_0(\mathring{g})+\alpha_1\theta_0\ewith\alpha_1\ \in\  \ker(\sff^1_{0\,*})[1]~,
\end{equation}
cf.\ Figure \ref{fig:HornFillerL1}. Here, $\sff^1_{0\,*}$ denotes the linearisation of $\sff^1_0$ at $\sfd^0_0(\mathring{g})$.\footnote{For brevity, we shall mostly write $\sff^p_{i\,*}$ instead of $\sff^p_{i\,*}|_{(\sfd^{p-1}_0\circ \cdots \circ \sfd^0_0)(\mathring{g})}$ in the following and likewise for the degeneracy maps.}  The expansion \eqref{eq:L1fix} follows directly from \eqref{eq:gFormExp-3} together with the fact that $g^{(1)}$ is a simplicial map and
\begin{equation}
\begin{gathered}
\sfd^0_0(\mathring{g})\ =\ g^{(0)}(0,0)\ =\ g^{(1)}(0,0)~,\\
 \mathring{g}\ =\ g^{(0)}(0)\ =\ g^{(1)}(0)\ =\ (\sff^1_0\circ g^{(1)})(\theta,0)\ =\ \mathring{g}+\sff^1_{0\,*}(\alpha_1)\theta_0~.
 \end{gathered}
\end{equation}
Altogether, $g^{(1)}$ is parametrised by $\bigcup_{\mathring{g}\in\CCG_0}\big\{\big(\mathring{g},\ker(\sff^1_{0\,*})[1]\big)\big\}$.

\vspace{10pt}
\begin{figure}[h]
\begin{center}
\tikzset{->-/.style={decoration={markings,mark=at position #1 with {\arrow{>}}},postaction={decorate}}}
\begin{tikzpicture}[scale=.8,every node/.style={scale=.8}]
   \draw[->-=.5] (0,0) -- (3,0);
   \draw[->-=.5] (0,0) -- (1.5,2.6);   
   \filldraw [black] (0,0) circle (2pt);
   \filldraw [black] (1.5,2.6) circle (2pt);
   \filldraw [black] (3,0) circle (2pt);
   \draw (-.8,0) node {$\begin{smallmatrix} g^{(2)}(0)\\=\,g^{(0)}(0)\end{smallmatrix}$};
   \draw (1.5,3.1) node {$\begin{smallmatrix} g^{(2)}(\theta_1)\\ =\,g^{(1)}(\theta_1)\end{smallmatrix}$};
   \draw (3.8,0) node {$\begin{smallmatrix}g^{(2)}(\theta_0)\\=\,g^{(1)}(\theta_0)\end{smallmatrix}$};
   \draw (1.5,-.5) node {$\begin{smallmatrix}g^{(2)}(\theta_0,0)\\=\,g^{(1)}(\theta_0,0)\end{smallmatrix}$};
   \draw (-.3,1.3) node {$\begin{smallmatrix}g^{(2)}(\theta_1,0)\\=\,g^{(1)}(\theta_1,0)\end{smallmatrix}$};
   \draw[->-=.5] (8,0) -- (11,0);
   \draw[->-=.5] (8,0) -- (9.5,2.6);   
   \draw[->-=.5] (9.5,2.6) -- (11,0);  
    \draw [-implies,double equal sign distance] (9.2,0.85) -- (10.06,1.35);
   \filldraw [black] (8,0) circle (2pt);
   \filldraw [black] (9.5,2.6) circle (2pt);
   \filldraw [black] (11,0) circle (2pt);
   \draw (7.2,0) node {$\begin{smallmatrix} g^{(2)}(0)\\=\,g^{(0)}(0)\end{smallmatrix}$};
   \draw (9.5,3.1) node {$\begin{smallmatrix} g^{(2)}(\theta_1)\\ =\,g^{(1)}(\theta_1)\end{smallmatrix}$};
   \draw (11.8,0) node {$\begin{smallmatrix}g^{(2)}(\theta_0)\\=\,g^{(1)}(\theta_0)\end{smallmatrix}$};
   \draw (7.7,1.3) node {$\begin{smallmatrix}g^{(2)}(\theta_1,0)\\=\,g^{(1)}(\theta_1,0)\end{smallmatrix}$};
   \draw (11.2,1.3) node {$\scriptstyle g^{(2)}(\theta_0,\theta_1)$};
   \draw (9.5,-.5) node {$\begin{smallmatrix}g^{(2)}(\theta_0,0)\\=\,g^{(1)}(\theta_0,0)\end{smallmatrix}$};
   \draw (9.5,.4) node {$\scriptstyle g^{(2)}(\theta_0,\theta_1,0)$};
\end{tikzpicture}

\begin{minipage}{14cm}
\caption{\label{fig:HornFillerL2}\small Construction of $g^{(2)}(\theta_0,\theta_1)$. The horn to the left, given by data constructed at levels zero and one, can be filled according to the right due to the Kan property.}
\end{minipage}
\end{center}
\end{figure}

At second level, the simplicial set $\CCZ^{(2)}$ consists of simplicial simplices with tuples of the form $(\theta_0,\ldots,\theta_0,\theta_1,\ldots,\theta_1,0,\ldots,0)$. From \eqref{eq:reccon}, we have
\begin{equation}
 g^{(2)}(\theta_0,\ldots,\theta_0,0,\ldots,0)\ =\ g^{(1)}(\theta_0,\ldots,\theta_0,0,\ldots,0)~.
\end{equation}
The horn depicted to the left in Figure \ref{fig:HornFillerL2} can now be filled by a simplicial 2-simplex
\begin{equation}
g^{(2)}(\theta_0,\theta_1,0)\ =\  \mathring{g}^{(2)}+\hat\gamma^{(2)}_0(\theta_0-\theta_1)+\hat\gamma^{(2)}_1\theta_1+\hat\gamma^{(2)}_{01}\theta_0\theta_1~.
\end{equation}
All components except for the top component are restricted, because
\begin{equation}
\begin{aligned}
 g^{(2)}(\theta_0,\theta_1,0)\ &=\ g^{(2)}(\theta_0,0,0)+g^{(2)}(\theta_1,\theta_1,0)-g^{(2)}(\theta_1,0,0)+\hat\gamma^{(2)}_{01}\theta_0\theta_1\\
 &=\ (\sfd^1_1\circ g^{(1)})(\theta_0,0)+(\sfd^1_0\circ g^{(1)})(\theta_1,0)-(\sfd^1_1\circ g^{(1)})(\theta_1,0)+\hat\gamma^{(2)}_{01}\theta_0\theta_1\\
  &=\ (\sfd^1_0\circ\sfd^0_0)(\mathring{g})+\sfd^1_{1\,*}(\alpha_1)(\theta_0-\theta_1)+\sfd^1_{0\,*}(\alpha_1)\theta_1+\hat\gamma^{(2)}_{01}\theta_0\theta_1~.
 \end{aligned}
\end{equation}
The top component $\hat\gamma^{(2)}_{01}$ is not completely arbitrary: since the action of all face maps $\sff^2_i$ with $i<2$ on $g^{(2)}(\theta_0,\theta_1,0)$ is fixed by $g^{(2)}$ being simplicial, the expressions $\sff^2_{i\,*}(\hat\gamma^{(2)}_{01})$ are fixed by Hessians of the maps $\sff^2_{i}$ for $i<2$. Thus, the remaining freedom in choosing $\hat\gamma^{(2)}_{01}$ is an element $\alpha_2\in \ker(\sff^2_{0\,*})[2]\cap \ker(\sff^2_{1\,*})[2] $. In summary, $g^{(2)}$ is parametrised by $\bigcup_{\mathring{g}\in\CCG_0}\big\{\big(\mathring{g},\ker(\sff^1_{0\,*})[1],\ker(\sff^2_{0*})[2]\cap \ker(\sff^2_{1\,*})[2]\big)\big\}$.

This step of constructing $g^{(2)}$ from $g^{(1)}$ is iterated as follows. The first term
\begin{subequations}
\begin{equation}
\rho^{(3)}_0(\theta_0,\theta_1,\theta_2)\ :=\ g^{(3)}(\theta_0,\theta_1,0,0)\ =\ (\sfd^{2}_{2}\circ g^{(2)})(\theta_0,\theta_1,0)
\end{equation}
agrees with $g^{(3)}(\theta_0,\theta_1,\theta_2,0)$ up to terms proportional to $\theta_2$. We want the next term to contribute these missing terms up to terms proportional to $\theta_1\theta_2$. We use the trivial formula $f(0,\theta)-f(0,0)=f(\theta,\theta)-f(\theta,0)$ to rewrite
\begin{equation}
\begin{aligned}
 \rho^{(3)}_1(\theta_0,\theta_1,\theta_2)
 \ :&\!\!=\ g^{(3)}(\theta_0,0,\theta_2,0)-g^{(3)}(\theta_0,0,0,0)\\
 &\!\!=\ g^{(3)}(\theta_0,\theta_2,\theta_2,0)-g^{(3)}(\theta_0,\theta_2,0,0)\\
 &\!\!=\ \sum_{i=0}^1 (-1)^{i+1} (\sfd^2_{2-i}\circ g^{(2)})(\theta_0,\theta_2,0)~.
\end{aligned}
\end{equation}
The last term contributes missing terms proportional to $\theta_1\theta_2$ up to terms proportional to $\theta_0\theta_1\theta_2$. We use again our above formula and obtain
\begin{equation}
\begin{aligned}
\rho^{(3)}_2(\theta_0,\theta_1,\theta_2)
\ :&\!\!=\ g^{(3)}(0,\theta_1,\theta_2,0)-g^{(3)}(0,0,\theta_2,0)-(g^{(3)}(0,\theta_1,0,0)-g^{(3)}(0,0,0,0))\\
 &\!\! =\ g^{(3)}(\theta_1,\theta_1,\theta_2,0)-g^{(3)}(\theta_1,0,\theta_2,0)-(g^{(3)}(0,\theta_1,0,0)-g^{(3)}(0,0,0,0))\\
 &\!\!=\ \sum_{i=0}^2(-1)^i(\sfd^2_{2-i}\circ g^{(2)})(\theta_1,\theta_2,0)-(\sfd^2_1\circ g^{(2)})(\theta_1,0,0)\,-\\ 
 &\kern1cm-(\sfd^2_{2}\circ g^{(2)})(0,\theta_1,0)+(\sfd^2_{2}\circ g^{(2)})(0,0,0)\\
 &\!\!=\ \sum_{i=0}^2(-1)^i(\sfd^2_{2-i}\circ g^{(2)})(\theta_1,\theta_2,0)-(\sfd^2_0\circ g^{(2)})(\theta_1,0,0)~.
\end{aligned}
\end{equation}
Altogether, we find
\begin{equation}
 g^{(3)}(\theta_0,\theta_1,\theta_2,0)\ =\ \rho^{(3)}_0(\theta_0,\theta_1,\theta_2)+\rho^{(3)}_1(\theta_0,\theta_1,\theta_2)+\rho^{(3)}_2(\theta_0,\theta_1,\theta_2)+\hat\gamma^{(3)}_{012}\theta_0\theta_1\theta_2~,
\end{equation}
\end{subequations}
where the action of the face maps $\sff_i^3$ with $i\leq 2$ on $g^{(3)}(\theta_0,\theta_1,\theta_2,0)$ is fixed, which determines the top component $\hat\gamma^{(3)}_{012}$ up to an element of $\alpha_{3}\in \bigcap_{i=0}^{2}\ker(\sff^{3}_{i\,*})[3]$. These observations are  generalised to the following lemma.

\begin{lemma}
We have
\begin{subequations}\label{eq:exp_g_0}
\begin{equation}
 g^{(k+1)}(\theta_0,\theta_1,\ldots,\theta_k,0)\ =\ \sum_{i=0}^k\rho^{(k+1)}_i(\theta_0,\theta_1,\ldots,\theta_k)+\hat \gamma^{(k+1)}_{0\cdots k}\theta_0\cdots \theta_k~,
\end{equation}
where
\begin{equation}
 \begin{aligned}
  \rho^{(k+1)}_0(\theta_0,\ldots,\theta_k)\ &:=\ (\sfd^{k}_{k}\circ g^{(k)})(\theta_0,\ldots,\theta_{k-1},0)~,\\
  \rho^{(k+1)}_1(\theta_0,\ldots,\theta_k)\ &:=\ \sum_{i=0}^1(-1)^{i+1}(\sfd^k_{k-i}\circ g^{(k)})(\theta_0,\ldots,\theta_{k-2},\theta_k,0)~,\\
  \rho^{(k+1)}_2(\theta_0,\ldots,\theta_k)\ &:=\ \sum_{i=0}^1(-1)^{i} (\sfd^k_{k-i}\circ g^{(k)})(\theta_0,\ldots,\theta_{k-3},\theta_{k-2},0)\,-\\
  &\hspace{1cm}-(\sfd^k_{k-2}\circ g^{(k)})(\theta_0,\ldots,\theta_{k-3},\theta_{k-2},0,0)~,\\
  &~~\vdots
 \end{aligned}
\end{equation}
\end{subequations}
Moreover, $\hat \gamma^{(k+1)}_{0\cdots k}$ is fixed by the $m$-th order derivatives of the maps $\sff^{k+1}_i$ for $2\leq m\leq k+1$,
\begin{equation}\label{eq:CondOnalpha}
 \sff^{k+1}_{i\,*}(\hat \gamma^{(k+1)}_{0\cdots k})\ =:\ \beta^{k+1}_i(\hat \gamma^{(0)}_{0},\ldots,\hat \gamma^{(k)}_{0\cdots (k-1)})~,
\end{equation}
where $\beta^{k+1}_i$ contains all these derivatives.\footnote{We shall discuss the terms $\beta^{k+1}_i$ more explicitly for Lie quasi-2-groups in Section \ref{ssec:ex_Lie2}.} Hence, the freedom in choosing $\hat \gamma^{(k+1)}_{0\cdots k}$ is an element $\alpha_{k+1}\in \bigcap_{i=0}^{k}\ker(\sff^{k+1}_{i\,*})[k+1]$. 
\end{lemma}

A direct consequence of this lemma and our previous discussion is the  following theorem.

\begin{thm}\label{thm:JetPara}
Let $\CCG$ be a Lie quasi-groupoid. Define $\sigma_{p}:\CCG_0\ \to\ \CCG_p$ for $p\geq1$ by $\sigma_{p}:=\sfd^{p-1}_{0}\circ\cdots\circ\sfd^1_0\circ\sfd^0_0$ and set ${\rm Ker}_\CCG[p]|_{\sigma_p(\mathring{g})}:=\bigcap_{i=0}^{p-1}\ker(\sff^p_{i\,*}|_{\sigma_p(\mathring{g})})[p]$ for some $\mathring{g}\in\CCG_0$, where $\sff^p_{i\,*}|_{\sigma_p(\mathring{g})}$ denotes the linearisation of $\sff^p_i:\CCG_p\to\CCG_{p-1}$ at $\sigma_p(\mathring{g})$. The 1-jet of $\CCG$ is parametrised by 
\begin{equation}\label{eq:Par1Jet}
 \sLie(\CCG)\ :=\ \bigtimes_{p\in\NN}\sigma^*_{p}{\rm Ker}_\CCG[p]\ \to\ \CCG_0~.
\end{equation}
Moreover, a simplicial map $g:N(\check\CC(X\times\FR^{0|1}\to X))\to\CCG$ has restrictions to $p$-simplices
\begin{equation}\label{eq:gp_from_gp}
 g^p(\theta_0,\ldots,\theta_p)\ =\ (\sff^{p+1}_{p+1}\circ g^{(p+1)})(\theta_0,\ldots,\theta_p,0)~,
\end{equation}
where the maps $g^{(p+1)}(\theta_0,\ldots,\theta_{p},0)$ are given by \eqref{eq:exp_g_0}.
\end{thm}

\noindent
If $\CCG$ is a Lie $n$-quasi-groupoid (i.e.\ all $p$-horn fillers for $p\geq n$ are unique) then the horn filler $g^{(n+1)}(\theta_0,\ldots,\theta_{n},0)$ of the horn $g^{(n+1)}(\theta_0,\ldots,\theta_{n})$ is unique, and, consequently, $\hat \gamma^{(n+1)}_{0\cdots n}$ is fully determined by $\{\hat \gamma^{(0)}_{0},\ldots,\hat \gamma^{(n)}_{0\cdots (n-1)}\}$. This implies that \eqref{eq:Par1Jet} reduces to \begin{equation}
 \sLie(\CCG)\ =\ \bigoplus_{p=1}^n\sigma^*_{p}{\rm Ker}_\CCG[p]\to\CCG_0~.
\end{equation}

\begin{rem}
Note that given a simplicial set $\CCX$, we can construct a complex $C^p:=\bigcap_{i=0}^{p-1} \ker(\sff^p_i)$ with differential $\dpar|_{C^p}:=\sff_p^p$ satisfying $\dpar|_{C^p}\circ\dpar|_{C^{p+1}}=0$. In order to have a well-defined cohomology theory, we restrict ourselves to a simplicial group $\CCX$, cf.\ Example \ref{exam:SimpGroup} and Remark \ref{rem:InftGrp}. The resulting complex is then known as the \uline{Moore complex} of $\CCX$. For example, the Moore complexes of the nerves of a semistrict Lie 2-algebra and of a strict 2-group are the categorically equivalent 2-term $L_\infty$-algebra and crossed module of groups, respectively.

For fixed $\mathring{g}\in\CCG_0$, the moduli space $\sLie(\CCG)$ from Theorem \ref{thm:JetPara} together with the differential
\begin{equation}
 \dpar_p\ :=\ \sff^p_{p\,*}|_{\sigma_p(\mathring{g})}\,:\,{\rm Ker}_\CCG[p]|_{\sigma_p(\mathring{g})}\ \to\ {\rm Ker}_\CCG[p-1]|_{\sigma_{p-1}(\mathring{g})}
\end{equation}
is the Moore complex of the (Abelian) simplicial group $\bigcup_{p\in \NN} (T[1])^p_{\sigma_p(\mathring{g})}\CCG_p$. 
\end{rem}

The Moore complex $\sLie(\CCG)$ is augmented to an $L_\infty$-algebroid by the homological vector field $Q$, see \eqref{eq:QonG}, corresponding to the induced action of $\inthom(\FR^{0|1},\FR^{0|1})$ on $\sLie(\CCG)$.

\begin{prop}\label{prop:Qvector}
The moduli space $\sLie(\CCG)$ from  Theorem \ref{thm:JetPara} becomes an N$Q$-manifold, that is, an $L_\infty$-algebroid, when endowed with the homological vector field
\begin{equation}
  Q\mathring{g}\ =\ -\sff^1_{1\,*}(\alpha_1)\eand Q \alpha_p\ =\ -\sff^{p+1}_{p+1\,*}(\alpha_{p+1})- R_{p+1}( \alpha_1,\ldots, \alpha_{p})
\end{equation}
for $\mathring{g}\in\CCG_0$ and $\alpha_p\in{\rm Ker}_\CCG[p]|_{\sigma_p(\mathring{g})}$ and $p\geq1$. Here $R_{p+1}$, contains all $m$-th order derivatives of $\sff^{p+1}_{p+1}$ for $2\leq m\leq p+1$ and all $n$-th order derivatives of $\sff^{k+1}_{i}$ for $i\leq k$ and $2\leq n\leq k+1$ and $1\leq k\leq p$. We call $\sLie(\CCG)$ the \uline{$L_\infty$-algebroid of the Lie quasi-groupoid} $\CCG$. If $\CCG_0$ is trivial, we obtain the \uline{$L_\infty$-algebra of the Lie quasi-group} $\CCG$. 
\end{prop}

\noindent 
{\it Proof:} Note that equation \eqref{eq:exp_g_0} implies that
\begin{equation}
g^{(p+1)}(\theta_0,\theta_1,\ldots,\theta_{p},0)\ =\ (\sfd^{p}_{p}\circ g^{(p)})(\theta_0,\ldots,\theta_{p-1},0)+(\cdots+\hat \gamma^{(p+1)}_{0\cdots p}\theta_0\cdots\theta_{p-1})\theta_p~.
\end{equation}
Consequently,
\begin{eqnarray}\label{eq:gpalphaexp}
  g^p(\theta_0,\ldots,\theta_p)\! &=&\! (\sff^{p+1}_{p+1}\circ g^{(p+1)})(\theta_0,\ldots,\theta_p,0)\notag\\
  &=&\!  g^{(p)}(\theta_0,\ldots,\theta_{p-1},0)+\cdots+\notag\\
   &&\kern1cm+\,\big[\sff^{p+1}_{p+1\,*}(\hat \gamma^{(p+1)}_{0\ldots p})+\hat R_{p+1}(\hat \gamma^{(0)}_{0},\ldots,\hat \gamma^{(k)}_{0\cdots (k-1)})\big]\theta_0\cdots\theta_p\notag\\
   &=&\! \sigma_p(\mathring{g})+\cdots+\hat\alpha_p\hat\theta_0\cdots\hat\theta_{p-1}+\cdots+\notag\\
   &&\kern1cm+\,\big[\sff^{p+1}_{p+1\,*}(\hat \gamma^{(p+1)}_{0\ldots p})+\hat R_{p+1}(\hat \gamma^{(0)}_{0},\ldots,\hat \gamma^{(k)}_{0\cdots (k-1)})\big]\hat\theta_0\cdots\hat\theta_p~,
\end{eqnarray}
where $\hat R_{p+1}$ contains all $m$-th derivatives of $\sff^{p+1}_{p+1}$ for $2\leq m\leq p+1$ and $\sigma_p$ was defined in Theorem \ref{thm:JetPara}. Therefore, the $\inthom(\FR^{0|1},\FR^{0|1})$-action on $g^p(\theta_0,\ldots,\theta_p)$ given in \eqref{eq:QonG} reads as
\begin{equation}
  Q\mathring{g}\ =\ -\sff^1_{1\,*}(\alpha_1)\eand Q \hat \alpha_p\ =\ -\sff^{p+1}_{p+1\,*}(\hat \gamma^{(p+1)}_{0\ldots p})-\hat R_{p+1}(\hat \gamma^{(0)}_{0},\ldots,\hat \gamma^{(k)}_{0\cdots (k-1)})~.
\end{equation}
Upon solving the linear equations \eqref{eq:CondOnalpha}, we write $\hat \gamma^{(k)}_{0\ldots k-1}$ (for $1\leq k\leq p+1$) as a linear combination of $\alpha_k\in \bigcap_{i=0}^{k-1}\ker(\sff^{k}_{i\,*})[k]$ and additional terms which depend on the higher derivatives of the face maps $\sff^k_i$ evaluated on $\{\alpha_1,\ldots,\alpha_{k-1}\}$. We may thus write
\begin{equation}
  Q\mathring{g}\ =\ -\sff^1_{1\,*}(\alpha_1)\eand Q \alpha_p\ =\ -\sff^{p+1}_{p+1\,*}(\alpha_{p+1})- R_{p+1}( \alpha_1,\ldots, \alpha_{p})~,
\end{equation}
where the higher derivative terms $\sff^k_i$ have been absorbed into $R_{p+1}$. \hfill $\Box$

\begin{rem}
In general, we may define $\mu_1(\alpha_p):=\sff^p_{p\,*}(\alpha_p)$, and $R_{p+1}$ will decompose as
\begin{equation}\label{eq:higher_homotopy_products}
 R_{p+1}( \alpha_1,\ldots, \alpha_{p})\ =\ \sum_{j=1}^{p}\sum_{\begin{smallmatrix} i_1+\cdots+ i_j={p}\\ 0<i_1\leq\cdots\leq i_j\end{smallmatrix}}\frac{1}{j!}\mu_j(\alpha_{i_1},\ldots,\alpha_{i_j})~,
\end{equation}
where the $\mu_j$ encode the homotopy products of the underlying $L_\infty$-algebra. We shall derive $\mu_2$ and $\mu_3$ explicitly below in the case of a Lie quasi-2-group.
\end{rem}

\begin{prop}
 Our definition of the $L_\infty$-algebra of a Lie quasi-group extends that of the Lie algebra of a Lie group.
\end{prop}
\begin{pf}
 If $\CCG=\big\{ \cdots \quadarrow \sG\times \sG\triplearrow \sG\doublearrow *\big\}$ is the nerve of the delooping $\sB\sG$ of a Lie group $\sG$ with $\unit_\sG=\sfd^0_0(*)=\sigma_1(*)$, then the moduli space of Theorem \ref{thm:JetPara} reads as
 \begin{equation}
 \sLie(\CCG)\ =\ \sigma^*_{1}{\rm Ker}_\CCG[1] \to *\ =\ \ker(\sff^1_0|_{\sigma_1(*)})[1]\ =\ T_{\unit_\sG}\sG[1]\ =\ \sLie(\sG)[1]~.
 \end{equation}
 The simplicial map $g$ is fully determined by its component $g^{1}(\theta_0,\theta_1)=\sff^2_2 g^{(2)}(\theta_0,\theta_1,0)$, which reads as
\begin{equation}
\begin{aligned}
g^{1}(\theta_0,\theta_1)\ &=\ \sff^2_2 g^{(2)}(\theta_0,\theta_1,0)\\
 \ &=\ \sff^2_2\big( (\sfd^1_0\circ\sfd^0_0)(*)+\sfd^1_{1\,*}(\alpha_1)(\theta_0-\theta_1)+\sfd^1_{0\,*}(\alpha_1)\theta_1+\hat\gamma^{(2)}_{01}\theta_0\theta_1\big)\\
 \ &=\ \unit_\sG+\alpha_1(\theta_0-\theta_1)+\tfrac12 \mu_2(\alpha_1,\alpha_1)\theta_0\theta_1~,
 \end{aligned}
\end{equation}
where we used the simplicial identities \eqref{eq:axioms_simplicial_set} and equation \eqref{eq:higher_homotopy_products} to rewrite the coefficient of $\theta_0\theta_1$. Proposition \ref{prop:Qvector} then states that
\begin{equation}
 Q\alpha_1=-R_2(\alpha_1,\alpha_1)=-\tfrac12\mu_2(\alpha_1,\alpha_1)~,
\end{equation}
and we recover the homological vector field of the Chevalley--Eilenberg description of the Lie algebra $\sLie(\sG)$ of the Lie group $\sG$.
\end{pf}

\subsection{The \texorpdfstring{$L_\infty$}{L-infinity}-algebra of a Lie quasi-2-group}\label{ssec:ex_Lie2}

To complement our general discussion, let us look at the concrete example of a Lie quasi-2-group, cf.\ Remark \ref{rem:W2PB-1}. Our findings will relate to those of \cite{Jurco:2014mva}, where a semistrict Lie 2-group was differentiated to a 2-term $L_\infty$-algebra. Our construction here, based on the language of simplicial manifolds, is significantly more straightforward and systematic than that the latter, which employed weak 2-categories.

A Lie quasi-2-group $\CCG$ has only one simplicial 0-simplex in $\CCG$, which we shall denote by $*$ in the following. Moreover, the $k$-horn fillers of $\CCG$ are unique for $k\geq 3$, which implies that the simplicial map $g:N(\check\CC(X\times\FR^{0|1}\to X)\to\CCG$ is completely fixed by the maps $g^{(k)}:\CCZ^{(k)}\rightarrow \CCG$ with $k< 3$. Let us compute these maps in the following. We start with a lemma.
\begin{lemma}
 For a Gra{\ss}mann-even function $g\ =\ a+a_i\hat\theta_i+\tfrac12 a_{ij}\hat\theta_i\hat\theta_j+\tfrac{1}{3!}a_{ijk}\hat\theta_i\hat\theta_j\hat\theta_k$, we have 
 \begin{equation}
  \begin{aligned}
    f\circ g\ &=\ f(a)+f'(a)a_i\theta_i+\tfrac12 [f'(a)a_{ij}-f''(a)a_ia_j]\hat\theta_i\hat\theta_j\,+\\ &\kern1cm+\tfrac{1}{3!}[f'(a)a_{ijk}-f''(a)(a_{ij} a_k+a_{jk}a_i+a_{ki}a_j)+f'''(a)a_ia_ja_k]\hat\theta_i\hat\theta_j\hat\theta_k~,
  \end{aligned}  
 \end{equation}
 where $f$ is another Gra{\ss}mann-even function.
\end{lemma}
\noindent Using this lemma and Theorem \ref{thm:JetPara}, we readily derive the relevant maps $g^{(k)}$:
\begin{subequations}
\begin{equation}
\begin{aligned}
g^{(0)}(0)\ &=\ *\\
 g^{(1)}(\theta_0,0)\ &=\ \sfd^0_0(g^{(0)}(0))+\alpha_1\theta_0\\
& =\ \sigma_0(*)+\alpha_1\hat\theta_0~,\\[5pt]
g^{(2)}(\theta_0,\theta_1,0)\ &=\ (\sfd^1_1\circ g^{(1)})(\theta_0,0)+(\sfd^1_0\circ g^{(1)})(\theta_1,0)-(\sfd^1_1\circ g^{(1)})(\theta_1,0)+\hat\gamma^{(2)}_{01}\theta_0\theta_1\\
&=\ \sigma_1(*)+\sfd^1_{1\,*}(\alpha_1)\hat\theta_0+\sfd^1_{0\,*}(\alpha_1)\hat\theta_1+\hat\gamma^{(2)}_{01}\hat\theta_0\hat\theta_1~,\\[5pt]
g^{(3)}(\theta_0,\theta_1,\theta_2,0)\ &=\ (\sfd^2_2\circ g^{(2)})(\theta_0,\theta_1,0)+\sum_{i=0}^1(-1)^{i+1}(\sfd^2_{2-i}\circ g^{(2)})(\theta_0,\theta_2,0)\,+\\
&\kern1cm+\sum_{i=0}^2(-1)^{i} (\sfd^2_{2-i}\circ g^{(2)})(\theta_1,\theta_2,0)-(\sfd^2_0\circ g^{(2)})(\theta_1,0,0)+\hat \gamma^{(3)}_{012}\theta_0\theta_1\theta_2\\
  &=\ \sigma_2(*)+(\sfd^2_2\circ \sfd^1_1)_*(\alpha_1)\hat\theta_0+
  (\sfd^2_2\circ\sfd^1_0)_*(\alpha_1)\hat\theta_1+(\sfd^2_0\circ\sfd^1_0)_*(\alpha_1)\hat\theta_2\,+\\
  &\kern1cm+\big[\sfd^2_{2*}(\hat\gamma^{(2)}_{01})+{\rm II}_{\sfd^2_2}(\sfd^1_{0\,*}(\alpha_1),\sfd^1_{1\,*}(\alpha_1))\big]\hat\theta_0\hat\theta_1+\\
  &\kern1cm+\big[\sfd^2_{1*}(\hat\gamma^{(2)}_{01})+{\rm II}_{\sfd^2_1}(\sfd^1_{0\,*}(\alpha_1),\sfd^1_{1\,*}(\alpha_1))\big]\hat\theta_0\hat\theta_2\,+\\
  &\kern1cm+\big[\sfd^2_{0*}(\hat\gamma^{(2)}_{01})+{\rm II}_{\sfd^2_0}(\sfd^1_{0\,*}(\alpha_1),\sfd^1_{1\,*}(\alpha_1))\big]\hat\theta_1\hat\theta_2+\hat \gamma^{(3)}_{012}\hat\theta_0\hat\theta_1\hat\theta_2~,
 \end{aligned}
\end{equation}
where
\begin{equation}\label{eq:condhatalpha2}
  \sff^2_{i\,*}(\hat\gamma^{(2)}_{01})+{\rm II}_{\sff^2_i}(\sfd^1_{0\,*}(\alpha_1),\sfd^1_{1\,*}(\alpha_1))\ =\ 0\efor i\ <\ 2~
\end{equation}
and
\begin{equation}\label{eq:condhatalpha3}
\begin{aligned}
  \sff^3_{i\,*}(\hat \gamma^{(3)}_{012})&+
  {\rm III}_{\sff^3_i}((\sfd^2_2\circ\sfd^1_1)_*(\alpha_1),(\sfd^2_2\circ\sfd^1_0)_*(\alpha_1),(\sfd^2_0\circ\sfd^1_0)_*(\alpha_1))\,- \\
  &-{\rm II}_{\sff^3_i}(\sfd^2_{2\,*}(\hat\gamma^{(2)}_{01}+{\rm II}_{\sfd^2_2}(\sfd^1_{0\,*}(\alpha_1),\sfd^1_{1\,*}(\alpha_1)),(\sfd^2_{0}\circ\sfd^1_0)_*(\alpha_1))\,-\\
  &-{\rm II}_{\sff^3_i}(\sfd^2_{1\,*}(\hat\gamma^{(2)}_{01}+{\rm II}_{\sfd^2_1}(\sfd^1_{0\,*}(\alpha_1),\sfd^1_{1\,*}(\alpha_1)),(\sfd^2_{2}\circ\sfd^1_1)_*(\alpha_1))\,+\\
  &+{\rm II}_{\sff^3_i}(\sfd^2_{0\,*}(\hat\gamma^{(2)}_{01}+{\rm II}_{\sfd^2_0}(\sfd^1_{0\,*}(\alpha_1),\sfd^1_{1\,*}(\alpha_1)),(\sfd^2_2\circ\sfd^1_0)_*(\alpha_1))\ =\ 0\efor i\ <\ 3~.
\end{aligned}
\end{equation}
\end{subequations}
Here, ${\rm II}_f$ indicates the second derivative (Hessian) of $f$ and ${\rm III}_f$ the third derivative. 

We now use again the notation introduced in \eqref{eq:CondOnalpha} and write \eqref{eq:condhatalpha2} as $\sff^2_{i\,*}(\hat\gamma^{(2)}_{01})=\beta^2_i$. Then \eqref{eq:condhatalpha2} is solved by
\begin{equation}\label{eq:solhatalpha2}
 \hat\gamma^{(2)}_{01}\ =\ \alpha_2+\sfd^1_{0\,*}(\beta^2_0)+\sfd^1_{1\,*}(\beta^2_1-\beta^2_0)
 \end{equation}
with $\alpha_2\in \ker(\sff^2_{0\,*})[2]\cap \ker(\sff^2_{1\,*})[2]$. Likewise, after writing \eqref{eq:condhatalpha3} as $\sff^3_{i\,*}(\hat \gamma^{(3)}_{012})=\beta^3_i$ and inserting \eqref{eq:solhatalpha2} into $\beta^3_i$, we solve \eqref{eq:condhatalpha2} by
\begin{equation}
 \hat \gamma^{(3)}_{012}\ =\ \sfd^2_{0\,*}(\beta^3_0)+\sfd^2_{1\,*}(\beta^3_1-\beta^3_0)+\sfd^2_{2\,*}\big(\beta^3_2-\beta^3_1+\beta^3_0-(\sfd^1_0\circ\sff^2_1)_*(\beta^3_0)\big)~.
\end{equation}
The parameters $\alpha_1$ and $\alpha_2$ fix the simplicial map $g$ completely, and the moduli space $\sLie(\CCG)$ is
\begin{equation}
\sLie(\CCG)\ =\ \ker(\sff^1_{0\,*}|_{\sigma_1(*)})[1]~\oplus~\ker(\sff^2_{0\,*}|_{\sigma_2(*)})[2]\cap\ker(\sff^2_{1\,*}|_{\sigma_2(*)})[2]~.
\end{equation}

We can now derive the expansion of the simplicial map $g$ from the $g^{(k)}$ using \eqref{eq:gp_from_gp} and $\sff^1_{1\,*}(\alpha_1)=0$. We have 
\begin{subequations}
\begin{equation}\label{eq:g0g1exp2G}
\begin{aligned}
g^0(\theta_0)\ &=\ (\sff^1_1\circ g^{(1)})(\theta_0,0)\ =\ *~,\\[5pt]
g^1(\theta_0,\theta_1)\ &=\ (\sff^2_2\circ g^{(2)})(\theta_0,\theta_1,0)\\
\ &=\ \sigma_0(*)+\alpha_1(\theta_0-\theta_1)+\big[\mu_1(\alpha_2)+\tfrac12\mu_2(\alpha_1,\alpha_1)\big]\theta_0\theta_1~,\\
 g^2(\theta_0,\theta_1,\theta_2)\ &=\ (\sff^3_3\circ g^{(3)})(\theta_0,\theta_1,\theta_2,0)\\
  &=\ \sigma_1(*)+\cdots+\hat\gamma^{(2)}_{01}\hat\theta_0\hat\theta_1+\cdots+\\
  &\kern.8cm+\big[ \sff^3_{3\,*}(\hat \gamma^{(3)}_{012})+
  {\rm III}_{\sff^3_3}((\sfd^2_2\circ\sfd^1_1)_*(\alpha_1),(\sfd^2_2\circ\sfd^1_0)_*(\alpha_1),(\sfd^2_0\circ\sfd^1_0)_*(\alpha_1))\,- \\
  &\kern1.5cm-{\rm II}_{\sff^3_3}(\sfd^2_{2\,*}(\hat\gamma^{(2)}_{01}+{\rm II}_{\sfd^2_2}(\sfd^1_{0\,*}(\alpha_1),\sfd^1_{1\,*}(\alpha_1)),(\sfd^2_{0}\circ\sfd^1_0)_*(\alpha_1))\,-\\
  &\kern1.5cm-{\rm II}_{\sff^3_3}(\sfd^2_{1\,*}(\hat\gamma^{(2)}_{01}+{\rm II}_{\sfd^2_1}(\sfd^1_{0\,*}(\alpha_1),\sfd^1_{1\,*}(\alpha_1)),(\sfd^2_{2}\circ\sfd^1_1)_*(\alpha_1))\,+\\
  &\kern1.5cm+{\rm II}_{\sff^3_3}(\sfd^2_{0\,*}(\hat\gamma^{(2)}_{01}+{\rm II}_{\sfd^2_0}(\sfd^1_{0\,*}(\alpha_1),\sfd^1_{1\,*}(\alpha_1)),(\sfd^2_{2}\circ\sfd^1_0)_*(\alpha_1))\big]\hat\theta_0\hat\theta_1\hat\theta_2~,
 \end{aligned}
\end{equation}
where
\begin{equation}\label{eq:mu1_mu2}
\begin{aligned}
 \mu_1(\alpha_2)\ &:=\ \sff^2_{2\,*}(\alpha_2)~,\\
 \tfrac12\mu_2(\alpha_1,\alpha_1)\ &:=\ \sfd^1_{0\,*}\big({\rm II}_{\sff^2_2}\big(\sfd^1_{0\,*}(\alpha_1),\sfd^1_{1\,*}(\alpha_1)\big)\big)-\sfd^1_{0\,*}\big({\rm II}_{\sff^2_0}\big(\sfd^1_{0\,*}(\alpha_1),\sfd^1_{1\,*}(\alpha_1)\big)\big)\,-\\
 &\kern1cm-\,\sfd^1_{1\,*}\big({\rm II}_{\sff^2_1}\big(\sfd^1_{0\,*}(\alpha_1),\sfd^1_{1\,*}(\alpha_1)\big)\big)+\sfd^1_{1\,*}\big({\rm II}_{\sff^2_0}\big(\sfd^1_{0\,*}(\alpha_1),\sfd^1_{1\,*}(\alpha_1)\big)\big)
\end{aligned}
\end{equation}
\end{subequations}
and we listed only the terms relevant to us in the following.

Using Proposition \ref{prop:Qvector} together with \eqref{eq:QonG}, we immediately infer from \eqref{eq:g0g1exp2G} that
\begin{equation}\label{eq:Qactionalpha1}
 Q\alpha_1\ =\ -\mu_1(\alpha_2)-\tfrac12\mu_2(\alpha_1,\alpha_1)~.
\end{equation}
as well as 
\begin{equation}
\begin{aligned}
 Q\hat\gamma^{(2)}_{01}\ &=\  -\sff^3_{3\,*}(\hat \gamma^{(3)}_{012})-
  {\rm III}_{\sff^3_3}((\sfd^2_2\circ\sfd^1_1)_*(\alpha_1),(\sfd^2_2\circ\sfd^1_0)_*(\alpha_1),(\sfd^2_0\circ\sfd^1_0)_*(\alpha_1))\,+ \\
  &\kern1cm+{\rm II}_{\sff^3_3}(\sfd^2_{2\,*}(\hat\gamma^{(2)}_{01}+{\rm II}_{\sfd^2_2}(\sfd^1_{0\,*}(\alpha_1),\sfd^1_{1\,*}(\alpha_1)),(\sfd^2_{0}\circ\sfd^1_0)_*(\alpha_1))\,+\\
  &\kern1cm+{\rm II}_{\sff^3_3}(\sfd^2_{1\,*}(\hat\gamma^{(2)}_{01}+{\rm II}_{\sfd^2_1}(\sfd^1_{0\,*}(\alpha_1),\sfd^1_{1\,*}(\alpha_1)),(\sfd^2_{2}\circ\sfd^1_1)_*(\alpha_1))\,-\\
  &\kern1cm-{\rm II}_{\sff^3_3}(\sfd^2_{0\,*}(\hat\gamma^{(2)}_{01}+{\rm II}_{\sfd^2_0}(\sfd^1_{0\,*}(\alpha_1),\sfd^1_{1\,*}(\alpha_1)),(\sfd^2_{2}\circ\sfd^1_0)_*(\alpha_1))
\end{aligned}
\end{equation}
by virtue of Proposition \ref{prop:Qvector} together with \eqref{eq:QonG}. Upon substituting \eqref{eq:condhatalpha2} and \eqref{eq:condhatalpha3} into this equation and using \eqref{eq:Qactionalpha1}, we obtain
\begin{subequations}\label{eq:mu2_mu3}
\begin{equation}
 Q\alpha_2\ =\ -\mu_2(\alpha_1,\alpha_2)-\tfrac{1}{3!}\mu_3(\alpha_1,\alpha_1,\alpha_1)
\end{equation}
with
\begin{equation}
\begin{aligned}
 \mu_2(\alpha_1,\alpha_2)\ &:=\ \Big[\sfd^0_{0\,*}(Q\beta^2_0)+\sfd^1_{1\,*}(Q\beta^2_1-Q\beta^2_0)+(\sff^3_3\circ\sfd^2_{0})_*(\beta^3_0)+(\sff^3_3\circ\sfd^2_{1})_*(\beta^3_1-\beta^3_0)\,+\\
 &\kern-1.5cm+\beta^3_2-\beta^3_1+\beta^3_0-(\sfd^1_0\circ\sff^2_1)_*(\beta^3_0)\Big]\Big|_{\alpha_1\alpha_2}-{\rm II}_{\sff^3_3}(\sfd^2_{2\,*}(\alpha_2),(\sfd^2_{0}\circ\sfd^1_0)_*(\alpha_1))\,-\\
 &\kern-1.5cm-{\rm II}_{\sff^3_3}(\sfd^2_{2\,*}(\alpha_2),(\sfd^2_{0}\circ\sfd^1_0)_*(\alpha_1))+{\rm II}_{\sff^3_3}(\sfd^2_{2\,*}(\alpha_2),(\sfd^2_{0}\circ\sfd^1_0)_*(\alpha_1))~,\\[5pt]
\end{aligned}
\end{equation}
and
\begin{equation}
\begin{aligned}
\tfrac{1}{3!}\mu_3(\alpha_1,\alpha_1,\alpha_1)\ &:=\ \big[\sfd^0_{0\,*}(Q\beta^2_0)+\sfd^1_{1\,*}(Q\beta^2_1-Q\beta^2_0)+(\sff^3_3\circ\sfd^2_{0})_*(\beta^3_0)\,+\\
 &\kern-2cm+(\sff^3_3\circ\sfd^2_{1})_*(\beta^3_1-\beta^3_0)+\beta^3_2-\beta^3_1+\beta^3_0-(\sfd^1_0\circ\sff^2_1)_*(\beta^3_0)\big]\big|_{\alpha_1\alpha_1\alpha_1}\,+\\
 &\kern-2cm+{\rm III}_{\sff^3_3}((\sfd^2_2\circ\sfd^1_1)_*(\alpha_1),(\sfd^2_2\circ\sfd^1_0)_*(\alpha_1),(\sfd^2_0\circ\sfd^1_0)_*(\alpha_1))\,-\\
  &\kern-2cm-{\rm II}_{\sff^3_3}(\sfd^2_{2\,*}(\sfd^1_{0\,*}(\beta^2_0)+\sfd^1_{1\,*}(\beta^2_1-\beta^2_0)+{\rm II}_{\sfd^2_2}(\sfd^1_{0\,*}(\alpha_1),\sfd^1_{1\,*}(\alpha_1)),(\sfd^2_{0}\circ\sfd^1_0)_*(\alpha_1))\,-\\
   &\kern-2cm-{\rm II}_{\sff^3_3}(\sfd^2_{1\,*}(\sfd^1_{0\,*}(\beta^2_0)+\sfd^1_{1\,*}(\beta^2_1-\beta^2_0)+{\rm II}_{\sfd^2_2}(\sfd^1_{0\,*}(\alpha_1),\sfd^1_{1\,*}(\alpha_1)),(\sfd^2_{2}\circ\sfd^1_1)_*(\alpha_1))\,+\\
    &\kern-2cm+{\rm II}_{\sff^3_3}(\sfd^2_{0\,*}(\sfd^1_{0\,*}(\beta^2_0)+\sfd^1_{1\,*}(\beta^2_1-\beta^2_0)+{\rm II}_{\sfd^2_2}(\sfd^1_{0\,*}(\alpha_1),\sfd^1_{1\,*}(\alpha_1)),(\sfd^2_{2}\circ\sfd^1_0)_*(\alpha_1))~.
\end{aligned}
\end{equation}
Here, the abbreviation $[\cdots]_{\alpha_1\alpha_2}$ stands for terms only involving expressions containing one $\alpha_1$ and one $\alpha_2$. Likewise, $[\cdots]_{\alpha_1\alpha_1\alpha_1}$  stands for terms only involving expressions containing three $\alpha_1$. For instance,
\begin{equation}
\begin{aligned}
{[Q\beta^2_i]}|_{\alpha_1\alpha_2}\ &=\ {\rm II}_{\sff^2_i}(\sfd^1_{0\,}(\mu_1(\alpha_2)),\sfd^1_{1\,*}(\alpha_1))- {\rm II}_{\sff^2_i}(\sfd^1_{0\,*}(\alpha_1),\sfd^1_{1\,}(\mu_1(\alpha_2)))~,\\
{[Q\beta^2_i]}|_{\alpha_1\alpha_1\alpha_1}\ &=\ \tfrac12{\rm II}_{\sff^2_i}(\sfd^1_{0\,}(\mu_2(\alpha_1,\alpha_1)),\sfd^1_{1\,*}(\alpha_1))-\tfrac12{\rm II}_{\sff^2_i}(\sfd^1_{0\,*}(\alpha_1),\sfd^1_{1\,}(\mu_2(\alpha_1,\alpha_1)))~.\\
\end{aligned}
\end{equation}
\end{subequations}

\hfill $\Box$

We summarise our results in the following theorem.
\begin{thm}\label{prop:Lie2}
 Let $\CCG$ be a Lie quasi-2-group. The moduli space $\sLie(\CCG)$ then is
 \begin{equation}\label{eq:1Jet2Groups}
\sLie(\CCG)\ =\ \ker(\sff^1_{0\,*}|_{\sigma_1(*)})[1]~\oplus~ \ker(\sff^2_{0\,*}|_{\sigma_2(*)})[2]\cap\ker(\sff^2_{1\,*}|_{\sigma_2(*)})[2]~,
\end{equation}
which is endowed with a homological vector field $Q$ whose evaluation on elements $\alpha_1\in \ker(\sff^1_{0\,*}|_{\sigma_1(*)})[1]$ and $\alpha_2\in \ker(\sff^2_{0\,*}|_{\sigma_2(*)})[2]\cap\ker(\sff^2_{1\,*}|_{\sigma_2(*)})[2]$ reads as
 \begin{equation}
\begin{aligned}
  Q\alpha_1\ &=\ -\mu_1(\alpha_2)-\tfrac12\mu_2(\alpha_1,\alpha_1)~,\\
  Q\alpha_2\ &=\ -\mu_2(\alpha_1,\alpha_2)-\tfrac{1}{3!}\mu_3(\alpha_1,\alpha_1,\alpha_1)~,
 \end{aligned}
\end{equation}
where the $\mu_i$ are given in \eqref{eq:mu1_mu2} and \eqref{eq:mu2_mu3}.
\end{thm}

\subsection{Equivalences between descent data}\label{sec:Equivalences}

Since the $L_\infty$-algebroid of a Lie quasi-groupoid $\CCG$ is given by the moduli space of $\CCG$-valued descent data for a principal $\CCG$-bundle subordinate to the surjective submersions $X\times\FR^{0|1}\to X$, it is natural to study the gauge equivalence relations between such data according to Definition \ref{def:EquivLQuasGrpBun}. The descent data is given by simplicial maps and therefore the gauge equivalence relations will take the form of simplicial homotopies. We are particularly interested in the effect of gauge transformations on the moduli describing the $L_\infty$-algebroid. Specifically, given two equivalent descent data parametrised by $\{\mathring{g},\alpha_1,\alpha_2,\ldots\}$ and $\{\tilde{\mathring{g}},\tilde \alpha_1,\tilde\alpha_2,\ldots\}$, we seek to understand the explicit relation between the two. This relation can be used to infer finite gauge transformation of connections on Lie quasi-groupoid bundles, which we shall discuss in Section \ref{sec:ConStr}.

Let $\CCZ$ be the simplicial manifold with simplicial $p$-simplices $\CCZ_p=X\times \FR^{0|p+1}$. According to Definition \ref{def:DefSimpHomo}, a simplicial homotopy between two simplicial maps $g,\tilde g\in \shom_\CatsSMfd(\CCZ,\CCG)$, is an element $h\in \shom_\CatsSMfd(\CCZ\times\Delta^1,\CCG)$ such that $h(z_p,(0,\ldots,0))=g(z_p)$ and $h(z_p,(1,\ldots,1))=\tilde g(z_p)$ for all simplicial $p$-simplices $z_p\in \CCZ_p$ for $p\geq0$. By Lemma \ref{lem:3SimpSets} and Remark \ref{rem:ComLemHomSimp}, we have
\begin{equation}
\shom_\CatsSMfd(\CCZ\times\Delta^1,\CCG)\ \cong\ \shom_\CatsSMfd(\CCZ,\inthom(\Delta^1,\CCG))~.
\end{equation}
By virtue of the discussion in the previous section, simplicial homotopies are thus parametrised by the 1-jet of $\inthom(\Delta^1,\CCG)$. Furthermore, by Lemma \ref{lem:KanHomXY} and Remark \ref{rem:ComLemHomSimp}, $\inthom(\Delta^1,\CCG)$ is a Lie quasi-groupoid, and, consequently, we can make direct use of Theorem \ref{thm:JetPara}. Specifically, we  arrive at the following result, suppressing again the dependence on $X$ in all formul\ae.

\begin{cor}\label{cor:HomotopyParametrisation}
For $p\geq1$, define $\sigma_{p}:\inthom_0(\Delta^1,\CCG)\ \to\ \inthom_p(\Delta^1,\CCG)$  by $\sigma_{p}:=\sfd^{p-1}_{0}\circ\cdots\circ\sfd^1_0\circ\sfd^0_0$ and set ${\rm Ker}_{\inthom(\Delta^1,\CCG)}[p]|_{\sigma_p(c^0)}:=\bigcap_{i=0}^{p-1}\ker(\sff^p_{i\,*}|_{\sigma_p(c^0)})[p]$, where $c^0\in\inthom_0(\Delta^1,\CCG)$ and $\sff^p_{i\,*}|_{\sigma_p(c^0)}$ denotes the linearisation of the face map $\sff^p_i:\inthom_p(\Delta^1,\CCG)\to \inthom_{p-1}(\Delta^1,\CCG)$ at $\sigma_p(c^0)$. Then $\shom_\CatsSMfd(\CCZ\times\Delta^1,\CCG)$ is parametrised by 
\begin{equation}
 \bigtimes_{p\in\NN}\sigma^*_{p}{\rm Ker}_{\inthom(\Delta^1,\CCG)}[p]\ \to\ \CCG_1~.
\end{equation}
Explicitly,
\begin{equation}\label{eq:hpExpansion}
\begin{aligned}
 \inthom_0(\Delta^1,\CCG)\ \ni\ h^0(\theta_0)\ &=\ c^0+\mu_1^{\inthom}(\chi^1)\theta_0~,\\ 
 \inthom_1(\Delta^1,\CCG)\ \ni\ h^1(\theta_0,\theta_1)\ &=\  \sfd^0_0(c^0)+\chi^1(\theta_0-\theta_1)+\sfd^0_{0\,*}|_{c^0}(\mu_1^{\inthom}(\chi^1))\theta_1+\\
  &\kern1cm+\big[\mu_1^{\inthom}(\chi^2)+\tfrac12\mu_2^{\inthom}(\chi^1,\chi^1)\big]\theta_0\theta_1~,\\
  &~\,\vdots
 \end{aligned}
\end{equation}
where $c^0\in\inthom_0(\Delta^1,\CCG)\cong\CCG_1$ and $\chi^p\in {\rm Ker}_{\inthom(\Delta^1,\CCG)}[p]|_{\sigma_p(c^0)}$.
\end{cor}
\noindent The face and degeneracy maps appearing in this corollary are the ones on $\inthom(\Delta^1,\CCG)$. To compare equivalent descent data parametrised by $\{\mathring{g},\alpha_1,\alpha_2,\ldots\}$ and $\{\tilde{\mathring{g}},\tilde \alpha_1,\tilde\alpha_2,\ldots\}$ we wish to express these in terms of the face and degeneracy maps on $\CCG$. Since $\shom(\Delta^1,\CCG)$ is in $\CatsSMfd$ by Remark \ref{rem:ComLemHomSimp}, we can apply Lemma \ref{lem:HomMap}.

\begin{lemma}\label{lem:HomoExpan}
For a simplicial homotopy $h\in\shom_\CatsSMfd(\CCZ,\inthom(\Delta^1,\CCG))$ as given in \eqref{eq:hpExpansion}, consider 
\begin{equation}
h^p_i(\theta_0,\ldots,\theta_p)\ :=\ h^p(\theta_0,\ldots,\theta_p)(s^p_i)\efor p\ \geq\ 0\eand 0\ \leq\ i\ \leq\ p~,
\end{equation}
where the $s^p_i$ are the non-degenerate simplicial $(p+1)$-simplices of $\Delta^p\times\Delta^1$ as introduced in \eqref{eq:NonDegSimDeltap1} and define 
 \begin{equation}
  c\ :=\ c^0(s^0_0)\eand\chi^p_i\ :=\ \chi^p(s^p_i)\efor p\ >\ 0\eand 0\ \leq\ i\ \leq\ p~.
  \end{equation}
Then,
\begin{subequations}\label{eq:hpi}
\begin{equation}\label{eq:hpiExpansion}
\begin{aligned}
 \CCG_1\ \ni\ h^0_0(\theta_0)\ &=\ c+ \sff^2_{2\,*}|_{\sfd^1_1(c)}(\chi^1_0)\theta_0~,\\
 \CCG_2\ \ni\ h^1_0(\theta_0,\theta_1)\ &=\ \sfd^1_1(c)+\chi^1_0(\theta_0-\theta_1)+ (\sfd^1_1\circ\sff^2_2)_*|_{\sfd^1_1(c)}(\chi^1_0)\theta_1\,+\\
 &\kern1cm+\big[\mu_1(\chi^2_0)+\tfrac12\mu_2(\chi^1_0,\chi^1_0)\big]\theta_0\theta_1~,\\
\CCG_2\ \ni\  h^1_1(\theta_0,\theta_1)\ &=\ \sfd^1_0(c)+\chi^1_1(\theta_0-\theta_1)+ (\sfd^1_0\circ\sff^2_2)_*|_{\sfd^1_1(c)}(\chi^1_0)\theta_1\,+\\
 &\kern1cm+\big[\mu_1(\chi^2_1)+\tfrac12\mu_2(\chi^1_0,\chi^1_1)\big]\theta_0\theta_1~,\\
&~\,\vdots
\end{aligned}
\end{equation}
with
\begin{equation}\label{eq:hpiConditions}
\begin{gathered}
\sff^2_{0\,*}|_{\sfd^1_0(c)}(\chi^1_1)\ =\ 0~,\\
\sff^3_{2\,*}|_{(\sfd^2_1\circ\sfd^1_1)(c)}(\chi^2_0)\ =\ 
\sff^3_{0\,*}|_{(\sfd^2_2\circ\sfd^1_0)(c)}(\chi^2_1)\ =\ 
\sff^3_{0\,*}|_{(\sfd^2_0\circ\sfd^1_0)(c)}(\chi^2_2)\ =\ 
\sff^3_{1\,*}|_{(\sfd^2_0\circ\sfd^1_0)(c)}(\chi^2_2)\ =\ 0\\
\vdots
\end{gathered}
\end{equation}
and
\begin{equation}\label{eq:hpiQ}
\begin{gathered}
 Qc\ =\ -\mu_1(\chi^1_0)~,\\
 Q\chi^1_0\ =\ -\mu_1(\chi^2_0)-\tfrac12\mu_2(\chi^1_0,\chi^1_0)\eand
  Q\chi^1_1\ =\ -\mu_1(\chi^2_1)-\tfrac12\mu_2(\chi^1_0,\chi^1_1)~,\\
  \vdots
\end{gathered}
\end{equation}
\end{subequations}
\end{lemma}

\noindent
{\it Proof:} Firstly, note that the equations in \eqref{eq:hpiQ} are a straightforward consequence  of Proposition \ref{prop:Qvector} when applied to the equations in \eqref{eq:hpiExpansion}.

To prove the expression for $h^0_0(\theta_0)$, we note that
\begin{equation}\label{eq:ActS10g0}
\begin{aligned}
\big(\sfd^0_0(c^0)\big)(s^1_0)\ &=\ c^0\big((\delta^0_0\times\id)(s^1_0)\big)\ =\ c^0\big(\sfd^1_1(s^0_0)\big)\ =\ \sfd^1_1\big(c^0(s^0_0)\big)\ =\ \sfd^1_1(c)~,\\
\big(\sfd^0_0(c^0)\big)(s^1_1)\ &=\ c^0\big((\delta^0_0\times\id)(s^1_1)\big)\ =\ c^0\big(\sfd^1_0(s^0_0)\big)\ =\ \sfd^1_0\big(c^0(s^0_0)\big)\ =\ \sfd^1_0(c)~.
\end{aligned}
\end{equation}
Following the arguments in the proof of Theorem \ref{thm:JetPara}, we have
\begin{equation}
 h^1(\theta_0,0)\ =\ \sfd^0_0(c^0)+\chi^1\theta_0~.
\end{equation}
Consequently, using \eqref{eq:ActS10g0}, we find
\begin{equation}
\begin{aligned}
 h^1_0(\theta_0,0)\ &=\ h^1(\theta_0,0)(s^1_0)\ =\ \sfd^0_0(c^0)(s^1_0)+\chi^1(s^1_0)\theta_0\ =\ \sfd^1_1(c)+\chi^1_0\theta_0~,\\
  h^1_1(\theta_0,0)\ &=\ h^1(\theta_0,0)(s^1_1)\ =\ \sfd^0_0(c^0)(s^1_1)+\chi^1(s^1_1)\theta_0\ =\ \sfd^1_0(c)+\chi^1_1\theta_0~.
 \end{aligned}
\end{equation}
Since $(\sff^2_0\circ h^1_1)(\theta_0,0)=(h^0_0\circ\sff^1_0)(\theta_0,0)=h^0_0(0)$, which follows from \eqref{eq:HomMap}, we conclude that the condition $\sff^1_{0\,*}|_{\sigma_1(c^0)}(\chi^1)=0$ translates into the requirement $\sff^1_{0\,*}|_{\sigma_1(c^0)}(\chi^1)(s^0_0)=0$, that is,
\begin{equation}
 \sff^2_{0\,*}|_{\sfd^1_0(c)}(\chi^1_1)\ =\ 0~.
 \end{equation}
Likewise, $(\sff^2_2\circ h^1_0)(\theta_0,0)=(h^0_0\circ\sff^1_1)(\theta_0,0)=h^0_0(\theta_0)$ yields $h^0_0(\theta_0)$ as given in \eqref{eq:hpiExpansion}.

To establish the expressions for $h^1_i(\theta_0,\theta_1)$, we proceed similarly. Indeed, considering
\begin{equation}
 h^2(\theta_0,\theta_1,0)\ =\ (\sfd^1_0\circ\sfd^0_0)(c^0)+\sfd^1_{1\,*}(\chi^1)(\theta_0-\theta_1)+\sfd^1_{0\,*}(\chi^1)\theta_1+\hat\chi^2\theta_0\theta_1~,
\end{equation}
which again can be read off from the proof of Theorem \ref{thm:JetPara}, together with
 \begin{equation}\label{eq:ActS2i}
 \begin{aligned}
 (\sfd^1_0\circ\sfd^0_0)(c^0)(s^2_0)\ &=\ (\sfd^2_1\circ\sfd^1_1)(c)~,\\
  (\sfd^1_0\circ\sfd^0_0)(c^0)(s^2_1)\ &=\ (\sfd^2_2\circ\sfd^1_0)(c)~,\\
 (\sfd^1_0\circ\sfd^0_0)(c^0)(s^2_2)\ &=\ (\sfd^2_0\circ\sfd^1_0)(c)~,
 \end{aligned}
 \end{equation}
which follow in a similar way to \eqref{eq:ActS10g0}, it is a straightforward exercise to show that
\begin{equation}
\begin{aligned}
 h^2_0(\theta_0,\theta_1,0)\ &=\ (\sfd^2_1\circ\sfd^1_1)(c)+\sfd^2_{2\,*}|_{\sfd^1_1(c)}(\chi^1_0)(\theta_0-\theta_1)+\sfd^2_{1\,*}|_{\sfd^1_1(c)}(\chi^1_0)\theta_1+\hat\chi^2_0\theta_0\theta_1~,\\
  h^2_1(\theta_0,\theta_1,0)\ &=\ (\sfd^2_2\circ\sfd^1_0)(c)+\sfd^2_{2\,*}|_{\sfd^1_0(c)}(\chi^1_1)(\theta_0-\theta_1)+\sfd^2_{0\,*}|_{\sfd^1_1(c)}(\chi^1_0)\theta_1+\hat\chi^2_1\theta_0\theta_1~,\\
    h^2_2(\theta_0,\theta_1,0)\ &=\ (\sfd^2_0\circ\sfd^1_0)(c)+\sfd^2_{1\,*}|_{\sfd^1_0(c)}(\chi^1_1)(\theta_0-\theta_1)+\sfd^2_{0\,*}|_{\sfd^1_0(c)}(\chi^1_1)\theta_1+\hat\chi^2_2\theta_0\theta_1~.
\end{aligned}
\end{equation}
Next, upon applying $\sff^3_2$ to $h^2_0(\theta_0,\theta_1,0)$, $\sff^3_0$ to $h^2_1(\theta_0,\theta_1,0)$, $\sff^3_0$ to $h^2_2(\theta_0,\theta_1,0)$, and $\sff^3_1$ to $h^2_2(\theta_0,\theta_1,0)$ and by making use of the identities \eqref{eq:HomMap}, we find
\begin{equation}\label{eq:HomotopyLS}
\begin{aligned}
\sff^3_{2\,*}|_{(\sfd^2_1\circ\sfd^1_1)(c)}(\hat\chi^2_0)\ &=\  -{\rm II}_{\sff^3_2}\big(\sfd^2_{1\,*}|_{\sfd^1_1(c)}(\chi^1_0),\sfd^2_{2\,*}|_{\sfd^1_1(c)}(\chi^1_0)\big)~,  \\
\sff^3_{0\,*}|_{(\sfd^2_2\circ\sfd^1_0)(c)}(\hat\chi^2_1)\ &=\ -{\rm II}_{\sff^3_0}\big(\sfd^2_{0\,*}|_{\sfd^1_1(c)}(\chi^1_0),\sfd^2_{2\,*}|_{\sfd^1_0(c)}(\chi^1_1)\big)~,   \\
\sff^3_{0\,*}|_{(\sfd^2_0\circ\sfd^1_0)(c)}(\hat\chi^2_2)\ &=\ -{\rm II}_{\sff^3_0}\big(\sfd^2_{0\,*}|_{\sfd^1_0(c)}(\chi^1_1),\sfd^2_{1\,*}|_{\sfd^1_0(c)}(\chi^1_1)\big)~,  \\
\sff^3_{1\,*}|_{(\sfd^2_0\circ\sfd^1_0)(c)}(\hat\chi^2_2)\ &=\  - {\rm II}_{\sff^3_1}\big(\sfd^2_{0\,*}|_{\sfd^1_0(c)}(\chi^1_1),\sfd^2_{1\,*}|_{\sfd^1_0(c)}(\chi^1_1)\big)~,
\end{aligned}
\end{equation}
so that the conditions $\sff^2_{0\,*}|_{\sigma_2(c^0)}(\chi^2)=\sff^2_{1\,*}|_{\sigma_2(c^0)}(\chi^2)=0$ translate into $\sff^2_{0\,*}|_{\sigma_2(c^0)}(\chi^2)(s^1_i)=\sff^2_{1\,*}|_{\sigma_2(c^0)}(\chi^2)(s^1_i)=0$. Explicitly, the $\chi^2_i$ represent the homogeneous solution to \eqref{eq:HomotopyLS}:
\begin{equation}
\sff^3_{2\,*}|_{(\sfd^2_1\circ\sfd^1_1)(c)}(\chi^2_0)\ =\ 
\sff^3_{0\,*}|_{(\sfd^2_2\circ\sfd^1_0)(c)}(\chi^2_1)\ =\ 
\sff^3_{0\,*}|_{(\sfd^2_0\circ\sfd^1_0)(c)}(\chi^2_2)\ =\ 
\sff^3_{1\,*}|_{(\sfd^2_0\circ\sfd^1_0)(c)}(\chi^2_2)\ =\ 0~.
\end{equation}
Furthermore,  using the identities $(\sff^3_3\circ h^2_i)(\theta_0,\theta_1,0)=(h^1_i\circ\sff^2_2)(\theta_0,\theta_1,0)=h^1_i(\theta_0,\theta_1)$ for $i=0,1$, we obtain
\begin{equation}\label{eq:SomeHomotEq}
\begin{aligned}
h^1_0(\theta_0,\theta_1)\ &=\ \sfd^1_1(c)+\chi^1_0(\theta_0-\theta_1)+(\sfd^1_1\circ\sff^2_2)_*|_{\sfd^1_1(c)}(\chi^1_0)\theta_1\,+\\
&\kern1cm+\Big[\sff^3_{3\,*}|_{(\sfd^2_1\circ\sfd^1_1)(c)}(\hat\chi^2_0)+{\rm II}_{\sff^3_3}\big(\sfd^2_{1\,*}|_{\sfd^1_1(c)}(\chi^1_0),\sfd^2_{2\,*}|_{\sfd^1_1(c)}(\chi^1_0)\big) \Big]\theta_0\theta_1\\
&=\ \sfd^1_1(c)+\chi^1_0(\theta_0-\theta_1)+(\sfd^1_1\circ\sff^2_2)_*|_{\sfd^1_1(c)}(\chi^1_0)\theta_1+\big[\mu_1(\chi^2_0)+\tfrac12\mu_2(\chi^1_0,\chi^1_0)\big]\theta_0\theta_1~,\\
h^1_1(\theta_0,\theta_1)\ &=\ \sfd^1_0(c)+\chi^1_1(\theta_0-\theta_1)+ (\sfd^1_0\circ\sff^2_2)_*|_{\sfd^1_1(c)}(\chi^1_0)\theta_1\,+\\
 &\kern1cm+\Big[\sff^3_{3\,*}|_{(\sfd^2_2\circ\sfd^1_0)(c)}(\hat\chi^2_1)+{\rm II}_{\sff^3_3}\big(\sfd^2_{0\,*}|_{\sfd^1_1(c)}(\chi^1_0),\sfd^2_{2\,*}|_{\sfd^1_0(c)}(\chi^1_1)\big) \Big]\theta_0\theta_1\\
 &=\ \sfd^1_0(c)+\chi^1_1(\theta_0-\theta_1)+ (\sfd^1_0\circ\sff^2_2)_*|_{\sfd^1_1(c)}(\chi^1_0)\theta_1+\big[\mu_1(\chi^2_1)+\tfrac12\mu_2(\chi^1_0,\chi^1_1)\big]\theta_0\theta_1~,
\end{aligned}
\end{equation} 
where in the expressions involving $\mu_i$ we have inserted the solution to \eqref{eq:HomotopyLS} (which is expressed in terms of the $\chi^2_i$).  \hfill $\Box$
 
\begin{thm}\label{thm:GaugeTrafo}
Equivalent descent data parametrised by $\{\mathring{g},\alpha_1,\alpha_2,\ldots\}$ and $\{\tilde{\mathring{g}},\tilde \alpha_1,\tilde \alpha_2,\ldots\}$ with $\mathring{g},\tilde{\mathring{g}}\in\CCG_0$ and $\alpha_p\in{\rm Ker}_\CCG[p]|_{\sigma_p(\mathring{g})}=\bigcap_{i=0}^{p-1}\ker(\sff^p_{i\,*}|_{\sigma_p(\mathring{g})})[p]$ and $\tilde\alpha_p\in{\rm Ker}_\CCG[p]|_{\sigma_p(\tilde{\mathring{g}})}=\bigcap_{i=0}^{p-1}\ker(\sff^p_{i\,*}|_{\sigma_p(\tilde{\mathring{g}})})[p]$ for $p\geq1$ are related by maps $c$ and $\chi^p_i$ as introduced in Lemma \ref{lem:HomoExpan} so that
\begin{subequations}\label{eq:GaugeTrafo}
\begin{equation}\label{eq:GaugeTrafo-L0}
 \sff^1_0(c)\ =\ \tilde{\mathring{g}} \eand \sff^1_1(c)\ =\ \mathring{g}~
\end{equation}
and
\begin{equation}\label{eq:GaugeTrafo-L1}
\begin{gathered}
 \sff^2_{2\,*}|_{\sfd^1_1(c)}(\chi^1_0)\ =\ -Qc\eand \sff^2_{0\,*}|_{\sfd^1_1(c)}(\chi^1_0)\ =\ \tilde\alpha_1~,\\
  \sff^2_{1\,*}|_{\sfd^1_1(c)}(\chi^1_0)\ =\ \sff^2_{1\,*}|_{\sfd^1_0(c)}(\chi^1_1)~,\\
   \sff^2_{2\,*}|_{\sfd^1_0(c)}(\chi^1_1)\ =\ \alpha_1\eand\sff^2_{0\,*}|_{\sfd^1_0(c)}(\chi^1_1)\ =\ 0
 \end{gathered}
\end{equation}
\end{subequations}
and higher relations for $p>1$.\footnote{For brevity, we have only displayed the relations for $p=0,1$ as we shall focus on those later on. We shall comment on the derivation of the higher relations in Remark \ref{rmk:HigherRelHomDer}.}
\end{thm} 

\noindent
{\it Proof:} 
We simultaneously prove \eqref{eq:GaugeTrafo-L0} and \eqref{eq:GaugeTrafo-L1}. To this end, consider
\begin{equation}
 h^1_0(\theta_0,0)\ =\ \sfd^1_1(c)+\chi^1_0\theta_0\eand
  h^1_1(\theta_0,0)\ =\ \sfd^1_0(c)+\chi^1_1\theta_0~.
\end{equation}
from the proof of the Lemma \ref{lem:HomoExpan}. Using the relations $\tilde g^1(\theta_0,0)=(\sff^2_0\circ h^1_0)(\theta_0,0)$ and $g^1(\theta_0,0)=(\sff^2_2\circ h^1_1)(\theta_0,0)$, which follow from \eqref{eq:HomMap-1}, we immediately obtain
\begin{equation}\label{eq:0LevelHomGam}
\begin{aligned}
  \sff^1_0(c)\ =\ \tilde{\mathring{g}} &\eand \sff^1_1(c)\ =\ \mathring{g}~,\\
  \sff^2_{0\,*}|_{\sfd^1_1(c)}(\chi^1_0)\ =\ \tilde \alpha_1 &\eand \sff^2_{2\,*}|_{\sfd^1_0(c)}(\chi^1_1)\ =\ \alpha_1~.
  \end{aligned}
\end{equation}
This establishes \eqref{eq:GaugeTrafo-L0} and partially also  \eqref{eq:GaugeTrafo-L1}. From \eqref{eq:HomMap-2} we have the identity $(\sff^2_1\circ h^1_1)(\theta_0,0)=(\sff^2_1\circ h^1_0)(\theta_0,0)$ which, in turn, implies that
\begin{equation}
  \sff^2_{1\,*}|_{\sfd^1_1(c)}(\chi^1_0)\ =\ \sff^2_{1\,*}|_{\sfd^1_0(c)}(\chi^1_1)~.
\end{equation}
Since also $Qc=- \sff^2_{2\,*}|_{\sfd^1_1(c)}(\chi^1_0)$ and $\sff^2_{0\,*}|_{\sfd^1_0(c)}(\chi^1_1)=0$ by virtue of \eqref{eq:hpi}, we have thus established \eqref{eq:GaugeTrafo-L1}. \hfill $\Box$

\begin{rem}\label{rmk:HigherRelHomDer}
Let us briefly comment on the derivations of the gauge relations for the next level. To this end, consider
\begin{equation}
\begin{aligned}
 h^2_0(\theta_0,\theta_1,0)\ &=\ (\sfd^2_1\circ\sfd^1_1)(c)+\sfd^2_{2\,*}|_{\sfd^1_1(c)}(\chi^1_0)(\theta_0-\theta_1)+\sfd^2_{1\,*}|_{\sfd^1_1(c)}(\chi^1_0)\theta_1+\hat\chi^2_0\theta_0\theta_1~,\\
  h^2_1(\theta_0,\theta_1,0)\ &=\ (\sfd^2_2\circ\sfd^1_0)(c)+\sfd^2_{2\,*}|_{\sfd^1_0(c)}(\chi^1_1)(\theta_0-\theta_1)+\sfd^2_{0\,*}|_{\sfd^1_1(c)}(\chi^1_0)\theta_1+\hat\chi^2_1\theta_0\theta_1~,\\
    h^2_2(\theta_0,\theta_1,0)\ &=\ (\sfd^2_0\circ\sfd^1_0)(c)+\sfd^2_{1\,*}|_{\sfd^1_0(c)}(\chi^1_1)(\theta_0-\theta_1)+\sfd^2_{0\,*}|_{\sfd^1_0(c)}(\chi^1_1)\theta_1+\hat\chi^2_2\theta_0\theta_1~.
\end{aligned}
\end{equation}
again taken from the proof of the Lemma \ref{lem:HomoExpan}. Using $\tilde g^2(\theta_0,\theta_1,0)=(\sff^3_0\circ h^2_0)(\theta_0,\theta_1,0)$ and $g^2(\theta_0,\theta_1,0)=(\sff^3_3\circ h^2_2)(\theta_0,0)$, we arrive after a short calculation at
\begin{equation}\label{eq:GThatgamma2}
\begin{aligned}
 \sff^3_{0\,*}|_{(\sfd^2_1\circ\sfd^1_1)(c)}(\hat\chi^2_0)+{\rm II}_{\sff^3_0}\big(\sfd^2_{1\,*}|_{\sfd^1_1(c)}(\chi^1_0),\sfd^2_{2\,*}|_{\sfd^1_1(c)}(\chi^1_0)\big)\ &=\ {\tilde{\hat\gamma}}^{(2)}_{01}~,\\
\sff^3_{3\,*}|_{(\sfd^2_0\circ\sfd^1_0)(c)}(\hat\chi^2_2)+{\rm II}_{\sff^3_3}\big(\sfd^2_{0\,*}|_{\sfd^1_0(c)}(\chi^1_1),\sfd^2_{1\,*}|_{\sfd^1_0(c)}(\chi^1_1)\big)\ &=\ \hat\gamma^{(2)}_{01}~,
\end{aligned}
\end{equation}
and using $(\sff^3_1\circ h^2_1)(\theta_0,\theta_1,0)=(\sff^3_1\circ h^2_0)(\theta_0,\theta_1,0)$ and $(\sff^3_2\circ h^2_2)(\theta_0,\theta_1,0)=(\sff^3_2\circ h^2_1)(\theta_0,\theta_1,0)$, we find
\begin{equation}
\begin{aligned}
\sff^3_{1\,*}|_{(\sfd^2_2\circ\sfd^1_0)(c)}(\hat\chi^2_1)+{\rm II}_{\sff^3_1}\big(\sfd^2_{0\,*}|_{\sfd^1_1(c)}(\chi^1_0),\sfd^2_{2\,*}|_{\sfd^1_0(c)}(\chi^1_1)\big)\ &=  \\
&\kern-5cm=\ \sff^3_{1\,*}|_{(\sfd^2_1\circ\sfd^1_1)(c)}(\hat\chi^2_0)+{\rm II}_{\sff^3_1}\big(\sfd^2_{1\,*}|_{\sfd^1_1(c)}(\chi^1_0),\sfd^2_{2\,*}|_{\sfd^1_1(c)}(\chi^1_0)\big)~, \\
\sff^3_{2\,*}|_{(\sfd^2_0\circ\sfd^1_0)(c)}(\hat\chi^2_2)+{\rm II}_{\sff^3_2}\big(\sfd^2_{0\,*}|_{\sfd^1_0(c)}(\chi^1_1),\sfd^2_{1\,*}|_{\sfd^1_0(c)}(\chi^1_1)\big)\ &= \\
&\kern-5cm=\ \sff^3_{2\,*}|_{(\sfd^2_2\circ\sfd^1_0)(c)}(\hat\chi^2_1)+{\rm II}_{\sff^3_2}\big(\sfd^2_{0\,*}|_{\sfd^1_1(c)}(\chi^1_0),\sfd^2_{2\,*}|_{\sfd^1_0(c)}(\chi^1_1)\big)~.
\end{aligned}
\end{equation}
From Lemma \ref{lem:HomoExpan}, we also know that
\begin{equation}\label{eq:HigherRelHomDer1}
\begin{aligned}
\sff^3_{2\,*}|_{(\sfd^2_1\circ\sfd^1_1)(c)}(\hat\chi^2_0)\ &=\  -{\rm II}_{\sff^3_2}\big(\sfd^2_{1\,*}|_{\sfd^1_1(c)}(\chi^1_0),\sfd^2_{2\,*}|_{\sfd^1_1(c)}(\chi^1_0)\big)~,  \\
\sff^3_{0\,*}|_{(\sfd^2_2\circ\sfd^1_0)(c)}(\hat\chi^2_1)\ &=\ -{\rm II}_{\sff^3_0}\big(\sfd^2_{0\,*}|_{\sfd^1_1(c)}(\chi^1_0),\sfd^2_{2\,*}|_{\sfd^1_0(c)}(\chi^1_1)\big)~,   \\
\sff^3_{0\,*}|_{(\sfd^2_0\circ\sfd^1_0)(c)}(\hat\chi^2_2)\ &=\ -{\rm II}_{\sff^3_0}\big(\sfd^2_{0\,*}|_{\sfd^1_0(c)}(\chi^1_1),\sfd^2_{1\,*}|_{\sfd^1_0(c)}(\chi^1_1)\big)~,  \\
\sff^3_{1\,*}|_{(\sfd^2_0\circ\sfd^1_0)(c)}(\hat\chi^2_2)\ &=\  - {\rm II}_{\sff^3_1}\big(\sfd^2_{0\,*}|_{\sfd^1_0(c)}(\chi^1_1),\sfd^2_{1\,*}|_{\sfd^1_0(c)}(\chi^1_1)\big)
\end{aligned}
\end{equation}
and
\begin{equation}\label{eq:HigherRelHomDer2}
\begin{aligned}
\sff^3_{3\,*}|_{(\sfd^2_1\circ\sfd^1_1)(c)}(\hat\chi^2_0)+{\rm II}_{\sff^3_3}\big(\sfd^2_{1\,*}|_{\sfd^1_1(c)}(\chi^1_0),\sfd^2_{2\,*}|_{\sfd^1_1(c)}(\chi^1_0)\big) \ &=\ -Q\chi^1_0~,\\
\sff^3_{3\,*}|_{(\sfd^2_2\circ\sfd^1_0)(c)}(\hat\chi^2_1)+{\rm II}_{\sff^3_3}\big(\sfd^2_{0\,*}|_{\sfd^1_1(c)}(\chi^1_0),\sfd^2_{2\,*}|_{\sfd^1_0(c)}(\chi^1_1)\big)\ &=\ -Q\chi^1_1~.
\end{aligned}
\end{equation}
Furthermore,
\begin{equation}
\begin{aligned}
 \hat\gamma^{(2)}_{01}\ &=\ \alpha_2-\sfd^1_{0\,*}|_{(\sfd^1_0\circ\sfd^0_0)(\mathring{g})}\Big({\rm II}_{\sff^2_0}\big(\sfd^1_{0\,*}|_{\sfd^0_0(\mathring{g})}(\alpha_1),\sfd^1_{1\,*}|_{\sfd^0_0(\mathring{g})}(\alpha_1)\big)\Big)\,-\\
 &\kern1.5cm-\,\sfd^1_{1\,*}|_{(\sfd^1_0\circ\sfd^0_0)(\mathring{g})}\Big({\rm II}_{\sff^2_1}\big(\sfd^1_{0\,*}|_{\sfd^0_0(\mathring{g})}(\alpha_1),\sfd^1_{1\,*}|_{\sfd^0_0(\mathring{g})}(\alpha_1)\big)\,-\\
 &\kern5cm-\,{\rm II}_{\sff^2_0}\big(\sfd^1_{0\,*}|_{\sfd^0_0(\mathring{g})}(\alpha_1),\sfd^1_{1\,*}|_{\sfd^0_0(\mathring{g})}(\alpha_1)\big)\Big)
 \end{aligned}
\end{equation}
and likewise for $\tilde{\hat{\gamma}}^{(2)}_{01}$; see equation \eqref{eq:solhatalpha2}. This is then to be substituted into \eqref{eq:GThatgamma2} to obtain the relations between $\alpha_2$ and $\tilde\alpha_2$ in terms of the $\chi^2_i$ (for $i=0,1,2$) which are the homogeneous solutions to \eqref{eq:HigherRelHomDer1}, \eqref{eq:HigherRelHomDer2}, and \eqref{eq:SomeHomotEq}.
\end{rem}

Finally, we may also discuss the equivalences between equivalences which will ultimately result in gauge transformations between gauge transformations. In particular, from Definition \ref{def:simplicial_k_homotopy}, Lemma \ref{lem:3SimpSets}, and Theorem \ref{thm:JetPara}, we can immediately infer the following result.

\begin{cor}
Let $\CCG$ be a Lie quasi-groupoid. For $p\geq1$, define the map\linebreak $\sigma_{p}:\inthom_0(\Delta^k,\CCG)\ \to\ \inthom_p(\Delta^k,\CCG)$  by $\sigma_{p}:=\sfd^{p-1}_{0}\circ\cdots\circ\sfd^1_0\circ\sfd^0_0$ and set\linebreak ${\rm Ker}_{\inthom(\Delta^k,\CCG)}[p]|_{\sigma_p(c^0_{(k)})}:=\bigcap_{i=0}^{p-1}\ker(\sff^p_{i\,*}|_{\sigma_p(c^0_{(k)})})[p]$  with $c^0_{(k)}\in\inthom_0(\Delta^k,\CCG)$ and $\sff^p_{i\,*}|_{\sigma_p(c^0_{(k)})}$ denotes the linearisation of $\sff^p_i:\inthom_p(\Delta^k,\CCG)\to \inthom_{p-1}(\Delta^k,\CCG)$ at $\sigma_p(c^0_{(k)})$. Then,\linebreak $\shom_\CatsSMfd(\CCZ\times\Delta^k,\CCG)$ is parametrised by 
\begin{equation}
 \bigtimes_{p\in\NN}\sigma^*_{p}{\rm Ker}_{\inthom(\Delta^k,\CCG)}[p]\ \to\ \inthom_0(\Delta^k,\CCG)~.
\end{equation}
\end{cor}

\subsection{A comment on non-flat moduli from d\'ecalage}\label{ssec:decalage}

The above construction of the supermanifold $\sLie(\CCG)$ representing the functor\linebreak $g:N(\check\CC(X\times\FR^{0|1}\to X))\to\CCG$ yields moduli $\alpha_p$ that are flat in the sense that 
\begin{equation}
\begin{aligned}
  Q\alpha_1+\mu_1(\alpha_2)+\tfrac12\mu_2(\alpha_1,\alpha_1)\ &=\ 0~,\\
  Q\alpha_2+\mu_2(\alpha_1,\alpha_2)+\tfrac{1}{3!}\mu_3(\alpha_1,\alpha_1,\alpha_1)\ &=\ 0~,
\end{aligned}
\end{equation}
etc., cf.\ Proposition \ref{prop:Lie2}. However, this construction can be generalised to obtain non-flat moduli by considering the so-called d\'ecalage of the nerve of the \v Cech groupoid of the surjective submersion $X\times \FR^{0|1}\rightarrow X$. Even though this is not the approach we intend to take in our subsequent discussion, let us briefly sketch this construction for the case of a Lie 1-quasi-group $\CCG=\sG\rightrightarrows *$.

\begin{definition}[\!\!{\cite{Illusie:1972aa}}]
 Given a simplical set $\CCX$, the \uline{d\'ecalage functor} yields a simplicial set $\Dec \CCX$ which has $n$-simplices $(\Dec\CCX)_p:=\CCX_{p+1}$ together with face maps $\sff_i^p:(\Dec\CCX)_p=\CCX_{p+1}\rightarrow \CCX_p$ for $i=0,\ldots,p$ and degeneracy maps $\sfd_i^p:(\Dec\CCX)_p=\CCX_{p+1}\rightarrow \CCX_{p+2}$ for $i=0,\ldots,p$.
\end{definition}

\noindent 
The d\'ecalage of the nerve of the relevant \v Cech groupoid $\CCD$ then has $p$-simplices
\begin{equation}
\CCD_p\ :=\ (\Dec N(\check\CC(X\times\FR^{0|1}\to X)))_p\ \cong\ X\times \FR^{0|p+2} 
\end{equation}
with face and degeneracy maps
\begin{equation}
\begin{aligned}
 \sff^p_i(x,\theta_0,\theta_1,\ldots,\theta_p,\theta_{p+1})\ &:=\ (x,\theta_0,\ldots,\theta_{i-1},\theta_{i+1},\ldots,\theta_p,\theta_{p+1})~,\\
 \sfd^p_i(x,\theta_0,\theta_1,\ldots,\theta_p,\theta_{p+1})\ &:=\ (x,\theta_0,\ldots,\theta_{i-1},\theta_i,\theta_i,\ldots,\theta_p,\theta_{p+1})
\end{aligned}
\end{equation}
for $x\in X$ and $\theta_i\in\FR^{0|1}$ for $i\in\{0,\ldots,p+1\}$. 

Let $\sG$ be a Lie group. A simplicial map $\CCD\rightarrow N(\sB\sG)$ is then fully determined by the component map $g^1(\theta_0,\theta_1,\theta_2):\CCD_1\rightarrow N_1(\sB\sG)=\sG\times\sG$. Its expansion in the $\theta_i$ is then constructed as in Section \ref{ssec:1jets}, and the final result is
\begin{equation}
  g^1(\theta_0,\theta_1,\theta_2)\ =\ \unit_\sG+\alpha_1(\theta_0-\theta_1)+\tfrac12 \mu_2(\alpha_1,\alpha_1)\theta_0\theta_1-\kappa_2(\theta_0-\theta_1)\theta_2-\mu_2(\alpha_1,\kappa_2)\theta_0\theta_1\theta_2~,
\end{equation}
where $\alpha_1\in \ker(f^1_{0*})[1]$ and $\kappa_2\in \ker(f^1_{0*})[2]$. The homological vector field $Q$ acts now on the moduli $\alpha_1$ and $\kappa_2$ according to
\begin{equation}
\begin{aligned}
 Q\alpha_1+\tfrac12\mu_2(\alpha_1,\alpha_1)\ &=\ \kappa_2~,\\
 Q\kappa_2+\mu_2(\alpha_1,\kappa_2)\ &=\ 0~.
\end{aligned}
\end{equation}
The first equation defines a curvatures $\kappa_2=Q\alpha_1+\tfrac12\mu_2(\alpha_1,\alpha_1)$ while the second equation amounts to the corresponding Bianchi identity. It is rather clear that this construction generalises to Lie $n$-quasi-groupoids leading to 
\begin{equation}
\begin{aligned}
  Q\alpha_1+\mu_1(\alpha_2)+\tfrac12\mu_2(\alpha_1,\alpha_1)\ &=\ \kappa_2~,\\
  Q\alpha_2+\mu_2(\alpha_1,\alpha_2)+\tfrac{1}{3!}\mu_3(\alpha_1,\alpha_1,\alpha_1)\ &=\ \kappa_3~,
\end{aligned}
\end{equation}
etc.\ with $\kappa_2\in \ker(f^1_{0*})[2]$. 

Note that the resulting N-manifold is a cone of the morphism $\id:\sLie(\CCG)\rightarrow \sLie(\CCG)$ as constructed in \cite[Section III.3.3.b)]{Gelfand:2003aa}.

\pagebreak
\section{Lie quasi-groupoid bundles with connection}\label{sec:ConStr}

\subsection{Local description and infinitesimal gauge transformations}

As explained in \cite{Jurco:2014mva}, the local description of the kinematical data of higher gauge theory can be derived from the homotopy Maurer--Cartan equation of an $L_\infty$-algebra. Specifically, one starts from the $L_\infty$-algebra given by the tensor product of the de Rham complex, regarded as a differential graded algebra, and the gauge $L_\infty$-algebra. The homotopy Maurer--Cartan equation for degree-1 elements of this $L_\infty$-algebra then amounts to a flat connection, and one can read off the natural definition of the corresponding higher curvature forms. Moreover, the infinitesimal symmetries of the Maurer--Cartan equation are well-known, and yield the infinitesimal gauge transformations of both higher connection and curvature forms. 

Here, however, we follow an alternative route. It was first noted by Atiyah \cite{Atiyah:1957} that flat connections over a manifold $X$ can conveniently be regarded as a splitting of the defining short exact sequence of what is now called the Atiyah Lie algebroid. There is a straightforward generalisation of this picture due to \cite{Bojowald:0406445,Kotov:2007nr,Gruetzmann:2014ica} (see also \cite{Sati:0801.3480,Fiorenza:2011jr} or \cite{Ritter:2015ymv} for details), which uses the convenient language of N$Q$-manifolds\footnote{cf.\ Section \ref{ssec:NQ_manifolds}}. We already described how to regard the $L_\infty$-algebroid of a Lie quasi-groupoid as an N$Q$-manifold. In fact, this form of an $L_\infty$-algebroid was the output of our differentiation method in Section \ref{ssec:1jets}. Note that we may also view the grade-shifted tangent bundle $T[1]X$ as an N$Q$-manifold. Here, the algebra of functions on $T[1]X$ is naturally identified with the differential forms $\Omega^\bullet_X$ and the homological vector field $Q$ is simply the de Rham differential.

Consider now a Lie quasi-groupoid $\CCG$ with corresponding $L_\infty$-algebroid $\sLie(\CCG)$ and a trivial principal $\CCG$-bundle over a manifold $X$. A higher gauged sigma model consists of a scalar field mapping $X$ to $\sLie(\CCG)_0=\CCG_0$ as well as a connection on the principal bundle. All of these fields are captured by a morphism of graded manifolds, 
\begin{equation}\label{eq:degree_preserving_map}
 \underline{a}\ =\ (\phi,a)\,:\,T[1]X\ \rightarrow\ \sLie(\CCG)~,
\end{equation}
cf.\ section \ref{ssec:presheaves_of_supermanifolds}. This can be understood as follows. We have a scalar field of a sigma model encoded in the morphism $\phi:X\rightarrow \CCG_0$ as well as a local morphism of graded rings $a:\CS_{\sLie(\CCG)}\rightarrow \phi_*\CS_{T[1]X}$. To extract the connection from the latter, recall that the morphism $a$ is fully characterised by its image of coordinate functions on $\sLie(\CCG)$. These images define $\sLie(\CCG)_p$-valued $p$-forms on $X$, which constitute a higher connection on the principal $\CCG$-bundle. 

To illustrate this perspective on connections, let us focus on the case of a principal $\sG$-bundle with $\sG$ a Lie group. In local coordinates $\xi^\alpha$ with respect to a basis $\tau_\alpha$ on $\sLie(\sG)[1]$, the image $a(\xi^\alpha)$ defines the 1-form gauge potential $A=a(\xi^\alpha)\tau_\alpha$.

This approach to gauged scalar fields and connections is particularly appealing as both the curvature and gauge transformations of the connection find very natural interpretations. The former is simply the failure of $\underline{a}$ to be a morphism of N$Q$-manifolds, and in the simple example of the principal $\sG$-bundle, we have
\begin{equation}
 F\ =\ F^\alpha\tau_\alpha~~~\mbox{with}~~~F^\alpha\ =\ (\dd \circ a-a\circ Q_{\sLie(\sG)})(\xi^\alpha)\ =\ \dd A^\alpha+\tfrac{1}{2}f_{\beta\gamma}{}^\alpha A^\beta\wedge A^\gamma~,
\end{equation}
where $Q_{\sLie(\sG)}:=-\tfrac12\xi^\beta\xi^\gamma f_{\beta\gamma}{}^\alpha \der{\xi^\alpha}$.

Note that to remain fully within the category of N$Q$-manifolds one may consider morphisms from $T[1]X$ to the Weil algebra $T[1]\sLie(\CCG)$ instead. This is the cone of $\id:\sLie(\CCG)\rightarrow \sLie(\CCG)$ and related to the discussion in Section \ref{ssec:decalage}.

Furthermore, gauge transformations between connections encoded in morphisms of graded manifolds $\underline{a}$ and $\underline{\tilde{a}}$ are simply flat homotopies between these \cite{Bojowald:0406445,Fiorenza:2010mh}, which, in turn, are morphisms of graded manifolds
\begin{equation}
\begin{gathered}
 \underline{\hat a}\ =\ (\hat\phi,\hat a)\,:\,T[1](X\times [0,1])\ \rightarrow\ \sLie(\CCG)~,\\
 \underline{\hat a}|_{T[1](X\times\{0\})}\ =\ \underline{a}\ =\ (\phi,a)~~~\mbox{and}~~~\underline{\hat a}|_{T[1](X\times\{1\})}\ =\ \underline{\tilde a}\ =\ (\tilde\phi,\tilde a)~,
\end{gathered}
\end{equation}
cf.\ diagram \eqref{diag:simplicial_homotopy}. Let us denote by $\hat F$ the failure of $\underline{\hat a}$ to be a morphism of N$Q$-manifolds with respect to the homological vector field given by the de Rham differential on $X\times[0,1]$. Then flat homotopies are those for which $\hat F$ has no components along the $[0,1]$-direction.

Let us again look at the instructive example of a principal $\sG$-bundle for $\sG$ a Lie group. We introduce local coordinates $x^\mu$ on $X$ and $r$ on $[0,1]$. In this case, we have $\hat F=\tfrac12\hat F_{\mu\nu} \dd x^\mu\wedge \dd x^\nu+\hat F_{\mu r}\dd x^\mu\wedge \dd r$ with
\begin{equation}
\hat F_{\mu r}\ =\ \der{x^\mu}\hat A_r(x,r)+[\hat A_\mu(x,r),\hat A_r(x,r)]-\der{r}\hat A_\mu(x,r)~.
\end{equation}
At infinitesimal level, $\hat F_{\mu r}=0$ yields gauge transformations with gauge parameter $A_r(x,0)$:
\begin{equation}
 \delta A\ :=\ \der{r} \left.\hat A(x,r)\right|_{r=0}\ =\ \dd A_r(x,0)+[A,A_r(x,0)]~.
\end{equation}
Integrating the differential equation $\hat F_{\mu r}=0$, one obtains the finite form of gauge transformations with gauge parameter valued in the identity component of $\sG$.

\subsection{Local description with finite gauge transformations}\label{ssec:local_finite_gauge}

The approach sketched in the previous section has a couple of drawbacks when in need of explicit formulas for higher principal bundles. First of all, integrating the differential equation arising from flat homotopies only yields gauge transformations in some identity component of the Lie quasi-groupoid. Secondly, the integration procedure for $L_\infty$-algebroid is in principle available \cite{Henriques:2006aa,Fiorenza:2010mh}, but difficult at best. Thirdly, even if the integration is successfully performed, the result of integrating the $L_\infty$-algebra of some higher Lie group $\CCG$ will not be $\CCG$, but a Lie quasi-groupoid equivalent to $\CCG$ in some suitable sense. For example, a strict Lie 2-algebra can be rather easily integrated as a crossed module of Lie algebras to a crossed module of Lie groups. When integrating it via the methods of \cite{Henriques:2006aa,Fiorenza:2010mh}, however, one obtains a Morita equivalent strict Lie 2-group \cite{Sheng:1109.4002}.

Fortunately, these issues can be circumvented by extending the results of Section \ref{ssec:1jets}: the finite gauge transformations can be gleaned from the action of the Chevalley--Eilenberg differential on the moduli parametrising the descent data, replacing the homological vector field $Q$ with the de Rham differential. This approach was first followed in \cite{Jurco:2014mva}. 

We now extend and detail this approach, making it more rigorous and combining in some sense the discussion in the previous section and the differentiation method of Section \ref{ssec:1jets}. Firstly, we have to replace the manifold $X\times \FR^{0|1}$ appearing in the surjective submersion underlying the differentiation in Section \ref{ssec:NQ_manifolds} by an appropriate N-manifold. Indeed, it turns out that the truncation of the $\RZ$-graded manifold $T[1]X\times \FR^{1}[-1]$ to the N-manifold $(T[1]X\times \FR^{1}[-1])_+$, whose structure sheaf is the subsheaf of structure sheaf of $T[1]X\times \FR^{1}[-1]$ generated by elements of non-negative grading, is the correct N-manifold for our purposes. On this N-manifold, there is a natural homological vector field $\breve Q$ given by the difference of the de Rham differential $\dd_X$ on $X$ and the infinitesimal action $Q$ of $\inthom(\FR^{0|1},\FR^{0|1})$, cf.~equation \eqref{eq:differential_on_moduli}. We can now define the local kinematical data of gauge theory as follows.

\begin{definition}
For a Lie quasi-groupoid $\CCG$, a \uline{connection} on a trivial principal $\CCG$-bundle over a manifold $X$ is a set of $\CCG$-valued descent data on $(T[1]X\times \FR^{1}[-1])_+\rightarrow T[1]X$, given by morphisms of $\NN$-graded simplicial manifolds. The \uline{local connection forms} are the moduli of this descent data. The \uline{curvature forms} of the local connection forms are the images of the connection forms under the action of the homological vector field $\breve Q$. \uline{Gauge transformations} between connections are given by simplicial homotopies between the simplicial maps encoding the connections.
\end{definition}

\noindent 
The morphisms of $\NN$-graded simplicial manifolds encoding the connection have an expansion in terms of the Gra{\ss}mann coordinates $\theta_i$ on the nerve of the \v Cech groupoid of $(T[1]X\times \FR^{1}[-1])_+\rightarrow T[1]X$ which is formally identical to that of Theorem \ref{thm:JetPara}. However, since we are now dealing with morphisms of $\NN$-graded manifolds, the coefficients $\alpha_i$ of the monomials $\theta_0,\ldots, \theta_{p-1}$ are now necessarily of degree $p$ and thus $p$-forms. The homological vector field $\breve Q$ is clearly of degree 1, since it increases the form degree by one and decreases the (negative) Gra\ss mann degree by one. Therefore, its induced action produces curvature forms for the $\alpha_p$ of degree $p+1$. Finally, it is clear that a reasonable notion of gauge equivalence of the connection has to arise from simplicial homotopies between the underlying simplicial maps. We will now discuss the example of a gauged sigma model to illustrate our definition.

Consider a manifold $Y$ together with a Lie group $\sG$ acting on $Y$. This gives rise to the action Lie groupoid $Y//\sG:=\sG\times Y\rightrightarrows Y$, where $\sfs(g,y)=y$ and $\sft(g,y)=gy$. Its Lie algebroid is the trivial fibration
\begin{equation}
 \CA\ :=\ \sLie(Y//\sG)\ =\ Y\times \sLie(\sG) \rightarrow Y~.
\end{equation}
The anchor map $\CA\rightarrow TY$ is given by $\sft_*|_\CA$, which we usually denote by $\mu_1$ in our discussion above. It encodes the infinitesimal action of $\sLie(\sG)$ on $Y$.

According to our definition, a connection is now given by descent data for a trivial $Y//\sG$-bundle over $T[1] X$ subordinate to the cover $(T[1]X\times \FR^{1}[-1])_+\rightarrow T[1]X$. Note that 
\begin{equation}
 (T[1]X\times \FR^{1}[-1])_+ \times_{T[1]X} (T[1]X\times \FR^{1}[-1])_+\ \cong\ (T[1]X\times \FR^{2}[-1])_+~,
\end{equation}
etc.\ We therefore have a simplicial map $g$, which is fully determined by the following component maps
\begin{subequations}
\begin{equation}
 \begin{gathered}
  g^0\,:\,(T[1]X\times \FR^{1}[-1])_+\ \rightarrow\ Y~,\\
  g^1\,:\,(T[1]X\times \FR^{2}[-1])_+\ \rightarrow\ Y\times\sLie(\sG)
 \end{gathered}
\end{equation}
with
\begin{equation}
 \begin{gathered}
  g^0(\theta_0)\ =\ \mathring{g}+\sft(A,\id_{\mathring{g}})\theta_0~,\\
  g^1(\theta_0,\theta_1)\ =\ \id_{\mathring{g}}+A(\theta_0-\theta_1)+\tfrac12 \mu_2(A,A)\theta_0\theta_1~,
 \end{gathered}
\end{equation}
\end{subequations}
where $\mathring{g}\in \shom_\CatDiff(X,Y)$, $A\in H^0(X,\Omega^1_X\otimes \sLie(\sG))$ and $\sft(A,\id_{\mathring{g}})$ is the linearisation of $\sft$ at $\id_{\mathring{g}}$. Applying now $\breve Q$ to $\mathring{g}$ and $A$, we obtain
\begin{equation}
\begin{gathered}
 F_{\mathring{g}}\ =\ \breve Q \mathring g\ =\ \dd \mathring g-\sft(A,\id_{\mathring{g}})\ =:\ \nabla \mathring{g}~,\\
 F_{A}\ =\ \breve Q A\ =\ \dd A-Q A\ =\ \dd A + \tfrac12 \mu_2(A,A)~.
\end{gathered}
\end{equation}
We thus obtain the expected fields and their curvatures. It only remains to work out the gauge transformations between connections $(\mathring{g},A)$ and $(\mathring{\tilde g},\tilde A)$, which are encoded in simplicial homotopies
\begin{equation}
 h\, :\, N((T[1]X\times \FR^{1}[-1])_+\times \Delta^1)\ \rightarrow\ N(Y//\sG)
\end{equation}
which have expansions
\begin{equation}
\begin{aligned}
 Y\times \sG\ \ni\ h^0_0(\theta_0)\ &=\ c+ \sff^2_{2\,*}|_{\sfd^1_1(c)}(\chi^1_0)\theta_0~,\\
 Y\times \sG\times \sG\ \ni\ h^1_0(\theta_0,\theta_1)\ &=\ \sfd^1_1(c)+\chi^1_0(\theta_0-\theta_1)+ (\sfd^1_1\circ\sff^2_2)_*|_{\sfd^1_1(c)}(\chi^1_0)\theta_1\,+\\
 &\kern1cm+\tfrac12\mu_2(\chi^1_0,\chi^1_0)\theta_0\theta_1~,\\
Y\times \sG\times \sG\ \ni\  h^1_1(\theta_0,\theta_1)\ &=\ \sfd^1_0(c)+\chi^1_1(\theta_0-\theta_1)+ (\sfd^1_0\circ\sff^2_2)_*|_{\sfd^1_1(c)}(\chi^1_0)\theta_1\,+\\
 &\kern1cm+\tfrac12\mu_2(\chi^1_0,\chi^1_1)\theta_0\theta_1~,
\end{aligned}
\end{equation}
cf.\ \eqref{eq:hpiExpansion}. Note that the horn fillers $\chi^1_{0,1}$ are uniquely determined and $c$ becomes a function of $Y$. Theorem \ref{thm:GaugeTrafo} then induces the following gauge transformations:
\begin{equation}
\begin{aligned}
\mathring{\tilde g}\ =\ c\mathring{g}\eand
 \tilde A\ =\ c^{-1}A c+c^{-1}\dd c~.
\end{aligned}
\end{equation}
This local treatment of gauged sigma models readily generalises to higher gauged sigma models.

\subsection{Global description}

The global description of the kinematical data of higher gauged sigma models is now obtained by gluing together the underlying local data using finite gauge transformations. This gluing procedure can be performed, even in the case of higher gauge theory, see \cite{Jurco:2014mva} for the case of semistrict principal 2-bundles. Whilst concrete cases are readily dealt with, the abstract discussion is rather cumbersome. We therefore only sketch the procedure in the following.

Let $\CCG$ be a Lie $n$-quasi-groupoid and let $Y\rightarrow X$ be a cover of a manifold $X$ with fibred products $Y^{[p]}:=Y\times_X Y\times_X\cdots\times_X Y$. Then, $N_p(\check \CC(Y\rightarrow X))=Y^{[p+1]}$. Consider a principal $\CCG$-bundle over $X$ as in Definition \ref{def:EquivLQuasGrpBun}. The \v Cech data of such a bundle consists of maps
\begin{equation}
  \phi\,:\,Y^{[1]}\ \to\ \CCG_0\eand g^p\,:\,Y^{[p+1]}\ \to\ \CCG_p\efor p\ \geq\ 1~.
\end{equation}
We supplement this data with a local connective structure consisting of
\begin{equation}
 \phi\,:\,Y^{[1]}\ \to\ \CCG_0\eand A_p\ \in\ H^0(Y^{[1]},\Omega^p_{Y^{[1]}}\otimes \phi^*\sLie_p(\CCG))\efor p\ \geq\ 1~.
\end{equation}
Note that $\phi$ appears in both the \v Cech data and in the local connective structure. This data then has to be supplemented by missing forms $\Lambda_{p,q}\in H^0(Y^{[q]}, \Omega^p_{Y^{[q]}}\otimes\phi^*\sLie_{p+q+1}(\CCG))$ to form the connective structure on a principal $\CCG$-bundle. The local components of the connective structure, that is, those components living on $Y^{[1]}$, are then glued together on double overlaps $Y^{[2]}$ via gauge transformations parametrised by data living on $Y^{[2]}$. The latter, in turn, are glued together by gauge transformations on triple overlaps $Y^{[3]}$ by gauge transformations parametrised by data on $Y^{[3]}$, etc. All ambiguities in translating local gauge transformations to gauge transformations over the $Y^{[p]}$ are completely fixed by considering transformations that should yield the identity. For triple overlaps $Y^{[3]}$, e.g., on can consider transformations from $\sff^2_0 Y^{[3]}$ to $\sff^2_1 Y^{[3]}$ to $\sff^2_2 Y^{[3]}$ back to $\sff^2_0 Y^{[3]}$, which has to amount to the identity.

\subsection{Higher base spaces}

Let us now describe how our constructions generalise to the case of higher base spaces. As models of the latter, we shall again use the inner Kan simplicial manifolds discussed in Sections \ref{sec:quasiCategory} and \ref{ssec:higher_base_1}. 

We start from a surjective submersion $\CCY\rightarrow \CCX$ of inner Kan simplicial manifolds as well as a vertically Kan bisimplicial manifold $\CCG$, which will provide the gauge structure. In a first step, we construct a principal $\CCG$-bundle over $\CCX$ subordinate to $\CCY\rightarrow \CCX$ as described in Section \ref{ssec:higher_base_1}. The next step is to derive the local higher connection forms, curvatures and their finite gauge transformations on  $\CCY\rightarrow \CCX$ . For this, we consider descend data for a principal $\CCG$-bundle over the simplicial N$Q$-manifold $T[1]\CCX$ subordinate to $T[1]\CCX\times \FR[-1]$. This map is parameterised by a simplicial set of differential forms, which constitute the higher connection. The corresponding curvatures are obtained from the induced action by the homological vector field $\breve Q$, which is the sum of the (simplicial) de Rham differential on $\CCX$ and the infinitesimal action $Q$ of $\inthom(\FR^{0|1},\FR^{0|1})$. The finite gauge transformations are then obtained from equivalence relations on such principal $\CCG$-bundles. In a last step, we use the finite gauge transformations to glue together the differential forms of the local connection to global objects, in a way which is consistent with the principal $\CCG$-bundle structure.

An explicit example of such a principal $\CCG$-bundle over a 2-space was constructed and discussed in \cite{Ritter:2015zur} and principal $\CCG$-bundles over certain generalised spaces were also discussed more recently in \cite{Kotov:2016lpx}.

Another example, we could think of could be a gauged sigma model with its source space being an orbifold $X//H$.  In this case the relevant higher base space would be the simplicial N$Q$-manifold $T[1]\CCX$ subordinate to $T[1]\CCX\times \FR[-1]$, with $\CCX$ being the nerve of action groupoid corresponding to $X//H$.

\section{Application: A simplicial Penrose--Ward transform}\label{sec:PenroseWard}

We shall now apply the above and discuss how principal $\CCG$-bundles for general Lie $k$-quasi-groupoids $\CCG$ can be combined with the language of twistors to formulate a rather general class of non-Abelian superconformal theories in six dimensions. In particular, we wish to generalise our previous results \cite{Saemann:2012uq,Saemann:2013pca,Jurco:2014mva}. Inspired by the Abelian setting of \cite{Saemann:2011nb,Mason:2011nw}, these works have established Penrose--Ward transforms between holomorphic principal $\CCG$-bundles over a twistor space for $\CCG$ a strict Lie 2-group, a semistrict Lie 2-group, or a Lie 3-group and certain non-Abelian supersymmetric self-dual tensor field theories in six dimensions. The discussion below will remove various restrictions on the gauge structure, and, consequently, provide the perhaps most general and concise framework for formulating self-dual non-Abelian tensor field theories within twistorial higher gauge theory.

For simplicity, we will only work in the complex setting, however, reality conditions can be imposed at any stage, cf.\ \cite{Saemann:2011nb}. Consequently, we will make use of the standard notation $\CO_X$ for the sheaf of holomorphic functions on a complex (super)manifold $X$ and $\Omega^p_X$ for the sheaf of holomorphic differential $p$-forms on $X$.

\subsection{Twistor space of chiral superspace}

The relevant twistor space for self-dual 3-forms in six dimensions was introduced in \cite{springerlink:10.1007/BF00132253,Hughston:1986hb,Hughston:1979TN,Hughston:1987aa,Hughston:1988nz,0198535651} and generalised supersymmetrically in \cite{Saemann:2012uq}. Let us consider the chiral complex superspace $M^{6|8n}:=\FC^{6|8n}:=(\FC^6,\CO_{\FC^6}\otimes \Lambda^\bullet\FC^{8n})$. Using the isomorphism $TM\cong S\wedge S$ between the tangent bundle and the anti-symmetric tensor product of the bundle of anti-chiral spinors $S$ on a spin manifold $M$,\footnote{Note that this decomposition is equivalent to choosing a conformal structure on $M$, see \cite[Remark 3.2]{Saemann:2011nb}.}, we can coordinatise $M^{6|8n}$ by $(x^{AB},\eta^A_I)$ with $x^{AB}=-x^{BA}$. Here, the $x^{AB}$s are the Gra{\ss}mann-even coordinates while the $\eta^A_I$s the Gra{\ss}mann-odd coordinates and $A,B,\ldots=1,\ldots,4$ are  (anti-)chiral spinor indices and $I,J,\ldots=1,\ldots,2n$ are the $R$-symmetry indices.  Anti-symmetric pairs of spinor indices can be raised and lowered by means of the Levi-Civita symbol: $x_{AB}=\frac12\varepsilon_{ABCD} x^{CD}$ $\Leftrightarrow$ $x^{AB}=\frac12\varepsilon^{ABCD} x_{CD}$. 

In this notation, a differential 1-form $A$ on $M^{6|0}$ has spinor components $A_{AB}=-A_{BA}$ and a differential 2-form $B$ is given by a traceless matrix $B_A{}^B$. A differential 3-form $H$ is given by a pair $(H_{AB},H^{AB})$, where $H_{AB}=H_{BA}$ contains the self-dual part of $H$ while $H^{AB}=H^{BA}$ contains the anti-self-dual part.

After decomposing $\FC^{8n}\cong\FC^4\otimes\FC^{2n}$ and choosing a holomorphic symplectic form $\Omega=(\Omega^{IJ})$ on $\FC^{2n}$, we  introduce the superspace derivatives
\begin{subequations}
\begin{equation}
  \dpar_{AB}\ :=\ \der{x^{AB}}\eand D^I_A\ :=\ \dpar^I_A-2\Omega^{IJ}\eta_J^B\dpar_{AB}
\end{equation}
with $\partial^I_A:=\partial/\partial \eta^A_I$. These derivatives generate the chiral $\CN=(2,0)$ supersymmetry algebra in six dimensions,
\begin{equation}
 [\dpar_{AB},D^I_C]\ =\ 0\eand \{D^I_A,D^J_B\}\ =\ -4\Omega^{IJ}\dpar_{AB}~.
\end{equation}
\end{subequations}

\begin{definition}
Let $\PP^3$ be the three-dimensional complex projective space with homogeneous coordinates $\lambda_A$.\footnote{In fact, here it is understood as a copy of the projectivised fibres of the dual of the anti-chiral spin bundle, hence, the chiral (i.e.~lower-script) spinor index.} Set $F^{9|8n}:=\FC^{6|8n}\times\PP^3$ which is called the \uline{correspondence space}. The \uline{twistor space}, denoted by $P^{6|2n}$ and coordinatised by $(z^A,\eta_I,\lambda_A)$, is the supermanifold that is defined by the double fibration
\begin{subequations}\label{eq:DoubleFib}
\begin{equation}
 \begin{picture}(50,40)
  \put(0.0,0.0){\makebox(0,0)[c]{$P^{6|2n}$}}
  \put(64.0,0.0){\makebox(0,0)[c]{$M^{6|8n}$}}
  \put(34.0,33.0){\makebox(0,0)[c]{$F^{9|8n}$}}
  \put(7.0,18.0){\makebox(0,0)[c]{$\pi_1$}}
  \put(55.0,18.0){\makebox(0,0)[c]{$\pi_2$}}
  \put(25.0,25.0){\vector(-1,-1){18}}
  \put(37.0,25.0){\vector(1,-1){18}}
 \end{picture}
\end{equation}
where $\pi_2$ is the trivial projection and $\pi_1$ is given by
\begin{equation}
\pi_1\,:\,(x^{AB},\eta^A_I,\lambda_A)\ \mapsto\ (z^A,\eta_I,\lambda_A)\ :=\ ((x^{AB}+\Omega^{IJ}\eta^A_I\eta^B_J)\lambda_B,\eta_I^A\lambda_A,\lambda_A)~.
\end{equation}
The equations 
\begin{equation}
 z^A\ =\ (x^{AB}+\Omega^{IJ}\eta^A_I\eta^B_J)\lambda_B\eand
 \eta_I\ =\ \eta_I^A\lambda_A
\end{equation}
\end{subequations}
are known as the \uline{incidence relations}.
\end{definition}

\noindent
By virtue of the incidence relations, we have 
\begin{equation}
 z^A\lambda_A-\Omega^{IJ}\eta_I\eta_J\ =\ 0~,
\end{equation}
which implies that $P^{6|2n}$ can be viewed as a quadric hypersurface embedded into the total space of the holomorphic fibration $\FC^{4|2n}\otimes \CO_{\PP^3}(1)\rightarrow \PP^3$ with fibre coordinates $z^A$ and $\eta_I$ and base coordinates $\lambda_A$. Here, $\CO_{\PP^3}(1)$ is the dual of the tautological bundle over $\PP^3$. As is standard, the double fibration \eqref{eq:DoubleFib} provides a geometric correspondence. In the case at hand, a point in $M^{6|8n}$ corresponds to an embedding of a three-dimensional complex projective space into $P^{6|2n}$ while a point in $M^{6|8n}$ corresponds to a $3|6n$-dimensional plane in $M^{6|8n}$ as a supersymmetric generalisation of a three-dimensional, self-dual and totally null plane embedded into $M^{6|0}=\FC^6$. Thus, the twistor space is the space of all such planes.

Note that we may also view the twistor space $P^{6|2n}$ as the leaf space of a foliation of the correspondence space $F^{9|8n}$ that is induced by an integrable rank-$3|6n$ distribution,  known as the \uline{twistor distribution}, and which is generated by the vector fields 
 \begin{equation}\label{eq:TwistDistVectFields}
  V^A\ :=\ \lambda_B\partial^{AB}\eand
  V^{IAB}\ :=\ \tfrac12\varepsilon^{ABCD}\lambda_C D^I_D~.
\end{equation}

\subsection{Penrose--Ward transform and self-dual fields}

The rough idea of relating data on twistor space to data on chiral superspace is as follows. For $\CCG$ a Lie $k$-quasi-groupoid, we shall consider a holomorphic principal $\CCG$-bundle over $P^{6|2n}$ subject to certain triviality conditions.  We then pull back this bundle to the correspondence space $F^{9|8n}$ along $\pi_1:F^{9|8n}\to P^{6|8n}$. Because of our choice of triviality, the pulled-back bundle will be holomorphically trivial on $F^{9|8n}$. We then perform a gauge transformation yielding a non-trivial higher connective structure relative to the fibration $\pi_1$. This connective structure allows for a push-down along $\pi_2:F^{9|8n}\to M^{6|8n}$ leading to a higher connective structure on $M^{6|8n}$ subject to a set of non-linear coupled partial differential equations constituting the superspace constraint equations for a non-Abelian self-dual tensor field theory. 

Let us make this now precise. Our starting point is the following class of principal quasi-groupoid bundles.
\begin{definition}
Let $\CCG$ be a Lie $k$-quasi-groupoid. A topologically trivial holomorphic principal $\CCG$-bundle over the twistor space that is holomorphically trivial on $\PP^3\cong \pi_1(\pi_2^{-1}(x))\hookrightarrow P^{6|2n}$ for all $x\in M^{6|8n}$ is called \uline{$M^{6|8n}$-trivial}.
\end{definition}

Such bundles will then be pulled back along the fibration $\pi_1$ and a trivialising gauge transformation will yield a connection consisting of differential forms relative to $\pi_1$. We now recall the relevant definitions.
\begin{definition}
The sheaf of \uline{holomorphic relative differential $p$-forms} on the correspondence space $F^{9|8n}$, denoted by $\Omega^p_{\pi_1}$,  is defined by the short exact sequence
\begin{subequations}
\begin{equation}
 0\ \longrightarrow\ \pi_1^*\Omega^1_{P^{6|2n}}\wedge\Omega^{p-1}_{F^{9|8n}}\ \longrightarrow\ \Omega^p_{F^{9|8n}}\ \longrightarrow\ \Omega^p_{\pi_1}\ \longrightarrow\ 0~.
\end{equation}
 Furthermore, let ${\rm pr}_{\pi_1}\!:\Omega_{F^{9|8n}}^p\to \Omega^p_{\pi_1}$ be the quotient mapping and $\dd$ be the holomorphic exterior derivative on  $F^{9|8n}$. The \uline{holomorphic relative exterior derivative}, denoted by $\dd_{\pi_1}$, is defined by 
\begin{equation}
 \dd_{\pi_1}\ :=\ {\rm pr}_{\pi_1}\circ\dd\,:\, \Omega^p_{\pi_1}\ \to\ \Omega^{p+1}_{\pi_1}~.
\end{equation}
\end{subequations}
\end{definition}

The holomorphic relative exterior derivative induces a complex known as the\linebreak \uline{holomorphic relative de Rham complex}. Specifically, it is the complex that is given in terms of an injective resolution of the topological inverse $\pi_1^{-1}\CO_{P^{6|2n}}$ of the sheaf $\CO_{P^{6|2n}}$ on the correspondence space $F^{9|8n}$,
\begin{equation}
 0\ \longrightarrow\ \pi_1^{-1}\CO_{P^{6|2n}}\ \longrightarrow\ \CO_{F^{9|8n}}\ \xrightarrow{\dd_{\pi_1}}\  \Omega^1_{\pi_1}\ \xrightarrow{\dd_{\pi_1}}\  \Omega^2_{\pi_1}\ \xrightarrow{\dd_{\pi_1}}\  \cdots ~.
\end{equation}
Here, the sheaf $\pi_1^{-1}\CO_{P^{6|2n}}$ consists of those holomorphic functions that are locally constant along the fibres of the fibration $\pi_1:F^{9|8n}\to P^{6|2n}$. 

More explicitly, we denote by $e_A$ and $e_{IAB}=\tfrac{1}{2}\eps_{ABCD}e_I^{CD}$ the generators of the dual of the twistor distribution. Since $\lambda_A V^A=\lambda_A V^{IAB}=0$ for the vector fields \eqref{eq:TwistDistVectFields}, these forms are defined modulo terms proportional to $\lambda_A$. The holomorphic relative exterior derivative then reads as
\begin{equation}\label{eq:RelDer}
  \dd_{\pi_1}\ =\ e_A V^A+e_{IAB} V^{IAB}\ =\ e_{[A}\lambda_{B]}\partial^{AB}+e^{AB}_I\lambda_A D^I_B~.
\end{equation}
The following lemma characterises relevant relative differential forms.

\begin{lemma}\label{lem:RelFormsExp}
Let  $\alpha_{p}\in H^0(F^{9|8n},\Omega^p_{\pi_1})$. In terms of $\lambda_A$, we have the following expansions for $p=1,2,3$:
\begin{equation}
\begin{aligned}
 \alpha_{1}\ &=\ e_{[A}\lambda_{B]}\,  A^{AB}+e^{AB}_I\lambda_A\, A^I_B~,\\
 \alpha_{2}\ &=\ -\tfrac14\varepsilon^{ABCD} e_A\wedge e_B\lambda_C\,  B_D{}^{E}\lambda_{E}+\tfrac12 \varepsilon^{ABCD}e_A\lambda_B\wedge e^{EF}_I\lambda_E\,B_{CD}{}^I_F~+\\
   &\kern1cm+\tfrac12 e^{CA}_I\lambda_C\wedge e^{DB}_J\lambda_D\, B^{IJ}_{AB}~,\\
\alpha_{3}\ &=\  -\tfrac13\varepsilon^{ABCD} e_A\wedge e_B\wedge e_C\lambda_D\,H^{EF}\lambda_E\lambda_F~+\\
 &\kern1cm -\tfrac14 \varepsilon^{ABCD} e_A\wedge e_B\lambda_C\, \wedge e^{EF}_I\lambda_E\, (H_{D}{}^G{}^I_F)_0\lambda_G~+\\
 &\kern2cm + \tfrac14 \varepsilon^{ABCD} e_A\lambda_B\wedge e^{EF}_I\lambda_E\wedge e^{GH}_J\lambda_G\,(H_{CD}{}^{IJ}_{FH})_0~+\\
 &\kern3cm +\tfrac16 e^{DA}_I\lambda_D\wedge e^{EB}_J\lambda_E\wedge e^{FC}_K\lambda_F\, H^{IJK}_{ABC}~.
 \end{aligned}
\end{equation}
The coefficients $A$, $B$, and $H$ depend only on $(x^{AB},\eta^A_I)\in M^{6|8n}$. Furthermore, $(H_{A}{}^B{}^I_C)_0$ is the totally trace-less part of $H_{A}{}^B{}^I_C$ while $(H_{AB}{}^{IJ}_{CD})_0$ denotes the part of $H_{AB}{}^{IJ}_{CD}$ which vanishes when contracted with $\varepsilon^{ABCD}$. 
\end{lemma}

\vspace{10pt}
\noindent
{\it Proof:}
This follows directly from the direct images of $\Omega^p_{\pi_1}$ under the projection $\pi_2:F^{9|8n}\to M^{6|8n}$. A detailed derivation can be found in \cite[Proposition 4.2]{Saemann:2011nb}. \hfill $\Box$

\

We can now state the main theorem of this section. 
\newpage
\begin{thm}\label{thm:PW_trafo}
Let $\CCG$ be a Lie $k$-quasi-groupoid and consider the double fibration \eqref{eq:DoubleFib}. There are bijections between 
  \begin{enumerate}[(i)]\setlength{\itemsep}{-1mm}
\item equivalence classes of $M^{6|8n}$-trivial principal $\CCG$-bundles over $P^{6|2n}$, 
\item equivalence classes of holomorphically trivial holomorphic principal $\CCG$-bundles over $F^{9|8n}$ that are equipped with a flat holomorphic relative connective structure, and
\item gauge equivalence classes of complex holomorphic solutions to constraint equations on $M^{6|8n}$ containing a 3-form curvature $H$ that is self-dual, that is, $H^{AB}=0$.
\end{enumerate}
\end{thm}
\noindent Particularly interesting is the following special case:
\begin{cor}
Consider the situation of Theorem \ref{thm:PW_trafo}. In the case $k=2$, we have the following fields on $M^{6|8n}$ 
\begin{subequations}\label{eq:ConstrainEq}
\begin{equation}
\begin{gathered}
\phi\,:\,M^{6|8n}\ \to\ \CCG_0~,\\
A\ \in\ H^0(M^{6|8n},\Omega^1_{M^{6|8n}}\otimes\phi^*\sLie_1(\CCG))~,\\
B\ \in\ H^0(M^{6|8n},\Omega^2_{M^{6|8n}}\otimes\phi^*\sLie_2(\CCG))~,\\
\phi^{IJ}\ =\ -\phi^{JI}\ \in\ H^0(M^{6|8n},\phi^*\sLie_2(\CCG))\ewith \Omega_{IJ}\phi^{IJ}\ =\ 0~,\\
\psi^I\ \in\ H^0(M^{6|8n},S^\vee\otimes \phi^*\sLie_2(\CCG))~,
\end{gathered}
\end{equation}
which satisfy the constraint equations
\begin{equation}
\begin{aligned}
\nabla_{AB}\phi\ &=\ \partial_{AB}\phi+\mu_1(A_{AB})\ =\ 0~,\\
\nabla_A^I\phi\ &=\ D^I_A\phi+\mu_1(A^I_A)\ =\ 0~,
\end{aligned}
\end{equation}
\begin{equation}
  \CF_A{}^B\ =\ 0~,\quad \CF_{AB}{}^I_C\ =\ 0~,\eand \CF^{IJ}_{AB}\ =\ 0~,
\end{equation}
and
\begin{equation}
\begin{aligned}
  H^{AB}\ &=\ 0~,\\
  H_{A}{}^B{}^I_C\ &=\ \delta^B_C\psi^I_A-\tfrac14\delta^B_A\psi^I_C~,\\
  H_{AB}{}^{IJ}_{CD}\ &=\ \varepsilon_{ABCD}\phi^{IJ}~,\\
  H^{IJK}_{ABC}\ &=\ 0~,
  \end{aligned}
\end{equation}
where the curvature 2- and 3-forms are defined as
\begin{equation}
\begin{aligned}
  \CF_A{}^B\ &=\ \partial^{BC} A_{CA}-\partial_{CA}A^{BC}+\mu_2(A^{BC},A_{CA})-\mu_1(B_A{}^B)~,\\
  \CF_{AB}{}^I_C\ &=\ ~\partial_{AB}A^I_C-D^I_CA_{AB}+\mu_2(A_{AB},A^I_C)-\mu_1(B_{AB}{}^I_C)~,\\
  \CF^{IJ}_{AB}\ &=\ D^I_AA^J_B+D^J_B A^I_A+\mu_2(A^I_A,A^J_B)+4\Omega^{IJ}A_{AB}-\mu_1(B^{IJ}_{AB})
  \end{aligned}
\end{equation}
and
\begin{equation}
\begin{aligned}
  H_{AB}\ &=\ \nabla_{C(A}B_{B)}{}^C+\mu_3(A_{C(A},A^{CD},A_{B)D})~,\\
  H^{AB}\ &=\ \nabla^{C(A}B_C{}^{B)}+\mu_3(A^{C(A},A_{CD},A^{B)D})~,\\
  H_{A}{}^B{}^I_C\ &=\ \nabla^I_CB_A{}^B-\nabla^{DB}B_{DA}{}^I_C+\nabla_{DA}B^{DB}{}^I_C-\mu_3(A^I_C,A^{BD},A_{DA})~,\\
  H_{AB}{}^{IJ}_{CD}\ &=\ \nabla_{AB}B^{IJ}_{CD}-\nabla^I_C B_{AB}{}^J_D-\nabla^J_D B_{AB}{}^I_C\,\,-\\
  &\kern1cm-2\Omega^{IJ}(\varepsilon_{ABF[C} B_{D]}{}^F-\varepsilon_{CDF[A} B_{B]}{}^F)-\mu_3(A_{AB},A^I_C,A^J_D)~,\\
  H^{IJK}_{ABC}\ &=\ \nabla^I_A B^{JK}_{BC}+\nabla^J_BB^{IK}_{AC}+\nabla^K_C B^{IJ}_{AB}\,+\\
     &\kern1cm+4\Omega^{IJ}B_{AB}{}^K_C+4\Omega^{IK}B_{AC}{}^J_B+4\Omega^{JK}B_{BC}{}^I_A-\mu_3(A^I_A,A^J_B,A^K_C)~.
  \end{aligned}
\end{equation}
The products $\mu_1$, $\mu_2$, and $\mu_3$ depend on the field $\phi$.
\end{subequations}
\end{cor}

\vspace{10pt}
\noindent
{\it Proof of the theorem:} First, note that the standard Stein cover on $\PP^3$ induces good covers $\hat\frU=\{\hat U_a\}$ on the holomorphic supervector bundle $P^{6|2n}\to\PP^3$ as well as $\frU':=\{U'_a:=\pi_1^{-1}(\hat U_a)\}$ on $F^{9|8n}$. 

(i) $\to$ (ii) A holomorphic principal $\CCG$-bundle over $P^{6|2n}$ is characterised by a holomorphic simplicial map $\hat g:N(\check\CC(\hat\frU\to P^{6|2n} ))\to\CCG$. Concretely, we have holomorphic maps $\hat g_a:\hat U_a\to \CCG_0$, $\hat g_{ab}:\hat U_a\cap \hat U_b\to \CCG_1$, $\hat g_{abc}:\hat U_a\cap \hat U_b\cap\hat U_c\to \CCG_2$ and so on subject to the simplicial identities. Upon pulling this bundle back\footnote{see Definition \ref{def:PullBackBundle}} to the correspondence space $F^{9|8n}$, we obtain a holomorphic simplicial map $g'=\pi_1^* \hat g:N(\check\CC(\frU'\to F^{9|8n} ))\to\CCG$ whose component maps are constant along $\pi_1$ and therefore satisfy
\begin{equation}\label{eq:Annihilation}
 \dd_{\pi_1}g'_a\ =\ 0~,\quad
  \dd_{\pi_1}g'_{ab}\ =\ 0~,\quad
   \dd_{\pi_1}g'_{abc}\ =\ 0~,\quad \ldots
\end{equation}

  \begin{figure}[h]
\begin{center}
\tikzset{->-/.style={decoration={markings,mark=at position #1 with {\arrow{>}}},postaction={decorate}}}
\begin{tikzpicture}[scale=0.7,every node/.style={scale=0.7}]
   \filldraw[pattern=north west lines, pattern color=green!20,draw=white] (0,0) -- (0,5) -- (5,8) -- cycle;
   \filldraw[pattern=north west lines, pattern color=blue!20,draw=white] (5,3) -- (10,5) -- (5,8) -- cycle;
   \filldraw[pattern=north west lines, pattern color=red!20,draw=white] (0,0) -- (5,3) -- (10,0) -- cycle;

   \draw (-0.2,-0.2) node {$0$};
   \draw (4.7,3.2) node {$0$};
   \draw (10.3,-0.2) node {$0$};
   \draw (-0.9,5.2) node {${\sff}^1_{1\,*}|_{\phi}(\alpha''_c)$};
   \draw (5,8.4) node {${\sff}^1_{1\,*}|_{\phi}(\alpha''_b)$};
   \draw (10.9,5.2) node {${\sff}^1_{1\,*}|_{\phi}(\alpha''_a)$};
   \draw [dashed,-implies,double equal sign distance,thick] (5.94,6.5) -- (6.5,5.6);
   \draw [dashed,-implies,double equal sign distance,thick] (8.7,2.08) -- (8.14,2.98);
   \draw (7.1,6) node {$\gamma''{}^1_{0,ab}$};
   \draw (7.6,2.6) node {$\gamma''{}^1_{1,ab}$};
   \draw [dashed,-implies,double equal sign distance,thick] (1,4.6) -- (2,4.2);
   \draw [dashed,-implies,double equal sign distance,thick] (4.2,3.32) -- (3.2,3.72);
   \draw (1,4.1) node {$\gamma''{}^1_{0,bc}$};
   \draw (4,4) node {$\gamma''{}^1_{1,bc}$};
   \draw [-implies,double equal sign distance,thick] (2,4) -- (3,3.5);
   \draw [-implies,double equal sign distance,thick] (7.5,1.25) -- (6.5,1.75);
   \draw (2.2,3.3) node {$\gamma''{}^1_{0,ac}$};
   \draw (6.5,1.3) node {$\gamma''{}^1_{1,ac}$};
   \draw [-implies,double equal sign distance,thick] (4.7,6.8) -- (4.7,5.8);
   \draw (5.2,6.3) node {$\tau$};
   \draw [dashed,-implies,double equal sign distance,thick] (5,1.8) -- (5,0.8);
   \draw (4.5,1.3) node {$0$};
   \draw (-0.6,2.5) node {$\alpha''_c$};
   \draw (5.5,5.5) node {$\alpha''_b$};
   \draw (10.6,2.5) node {$\alpha''_a$};
   %\draw (5.5,2.3) node {$h_{ac,01}$};
   %\draw (7.3,4.3) node {$h_{ab,01}$};
   %\draw (3.3,4.4) node {$h_{bc,01}$};
   %
   \draw (1.4,6.8) node {$-\dd_{\pi_1}\sigma_1(\phi)$};
   \draw (8.4,6.8) node { $-\dd_{\pi_1}\sigma_1(\phi)$};
   \draw (4.9,4.7) node {$-\dd_{\pi_1}\sigma_1(\phi)$};
   \draw (2.2,1.8) node {$0$};
   \draw (7.8,1.8) node {$0$};
   \draw (5,-.3) node {$0$};
   %prism
   \draw[->-=.5] (0,0) -- (0,5); 
   \draw[->-=.5] (0,5) -- (5,8); 
   \draw[->-=.5]  (5,8) -- (10,5);
   \draw[->-=.5] (0,5) -- (10,5);
   \draw[->-=.5] (0,0) -- (10,5);
   \draw[dashed,->-=.5] (0,0) -- (5,8);   
   \draw[->-=.5] (0,0) -- (10,0);   
   \draw[dashed,->-=.5] (0,0) -- (5,3);   
   \draw[dashed,->-=.5] (5,3) -- (10,0);   
   \draw[->-=.5] (10,0) -- (10,5);   
   \draw[dashed,->-=.5] (5,3) -- (10,5);   
   \draw[dashed,->-=.5] (5,3) -- (5,8);
   \filldraw [black] (0,0) circle (2pt);
   \filldraw [black] (0,5) circle (2pt);
   \filldraw [black] (5,8) circle (2pt);
   \filldraw [black] (10,5) circle (2pt);
   \filldraw [black] (10,0) circle (2pt);
   \filldraw [black] (5,3) circle (2pt);
\end{tikzpicture}
\begin{minipage}{14cm}
\caption{\small \label{fig:TripleGauge} Compatibility conditions for gauge transformation of the one-form gauge potential on triple overlaps. The horn filler $\tau$ of the top simplex exists trivially.}
\end{minipage}
\end{center}
\end{figure}
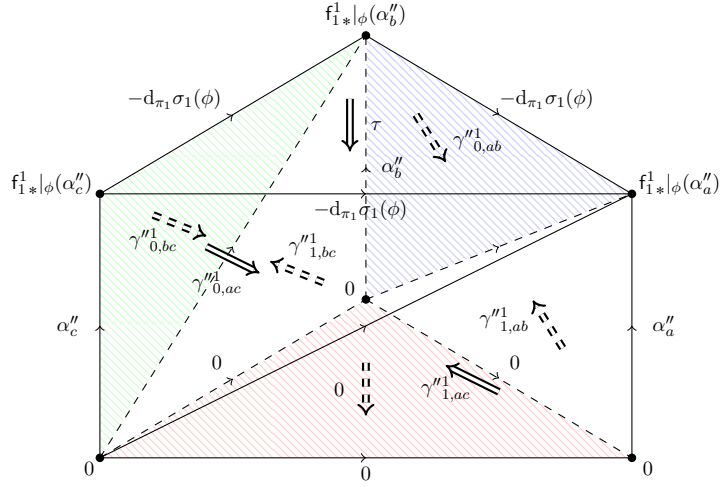

Since, by assumption, the bundle over $P^{6|2n}$ is $M^{6|8n}$-trivial, the pullback is trivial on all of $F^{9|8n}$. Thus, there exists a coboundary transformation trivialising $g'$; see Definition \ref{def:TrivialBundle}. That is, there is a homotopy\footnote{see Definition \ref{def:DefSimpHomo} and Lemma \ref{lem:HomMap} together with the comment above Lemma \ref{lem:HomoExpan}} $h':N(\check\CC(\frU'\to F^{9|8n} ))\to\inthom(\Delta^1,\CCG)$ between $g':N(\check\CC(\frU'\to F^{9|8n} ))\to\CCG$ and another simplical map $g'':N(\check\CC(\frU'\to F^{9|8n} ))\to\CCG$ such that 
\begin{equation}\label{eq:TrivBunGauge}
 g''_{a_0\cdots a_{p}}\ =\ (\sfd^{p-1}_{0}\circ\cdots\circ\sfd^1_0\circ\sfd^0_0)(g''_{a_0})\, :\, U'_{a_0}\cap\cdots\cap U'_{a_{p}}\ \to\ \CCG_p
 \end{equation}
  for $p\geq1$. In particular, this implies that there exists a global holomorphic map $\phi: F^{9|8n}\to \CCG_0$ with $g''_a=\phi|_{U'_a}$.\footnote{Note that since $\PP^3$ is compact, we actually have $\phi:M^{6|8n}\to \CCG_0$. We shall come back to this shortly when deriving the superspace constraint equations on $M^{6|8n}$.} Furthermore, by virtue of \eqref{eq:Annihilation} together with \eqref{eq:TrivBunGauge}, we know that there exist
    \begin{subequations}
  \begin{equation}
  \begin{gathered}
  \alpha''_{p,a}\ \in\  H^0(U'_a,\Omega^p_{\pi_1}\otimes \phi^*\sLie_p(\CCG))~,\\
  \gamma''{}^p_{i,a_0\cdots a_p}\ \in\ H^0(U'_{a_0}\cap\cdots\cap U'_{a_{p}}, \Omega^p_{\pi_1}\otimes \phi^*\sLie_{p+1}(\CCG))
  \end{gathered}
  \end{equation}
  for $p\geq1$ and $0\leq i\leq p$. 
  
These maps satisfy various cocycle conditions as implied by the relations displayed in Lemma \ref{lem:HomMap}; see also Theorem \ref{thm:GaugeTrafo}. For instance, the cocycle conditions for $\alpha''_{1,a}$ read as 
\begin{equation}\label{eq:cocycle_cond_PW}
\begin{gathered}
 \sff^2_{2\,*}|_{\sigma_2(\phi)}(\gamma''{}^1_{0,ab})\ =\ -\dd_{\pi_1}\sigma_1(\phi)\eand \sff^2_{0\,*}|_{\sigma_2(\phi)}(\gamma''{}^1_{0,ab})\ =\ \alpha''_{1,b}~,\\
  \sff^2_{1\,*}|_{\sigma_2(\phi)}(\gamma''{}^1_{0,ab})\ =\ \sff^2_{1\,*}|_{\sigma_2(\phi)}(\gamma''{}^1_{1,ab})~,\\
   \sff^2_{2\,*}|_{\sigma_2(\phi)}(\gamma''{}^1_{1,ab})\ =\ \alpha''_{1,a}\eand\sff^2_{0\,*}|_{\sigma_2(\phi)}(\gamma''{}^1_{1,ab})\ =\ 0
 \end{gathered}
\end{equation}
\end{subequations} 
with $\sigma_p:=\sfd_0^{p-1}\circ\cdots\circ\sfd^1_0\circ\sfd^0_0$ as defined in Theorem \ref{thm:JetPara}. Using the simplicial identities \eqref{eq:axioms_simplicial_set}, it is rather easy to see that these equations imply that ${\sff}^1_{1\,*}|_{\phi}(\alpha''_a)={\sff}^1_{1\,*}|_{\phi}(\alpha''_b)$. In addition, in Figure \ref{fig:TripleGauge}  we have displayed the compatibility conditions on non-empty triple overlaps $U'_a\cap U'_b\cap U'_c$. Clearly, the  top and bottom simplicial 2-simplices trivially exist as their faces coincide. 

Specialising to the case of a Lie 2-quasi-groupoid with $k=2$ for a moment, only $\gamma''{}^1_{i,ab}$ is non-trivial. All higher $\gamma''{}^p_{i,a_0\cdots a_p}$ for $p>1$ are expressed in terms of the $\gamma''{}^1_{i,ab}$ as for $k=2$ all horns of $\CCG$ can be filled uniquely for $p>2$, see again the discussion in Theorem \ref{thm:GaugeTrafo}. Note that the \v Cech 1-cochain defined by 
\begin{equation}
 \Lambda_{ab}\ =\ \gamma''{}^1_{1,ab}-\gamma''{}^1_{0,ab}-\dd_\pi\sigma_2(\phi)
\end{equation}
on $U'_a\cap U'_b$ satisfies the equation
\begin{equation}
 \sff^2_{i*}|_{\sigma_2(\phi)}(\Lambda_{bc}-\Lambda_{ac}+\Lambda_{ab})\ =\ 0~,
\end{equation}
which follows from \eqref{eq:cocycle_cond_PW}. An analysis for the cocycle conditions at one level higher reveals that in fact
\begin{equation}
 \Lambda_{bc}-\Lambda_{ac}+\Lambda_{ab}\ =\ 0~.
\end{equation}
Thus, $\Lambda$ defines a \v Cech 1-cocycle with values in $\Omega^1_{\pi_1}\otimes\phi^*\sLie_3(\CCG)$. The cohomology group\linebreak $H^1(F^{9|8n},\Omega^1_{\pi_1}\otimes \phi^*\sLie_{3}(\CCG))$, however, vanishes as is implied by the discussion of  \cite{Saemann:2011nb,Saemann:2012uq}, see also \cite{Mason:2011nw}. We may therefore split the cocycle into a cochain $\lambda$:
\begin{equation}\label{eq:splitting_PW}
 \Lambda_{ab}\ =\ \Lambda_a-\Lambda_b~.
\end{equation}
From
\begin{equation}
\sff^2_{2*}|_{\sigma_2(\phi)}(\Lambda_{ab})\ =\ \alpha''_{1,a}~,~~~\sff^2_{1*}|_{\sigma_2(\phi)}(\Lambda_{ab})\ =\ 0\eand \sff^2_{0*}|_{\sigma_2(\phi)}(\Lambda_{ab})\ =\ -\alpha''_{1,b}~,
\end{equation}
we conclude that 
\begin{equation}
\sff^2_{2*}|_{\sigma_2(\phi)}(\Lambda_{a})\ =\ \alpha''_{1,a}+\zeta_2~,~~~\sff^2_{1*}|_{\sigma_2(\phi)}(\Lambda_{a})\ =\ \zeta_1\eand \sff^2_{2*}|_{\sigma_2(\phi)}(\Lambda_{a})\ =\ \zeta_0~,
\end{equation}
where $\zeta_i$ are elements of $H^0(F^{9|8n},\Omega^1_{\pi_1}\otimes \phi^*\sLie_{3}(\CCG))$. The forms $\zeta_{1,2}$ are now redundancies in the splitting \eqref{eq:splitting_PW}, but the form $\zeta_0$ is a globally defined 1-form $\alpha'''_{1}\in H^0(F^{9|8n},\Omega^1_{\pi_1}\otimes \phi^*\sLie_p(\CCG))$. The $\Lambda_a$ can also be regarded as gauge parameters for a global gauge trivial 2-form potential and therefore define a global 2-form $\alpha'''_{2}\in H^0(F^{9|8n},\Omega^1_{\pi_1}\otimes \phi^*\sLie_p(\CCG))$. For a more detailed discussion of this point, see \cite[Section 3.2]{Saemann:2012uq}. By construction, these forms together with $\phi$ define a flat holomorphic relative connective structure, that is,
 \begin{equation}\label{eq:RelFlatEq}
\begin{aligned}
 \dd_{\pi_1}\phi+\mu_1(\alpha'''_1)\ &=\ 0~,\\
  \dd_{\pi_1}\alpha'''_1+\mu_1(\alpha'''_2)+\tfrac12\mu_2(\alpha'''_1,\alpha'''_1)\ &=\ 0~,\\
  \dd_{\pi_1}\alpha'''_2+\mu_2(\alpha'''_1,\alpha'''_2)+\tfrac{1}{3!}\mu_3(\alpha'''_1,\alpha'''_1,\alpha'''_1)\ &=\ 0~.
 \end{aligned}
\end{equation}
Note that the products $\mu_1$ and $\mu_2$ both depend on $\phi$.

It is clear now how this generalises for $2<k<\infty$. One studies the compatibility conditions for the faces of  $\gamma''{}^{k-1}_{i,a_0\cdots a_{k-1}}$ on $U'_{a_0}\cap\cdots\cap U'_{a_{k}}$. Then, one finds linear combination of the corresponding simplices that form elements of the cohomology groups  $H^k(F^{9|8n},\Omega^{k-1}_{\pi_1}\otimes \phi^*\sLie_{k+1}(\CCG))$. This cohomology group vanishes for any $k\geq1$, and the cocycles can be split into coboundaries, which defined global $p$-forms. As before, the transition functions \eqref{eq:TrivBunGauge}  remain untouched by the splitting.
  
(ii) $\to$ (iii) Firstly, note that $\phi: F^{9|8n}\to \CCG_0$ is a global holomorphic map and since $\PP^3$ is compact it does not depend on the coordinates $\lambda_A$ and thus can be viewed as a global holomorphic map $\phi:M^{6|8n}\to \CCG_0$. The rest then follows straightforwardly from \eqref{eq:RelFlatEq} (and similarly for $k>2$) together with the expansions displayed in Lemma \ref{lem:RelFormsExp} (and similarly for $k>2$). The constraint $ \Omega_{IJ}\phi^{IJ}=0$ follows after some algebraic manipulations from the purely Gra{\ss}mann-odd Bianchi identity of the 3-form $H$. 

(iii) $\to$ (ii) $\to$ (i) Having found a solution to the constraint equations (e.g.~\eqref{eq:ConstrainEq} for $k=2$), one can always construct global relative gauge potentials by means of the expansions in Lemma \ref{lem:RelFormsExp} (and likewise for higher differential forms). These, in turn, define a flat holomorphic relative connection for a trivial holomorphic principal $\CCG$-bundle on the correspondence space $F^{9|8n}$. Next, we use a generalisation of the higher non-Abelian Poincar\'e lemma for strict Lie 2-groups \cite{Demessie:2014ewa} to gauge away the flat relative connection, which leaves us with a topologically and holomorphically trivial principal $\CCG$-bundle on $F^{9|8n}$ whose transition functions are constant along the fibres of $\pi_1$. Therefore, this bundle descends to an $M^{6|8n}$-trivial principal $\CCG$-bundle on $P^{6|2n}$. \hfill $\Box$

\acknowledgements

We would like to thank James Grant, Patricia Ritter, Urs Schreiber, and Danny Stevenson for fruitful discussions and comments, and we are grateful to Severin Bunk for commenting on a first iteration of Section \ref{sec:SimpMan}. We also thank the many contributors of the \href{http://www.ncatlab.org}{\tt nLab} for sharing their knowledge and ideas. B.J. expresses his gratitude to the Max Planck Institute for Mathematics in Bonn and the Tohoku Forum for Creativity for their kind hospitality. B.J.~and C.S.~would like to thank the organisers of the workshop {\it Higher Structures in String Theory} at the at the Erwin Schr{\"o}dinger International Institute for Mathematical Physics and the institute for hospitality.  B.J. was supported by the grant GA\v CR P201/12/G028. B.J.~and M.W.~acknowledge support from the LMS by a Research in Pairs Grant (Scheme 4).  C.S.~was supported in part by the STFC Consolidated Grant ST/L000334/1 {\it Particle Theory at the Higgs Centre}. M.W.~was supported in part by the STFC Consolidated Grant ST/L000490/1 {\it Fundamental Implications of Fields, Strings, and Gravity}. This work was supported in part by the Action MP1405 QSPACE from COST.

\datamanagement

No additional research data beyond the data presented and cited in this work are needed to validate the research findings in this work.

\appendices

\subsection{The Duskin nerve of a weak 2-category}\label{app:Duskin_nerve}
 
Quasi-categories form models for higher categories whose $n$-morphisms for $n \geq 2$ are equivalences in the sense that the natural nerves for those higher categories are indeed quasi-categories. While a category can be reconstructed from its nerve, these higher categories can only be recovered up to equivalences. To illustrate this point, we briefly sketch the example of weak 2-categories.

Recall that a weak 2-category or bicategory $\CB$ consists of a set $\CB_0$ of objects, together with categories of morphisms $\CB_1(\alpha,\beta)$ for each pair of objects $\alpha,\beta\in \CB_0$. Objects and morphisms in the categories $\CB_1(\alpha,\beta)$ are called 1-morphisms and 2-morphisms, respectively. There is a well-defined functorial composition of 1-morphisms and 2-morphisms in $\CB_1(\alpha,\beta)$ and $\CB_1(\beta,\gamma)$ for $\alpha,\beta,\gamma\in \CB_0$ known as the horizontal composition and denoted by $\otimes$. Additional natural isomorphisms describe the failure of this composition to be associative and left- and right-unital. In addition, two 2-morphisms of $\CB_1(\alpha,\beta)$ can also be composed, which is called vertical composition and denoted by $\circ$. See, e.g.~\cite{Jurco:2014mva} for a detailed definition of weak 2-categories, functors between them, and some of their applications in the context of self-dual higher gauge theory. A reasonable notion of the nerve of a weak 2-category is then the following, cf.~\cite[Section 6]{Duskin02simplicialmatrices}:

\begin{definition}
The \uline{Duskin nerve}  $N(\CB)$ of a weak 2-category $\CB$ is the simplicial set $N(\CB):=\bigcup_{p\in\NN_0} N_p(\CB)$ with $N_{p}(\CB):=\CatFun_{2,{\rm nlax}}([p],\CB)$, where $\CatFun_{2,{\rm nlax}}$ is the 2-category of normalised lax 2-functors.
\end{definition}

\noindent 
Explicitly, the 0-simplices are given by the objects of $\CB$, and the 1-simplices are the 1-morphisms of $\CB$ with evident face maps. The 2-simplices of the nerve consist of 2-morphisms $x_{\alpha\beta\gamma}:x_{\alpha\beta}\otimes x_{\beta\gamma}\Rightarrow x_{\alpha\gamma}$ whose faces are $(x_{\beta\gamma},x_{\alpha\gamma},x_{\alpha\beta})$. The 3-simplices are then given by natural isomorphisms including the associators and the unitors of the weak 2-category.

For an ordinary category $\CCC$ regarded as a weak 2-category $\CB$, $\CatFun_{2,{\rm nlax}}([p],\CB)=\CatFun([p],\CCC)$ and the Duskin nerve reduces to the ordinary nerve. 

The reason for the restriction to weak 2-categories in which all 2-morphisms are isomorphisms is now easy to understand. Given an inner 2-horn consisting of two 1-simplices, we have a priori many horn fillers, but not a unique one. In particular, we can reconstruct the horizontal composition of the 1-morphisms only up to equivalent choices. Using 3-horn fillers, one can show that any two 2-horn fillers are related by a (unique) 2-simplex or 2-morphism. We therefore have to declare all horn fillers isomorphic, which, in turn, implies that all 2-morphisms in the weak 2-category are isomorphisms.

The following statement, which characterises a quasi-category which is the Duskin nerve of a weak 2-category, is now very plausible.

\begin{prop}[\!\!{\cite[Theorem 8.6]{Duskin02simplicialmatrices}}]\label{prop:Duskin}
A simplicial set $\CCX$ is the Duskin nerve of a weak 2-category in which all 2-morphisms are isomorphisms if and only if $\CCX$ is a 2-quasi-category.  A simplicial set $\CCX$ is the Duskin nerve of a weak 2-groupoid  if and only if $\CCX$ is a 2-quasi-groupoid.   
\end{prop}

%\bibliographystyle{latexeu}
%\bibliography{references}
%\bibliography{refs}

\end{document}